\documentclass{jfm}

\usepackage{graphicx}
\graphicspath{{./pictures/}}
\usepackage{placeins}
\usepackage{xcolor}
\usepackage{newtxtext}
\usepackage{newtxmath}
\usepackage{natbib}
\usepackage{hyperref}
\hypersetup{
    colorlinks = false,
    urlcolor   = black,
    citecolor  = black,
}

\newcommand{\RomanNumeralCaps}[1]
\linenumbers

\usepackage{tikz,xcolor,hyperref}

\definecolor{lime}{HTML}{A6CE39}
\DeclareRobustCommand{\orcidicon}{
	\begin{tikzpicture}
	\draw[lime, fill=lime] (0,0) 
	circle [radius=0.16] 
	node[white] {{\fontfamily{qag}\selectfont \tiny ID}};
	\draw[white, fill=white] (-0.0625,0.095) 
	circle [radius=0.007];
	\end{tikzpicture}
	\hspace{-2mm}
}
	
\foreach \x in {A, ..., Z}{\expandafter\xdef\csname orcid\x\endcsname{\noexpand\href{https://orcid.org/\csname orcidauthor\x\endcsname}
			{\noexpand\orcidicon}}
}

\title{Dynamics and fluid-structure interaction in turbulent flows within and above flexible canopies}

\author{Giulio Foggi Rota\aff{1}, Alessandro Monti\aff{1}, Stefano Olivieri\aff{1,2} and Marco Edoardo Rosti\aff{1}\corresp{\email{marco.rosti@oist.jp}}}

\affiliation{
\aff{1}Complex Fluids and Flows Unit, Okinawa Institute of Science and Technology Graduate University (OIST), 1919-1 Tancha, Onna-son, Okinawa 904-0495, Japan
\aff{2}Currently at Universidad Carlos III de Madrid, Leganés, 28911, Madrid, Spain
}

\begin{document}
\maketitle

\begin{abstract}
Flexible canopy flows are often encountered in natural scenarios, e.g., when crops sway in the wind or when submerged kelp forests are agitated by marine currents.
Here, we provide a detailed characterisation of the turbulent flow developed above and between the flexible filaments of a fully submerged dense canopy and we describe their dynamical response to the turbulent forcing.
We investigate a wide range of flexibilities, encompassing the case in which the filaments are completely rigid and standing upright as well as that where they are fully compliant to the flow and deflected in the streamwise direction.
We are thus able to isolate the effect of the canopy flexibility on the drag and on the inner-outer flow interactions, as well as the two flapping regimes of the filaments already identified for a single fiber.
Furthermore, we offer a detailed description of the Reynolds stresses throughout the wall-normal direction resorting to the Lumley triangle formalism, and we show the multi-layer nature of turbulence inside and above the canopy.
The relevance of our investigation is thus twofold: the fundamental physical understanding developed here paves the way towards the investigation of more complex and realistic scenarios, while we also provide a thorough characterisation of the turbulent state that can prove useful in the development of accurate turbulence models for RANS and LES.

\end{abstract}

\section{Introduction}
\label{sec:intro}

Canopy flows often occur in nature, when a wall-bounded flow interacts with a multitude of slender objects protruding from a supporting surface.
In the atmospheric boundary layer, various types of obstacles arranged in different patterns (e.g., threes in forests, plants in cultivated fields, wind turbines in wind farms) are exposed to surface winds and significantly alter its dynamics.
As outlined by \cite{belcher-harman-finnigan-2012}, forests play a fundamental role in promoting turbulence and enhancing mixing. They also shade the surface of the Earth and favour the vertical transport of multiple species through the lower layers of the atmosphere, affecting the surface ozone levels \citep{makar-etal-2017}. 
Noticeably, the complex updraft generated by the canopy promotes seeds dispersal \citep{quin-etal-2022}, thus regulating the distribution of vegetation.
In water, marine currents frequently interact with seagrass meadows \citep{mossa-etal-2017} and different animal furs are associated to different swimming performances \citep{bushnell-moore-1991}.
Furthermore, from an anatomical perspective, mucus is transported by ciliated surfaces in the bronchial epithelium \citep{loiseau-etal-2020}, intestinal villi are responsible for the absorption of nutrients in the body, and cells or small invertebrates often employ cilia to propel themselves \citep{dauptain-favier-bottaro-2008}. 
The study of canopy flows is therefore motivated by their ubiquity and by the number of nodal functions they absolve to. 
While in this work we tackle the problem from a fundamental perspective, the way to multiple engineering applications is being paved.
For example, \cite{wang-etal-2022-2} show how meta-surfaces covered in cilia can be employed for microfluidics manipulation, while \cite{zhu-etal-2022} consider the use of submerged canopies for the purpose of costal protection, based on their ability to affect the movement of sediments \citep{nepf-2012-2,zhao-nepf-2021}.
As noted by \cite{luhar-rominger-nepf-2008}, dense meadows can promote sediment retention, stabilising the bed and promoting their own persistence. Conversely, a reduction in canopy density leads to increased flow and stress near the bed, which can lead to further canopy deterioration.
Unravelling the complex dynamics of the flow and of the canopy elements can therefore not only offer relevant insight on multiple natural phenomena, but also provide solid grounds for innovative engineering solutions.

The chaotic motion of air in and above plant canopies has been systematically investigated and modelled from the second half of the 20\textsuperscript{th} century. 
In their seminal review, \cite{raupach-thom-1981} laid solid foundations for the study of turbulence and transport in canopy flows, providing a first characterisation of those phenomena and reviewing the most successful approaches for the prediction of the mean flow and turbulence intensity.
An exhaustive description of the mean flow features and a detailed characterisation of the key turbulent quantities was later offered by \cite{finnigan-2000}. \cite{finnigan-2000} also introduced a phenomenological model for the sustainment of turbulence, reliant on  the inviscid instability of the shear layer generated by the drag discontinuity at the tip of a dense submerged canopy.
The relevance of such shear layer was first observed by \cite{raupach-finnigan-brunei-1996}, while its peculiar nature was successively highlighted by \cite{ghisalberti-nepf-2004}, who noted that it does not grow continuously downstream as a free shear layer, but rather reaches a finite thickness set by the rate of momentum exchange between the flow and the solid structures.
Notwithstanding this noticeable difference, it is yet passible of a Kelvin-Helmholtz like instability which induces the formation of elongated spanwise vortices (``rollers'') controlling the exchange of mass and momentum between the canopy and the outer flow \citep{nepf-2012-1, chowdhuri-ghannam-banerjee-2022}. The further instability of those rollers is held responsible for  the formation of secondary vortices \citep{finnigan-shaw-patton-2009}, organised in trains of head-up and head-down hairpins aligned with the flow, causing intense sweeps often observed to penetrate the canopy. 
The slow fluid drawn from inside the canopy in between the legs of the hairpins, instead, gives rise to elongated regions of low streamwise velocity close to the canopy tip. 
The rollers, hairpins and velocity streaks generated by the shear layer instability dictate the structure of turbulence throughout the flow, as happens also in highly permeable porous media \cite{manes-poggi-ridolfi-2011}.

Alongside the theoretical approach, there has been a flourishing of models to predict various quantities of interest based on experimental measurements. In particular, the canopy drag coefficient is relevant in most engineering applications and exhibits a direct dependence from the Reynolds and the Froude numbers, as demonstrated by \cite{liu-zeng-2016}, \cite{mossa-etal-2021} and \cite{rubol-ling-battiato-2018}, who further investigated its dependence from the canopy permeability.
Additional models for other flow quantities are summarised by \cite{brunet-2020}, who offers a physical overview and presents a historical summary of the evolution of the field. 
Recently, \cite{vieira-allshouse-mahadevan-2023} have introduced a simplified model of the unidirectional flow over a canopy capable of accounting for the shear layer instability above its tip, while \cite{condefrias-etal-2023} have developed an experimentally-validated approach to predict the boundary layer thickness at the seafloor based upon its negative correlation with the turbulent kinetic energy within the layer. 

A radical distinction can be made between canopies constituted by practically rigid elements and flexible ones. The study of the flow within an array of rigid pillars is in facts a purely fluid-dynamical problem, while the flow over and between the flexible filaments of an hairy surface is characterised by the complex interaction between the fluid and the structure.
Historically, experiments have been able to tackle both cases from the very beginning of the field, while simulations have mainly focused on rigid canopies. The first computations accounting for the fully coupled dynamics of a turbulent flow with an array of flexible elements have only recently made their appearance due to their outstanding computational cost \citep{tschingale-etal-2021,he-liu-shen-2022,wang-etal-2022-1,lohrer-frohlich-2023,monti-olivieri-rosti-2023}. 

In the case of rigid slender elements (either clamped or freely dispersed in a turbulent flow), the most immediate effect is a modification of the classical energy cascade.
\cite{olivieri-etal-2020-2} observed in homogeneous isotropic turbulence how the fibres remove energy from the largest eddies and divert it towards finer ones via a ``spectral short cut''  mechanism, first proposed by \cite{finnigan-2000}. Large-scale mixing is therefore depleted in favour of the small-scale one.
On top of this mechanism, the flow is modified by the canopy according to the ``tightness of its packing'' (i.e., the solidity, \cite{luhar-rominger-nepf-2008, monti-etal-2022, nicholas-etal-2023}), ranging from the sparse to the dense regime. However, the solidity constitutes a non-exhaustive parameter for the characterisation of a turbulent canopy flow, as it does not account for the orientation of the filaments (which can significantly affect the resulting flow) and outer quantities would provide a better alternative \citep{monti-etal-2022}. Nevertheless, we stick to it for the time being due to historical reasons.
\cite{sharma-garciamayoral-2018} investigated the flow over a sparse rigid canopy by means of direct numerical simulations. They noted that a sparse canopy does not significantly disturb the near-wall turbulence cycle, but causes its rescaling to an intensity consistent with a lower friction velocity within the canopy: an effect similar to that of $k$-type roughness. They also found evidences of the formation of Kelvin–Helmholtz like instabilities at the canopy tip.
The large-eddy simulations performed by \cite{monti-omidyeganeh-pinelli-2019} in the marginally dense regime confirmed the existence of spanwise rollers generated by the shear layer at the canopy tip and highlighted how those are modulated by outer streamwise vortices penetrating the canopy.
The effects of the spacing between the elements and their height were assessed by \cite{monti-etal-2020} and \cite{sharma-garciamayoral-2020-2} for a dense canopy, where drag sets the shape of the mean velocity profile and is held responsible for the inviscid shear-layer instability at the tip. 
The intense Kelvin–Helmholtz like instability also dominates within the canopy, projecting its footprint, while the outer flow resembles those attained on top of rough walls and densely packed porous media, extensively discussed in literature \citep{jimenez-2004,wood-he-apte-2020}.
Remarkably, notwithstanding the strong anisotropy of the medium, the experiments of \cite{shnapp-etal-2020} highlighted how short-time Lagrangian statistics remain quasi-homogeneous due to the intense dissipation associated to the turbulent fluctuations. 
The last regime to be tackled numerically and, arguably, the most challenging was the transitional one, where physical characteristics unique to the sparse and dense scenarios coexist. That was investigated by \cite{monti-etal-2020} and \cite{nicholas-omidyeganeh-pinelli-2022}. 

The picture becomes more complex in the case of flexible slender elements, as their structural dynamics needs to be accounted for, and a significant portion of the turbulent kinetic energy is generated by the ``waving contribution" originated from the correlation between the hydrodynamic drag and the waving motion of the stems \citep{he-liu-shen-2022}. 
\cite{jin-ji-chamorro-2016} laid the foundations for understanding the dynamical response of flexible and slender elements forced by a turbulent flow.
On top of that, \cite{rosti-etal-2018} developed a phenomenological theory to describe the dynamics of free flexible fibres in homogeneous isotropic turbulence and validated it numerically. They identified two regimes of motion: one in which the fibres are slaved to the turbulent fluctuations of the flow and one in which they exhibit their natural response; \cite{olivieri-mazzino-rosti-2021} later supported the result with a wider set of numerical simulations.
We confirmed the existence of those two regimes also in the case of a clamped flexible fibre in wall turbulence \citep{foggirota-etal-2024}, nevertheless noticing a significant difference in the dominant oscillation frequency for the turbulence dominated regime. 
In this case, in facts, the flapping state of the fiber non-trivially relates to the largest scale of the flow and not to the turbulent eddies of comparable size, as found by \cite{olivieri-mazzino-rosti-2021}.
In the case of a flexible canopy, the individual dynamics of the elements is altered \citep{fu-etal-2023} and a collective dynamics (honami/monami) emerges on top of that, as measured experimentally by \cite{py-delangre-moulia-2006} in the case of a crop field driven by the wind.
The state-of-the-art direct numerical simulations (DNS) performed by \cite{monti-olivieri-rosti-2023} shed light on the topic, highlighting how such collective motion is decoupled from the structural natural response of the filaments and independent from their rigidity, being driven by the turbulent fluctuations of the flow only. 
Such causal dependence is coherent with the previous results of the large eddy simulations from \cite{tschingale-etal-2021} and \cite{wang-etal-2022-1}.

\begin{figure}
\centering
\includegraphics[width=\textwidth]{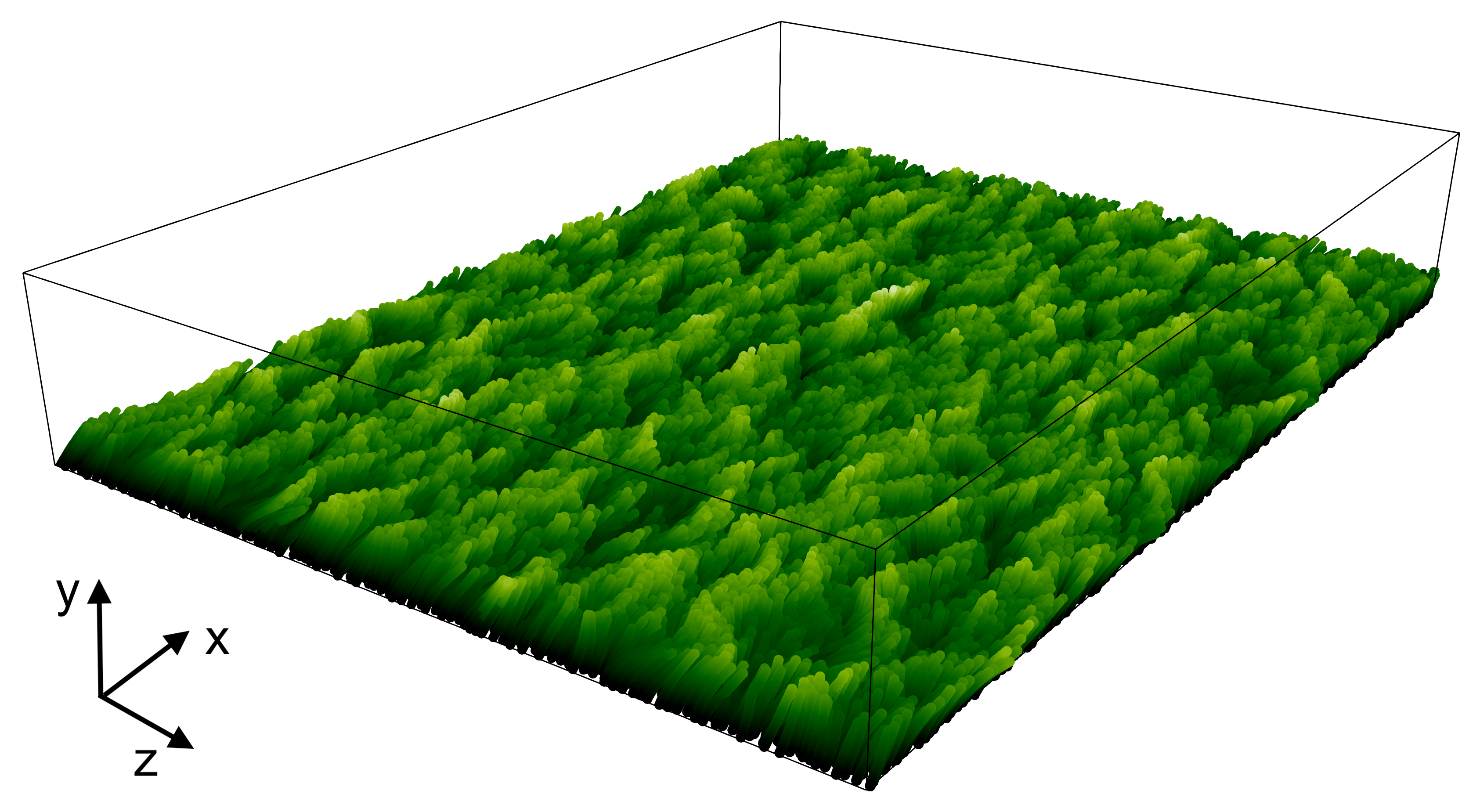}
\caption{Representation of our computational domain, with the empty region occupied by the fluid and the flexible filaments constituting the canopy coloured in shades of green, varying from dark to light with the elevation. The mean flow is aligned with the $x$ axis, while the $y$ axis corresponds to the wall-normal direction.} 
\label{fig:domain}
\end{figure}

In this work we investigate the turbulent motion of the fluid and the dynamics of the flexible filaments constituting a dense submerged canopy. We build on top of an extended version of the DNS database generated by \cite{monti-olivieri-rosti-2023}, where each filament is modelled individually by means of an immersed boundary method \citep[first introduced by][]{goldstein-handler-sirovich-1993} in the Lagrangian formulation described by \cite{yu-2005} and \cite{huang-etal-2007}. Consistently with such approach, the filaments are therefore modelled as inextensible beams.
We provide a detailed characterisation of the flow above and within the canopy, as well as describing the interaction between the two and comparing them to experimental measurements. We also assess the consequences on the flow of a variation in the filament density and analyse the effects induced by their motion. 
From a structural standpoint, we characterise the individual motion of the filaments in the canopy and reconcile it with the picture we previously described for single isolated filaments \citep{foggirota-etal-2024}. 
The remaining part of this paper is organised as follows: in \S\ref{sec:setupnmethods} we describe the setup of our simulations and the numerical methods employed. The dynamics of the flow is characterised in \S\ref{sec:fluiddyn} where, after reporting relevant mean quantities (\S\ref{sec:meanflow}) and comparing them to available experimental results (\S\ref{sec:experiments}), we investigate the energy spectra (\S\ref{sec:spectra}) and the Lumley triangle (\S\ref{sec:lumley}) throughout the whole domain. We further explore the interaction between the inner and the outer flow by carrying out a quadrant analysis (\S\ref{sec:quadrant}) at the canopy tip. 
The consequences on the flow of a variation in the density of the filaments are also assessed (\S\ref{sec:drho}) and the effects of their flapping motion are explored by ``freezing" them in their instantaneous deflected configuration (\S\ref{sec:freeze}).
In \S\ref{sec:fildyn} we characterise the structural dynamics of the filaments by analysing their individual motion (\S\ref{sec:filmotion}) and describing their flapping state (\S\ref{sec:fmap}).
Finally, in \S\ref{sec:conclusions}, we summarise the main outcomes of our investigation and critically discuss their implications, with few remarks on future developments.

\section{Setup and methods}
\label{sec:setupnmethods}

Our simulations are carried out in an open channel described by means of a right handed Cartesian reference frame with the $x$ axis oriented along the streamwise direction and the $y$ axis perpendicular to the bottom wall. 
The computational domain is therefore a box of volume $L_x\times L_y\times L_z=2\pi H \times H \times 1.5\pi H$, where we enforce the no-slip and no-penetration boundary conditions at the bottom face; no-penetration and stress-free conditions are instead imposed at the top face in the same fashion of \cite{calmet-magnaudet-2003}, while the wall-parallel directions are treated as periodic. 
The stems of the flexible filaments constituting the canopy are vertically clamped to the bottom wall and protrude upward into the flow as in figure \ref{fig:domain}.
To discretise the fluid flow we adopt an Eulerian grid made of $N_x\times N_y\times N_z=1152\times 384\times 864$ points homogeneously distributed along the periodic directions, while a non homogeneous stretched distribution is adopted along the $y$ axis in order to correctly capture the sharp variation of the velocity at the canopy tip. In particular, we employ a finer and locally uniform resolution in the region containing the canopy, achieving a constant wall-normal spacing $\Delta y/H = 0.002$ for $y/H \in [0.0,0.3]$, and smoothly transition to a wider wall-normal spacing above that, attaining $\Delta y/H = 0.004$ at $y/H =1$.

We consider the motion of an incompressible Newtonian fluid, described by the mass (\ref{eq:mass}) and momentum (\ref{eq:momentum}) balances.
Denoting with $\mathbf{u}(\mathbf{x},t)$ and $p(\mathbf{x},t)$ the velocity and pressure fields, both function of the spatial coordinates $\mathbf{x}$ and time $t$, with $\rho _f$ the volumetric fluid density and with $\nu$ the kinematic viscosity, the governing equations are
\begin{eqnarray}
   &\nabla \cdot \mathbf{u} = 0,
   \label{eq:mass}\\
   &\displaystyle \frac{\partial \mathbf{u}}{\partial t} + \nabla \cdot (\mathbf{u} \mathbf{u}) = - \frac{1}{\rho_f} \nabla p + \nu \nabla^2 \mathbf{u} + \mathbf{f}_{\mathbf{\rm{fib}}} + \mathbf{f}_{\mathbf{\rm{for}}},
   \label{eq:momentum}
\end{eqnarray}
where two forcing terms have been introduced: $\mathbf{f}_{\mathbf{\rm{fib}}}$, better defined in the following,  is the force field acting on the fluid computed with a Lagrangian immersed boundary method (IBM) \citep{peskin-2002, huang-etal-2007, banaei-rosti-brandt-2020,olivieri-etal-2020-2} to account for the presence of the filaments in the fluid, while $\mathbf{f}_{\mathbf{\rm{for}}}$ is the homogeneous force field equally applied to all grid points to attain at every time instant the desired flow rate in the streamwise direction. 
Averaging the streamwise velocity over the domain volume $V$, we define the mean velocity $\hat{U}=\frac{1}{V}\iiint_{V} u dV$.
In order to attain the desired value of the mean velocity $U_b$, here set to unity for simplicity, the term $\mathbf{f}_{\mathbf{\rm{for}}}$ is computed as $\mathbf{f}_{\mathbf{\rm{for}}}=[(U_b-\hat{U})/dt]\mathbf{\hat{e}_x}$, where $dt$ is the time step of the simulation and $\mathbf{\hat{e}_x}$ is the versor denoting the streamwise direction. 
This kind of forcing is customary in the DNS of turbulent channel flows, as discussed in literature \citep{hasegawa-quadrio-frohnapfel-2014}.
The value of the bulk Reynolds number $Re_b=U_b H/\nu$ is thus constant and imposed by an appropriate choice of the kinematic viscosity of the fluid $\nu$. Here, we set $Re_b=5000$ in order to ensure a fully developed turbulent flow above the canopy.
The balance equations, along with the set of boundary conditions highlighted above, constitute a well posed problem that we tackle numerically by means of our in-house solver, \textit{Fujin} (\href{https://groups.oist.jp/cffu/code}{https://groups.oist.jp/cffu/code}). 
We adopt second order central finite differences to discretise the velocity and the pressure on a staggered Cartesian grid, resorting to a second order Adams-Bashforth scheme for time stepping within a projection-correction approach \citep{kim-moin-1985}.
The Poisson equation is efficiently solved with a Fast Fourier Transform (FFT) based algorithm \citep{dorr-1970} and the whole code is parallelised using the Message Passing Interface (MPI) and the \textit{2decomp} library.

The canopy is constituted by 15552 filaments of length $h = 0.25 H$ and diameter $d \approx 2\cdot10^{-2} H$, placed in a semi-random arrangement to avoid preferential flow channeling effects.
This value of $d$ yields a local Reynolds number $Re_d=U_{mc} d/\nu\approx10$ based on the mean velocity scale $U_{mc}$ below the canopy tip $(y_{out})$, $U_{mc}=({1}/{y_{out}})\int_0^{y_{out}}{\langle u \rangle}dy$, with $\langle u \rangle$ the mean streamwise velocity profile.
To arrange the filaments, we divide the bottom wall of the channel into a grid of $n_x \times n_z = 144 \times 108$ rectangular tiles of area $\Delta S^2=(L_x/n_x)\times(L_z/n_z)$, and randomly place each of them within each tile sampling a uniform distribution. 
This tiling is not the numerical grid, and it is employed only to achieve the desired distribution of the filaments while maintaining control over the canopy parameters.
We thus ensure a nominal solidity value of $\lambda = h d / \Delta S^2 \approx 1.43$, laying well within the dense canopy regime \citep{monti-etal-2020}.
Different canopies are produced on varying the Cauchy number, $Ca = (\rho_f d h^3 U_b^2)/(2\gamma)$ (representing the ratio between the deforming force exerted by the fluid and the elastic restoring force opposed by the filaments) and the volume density ratio between the filaments and the fluid, $\rho_s/\rho_f$.
Here, $\gamma$ is the bending rigidity of the filaments, given by the product of the bending modulus with the moment of inertia of the filament cross section.
Our study considers seven canopies characterised by $\rho_s/\rho_f=1.0+1.46\cdot10^{-3}$ and spanning $Ca\in\{ 0,1,10,25,50,100,500\}$.
Nonetheless, $Ca=500$ is only referred to when investigating the dynamics of the filaments (\S\ref{sec:fildyn}), as the dynamics of the fluid (\S\ref{sec:fluiddyn}) shows minimal changes compared to $Ca=100$. 
For few specific purposes discussed in \S\ref{sec:drho}, we also consider two additional canopies with a different density ratio, $\rho_s/\rho_f\in\{1.0+1.46\cdot10^{-1},1.0+1.46\cdot10^{-2}\}$ at $Ca=25$.

The filaments are represented as mono-dimensional entities, discretised into a line of Lagrangian points, that obey a generalisation of the Euler-Bernoulli beam model allowing for finite deflections, but retaining the inextensibility constraint. In the rigid canopy case (i.e., $Ca=0$) each filament is made of $n_L=81$ Lagrangian points, attaining a spatial resolution $\Delta s = h/(n_L-1)$ comparable to the Eulerian grid spacing in the wall-normal direction. 
In the flexible canopy cases, such spatial resolution would impose a too strict constraint on the time step and we therefore reduce $n_L$ to 32: this proves acceptable as the flexibility makes the filaments more compliant to the flow and the velocity difference between the two phases is therefore reduced. 
Different discretisations were tested by \cite{monti-olivieri-rosti-2023} and \cite{foggirota-etal-2024}, without significant variations in the filament dynamics for the parameters considered here.
We use the same approach by \cite{banaei-rosti-brandt-2020} to model the dynamics of flexible and inextensible filaments. It consists of an extended version of the distributed-Lagrange-multiplier/fictitious-domain (DLM/FD) formulation of the continuum equations introduced by \cite{yu-2005}.
Denoting with $\mathbf{X}(s,t)$ the position of a point on the neutral axis of a filament as a function of the curvilinear abscissa $s$ and time $t$, and introducing the linear density difference between the filament and the fluid $ \Delta \Tilde{\rho} = (\rho_s-\rho_f)\pi d^2/4 $, its structural dynamics is described by 
\begin{eqnarray}
    &\Delta \Tilde{\rho} \displaystyle \frac{\partial^2 \mathbf{X}}{\partial t^2} = 
    \displaystyle \frac{\partial}{\partial s}\left(T\displaystyle \frac{\partial \mathbf{X}}{\partial s}\right) - \gamma \displaystyle \frac{\partial^4 \mathbf{X}}{\partial s^4} - \mathbf{F}, 
    \label{eq:eulerBernoulli}\\
    &\displaystyle \frac{\partial \mathbf{X}}{\partial s} \cdot \displaystyle \frac{\partial \mathbf{X}}{\partial s} = 1,
    \label{eq:inextensibility}
\end{eqnarray}
where $T$ is the tension enforcing the inextensibility and $\mathbf{F}$ is the force acting on the filaments computed by the Lagrangian IBM to couple them with the fluid, as described later.
We complement equations (\ref{eq:eulerBernoulli},\ref{eq:inextensibility}) with an appropriate set of boundary conditions, imposing $\mathbf{X}\rvert_{s=0}=\mathbf{X_0}$ along with ${\partial\mathbf{X}}/{\partial s}\rvert_{s=0}=(0,1,0)$ at the clamp and ${\partial^3\mathbf{X}}/{\partial s^3}\rvert_{s=h}={\partial^2\mathbf{X}}/{\partial s^2}\rvert_{s=h}=\mathbf{0}$ along with $T\rvert_{s=h}=0$ at the free end, and solve them following the approach of \cite{huang-etal-2007}. 
Nevertheless, here, the bending term is treated implicitly as in \cite{banaei-rosti-brandt-2020} to allow for a larger time step. 
The set of Lagrangian equations introduced above, in the absence of any external forcing, is passible of a normal mode analysis yielding the natural frequency $f_{nat}=({\beta_1}/{(2\pi h^2)})\sqrt{{\gamma}/{\tilde{\rho_s}}}$, related to the natural pulsation $\omega_1$ by $f_{nat}={\omega_1}/{(2\pi)}$.
$\tilde{\rho_s}$ is the filament density per unit length, while $\beta_1$ is a coefficient approximately equal to $3.516$, determined through the analysis.
Writing $\tilde{\rho_s}=\rho_s \pi d^2 /4$, where $\rho_s$ is the filament density per unit volume, there follows $f_{nat}\approx({3.516}/{(d h^2)})\sqrt{{\gamma}/{(\rho_s \pi^3)}}$. 
$f_{nat}$, as noticed by \cite{foggirota-etal-2024}, plays a significant role in determining the dynamical response of the filaments to the fluid.
As in most cases the filaments are flexible and swaying in the flow, they might collide with the wall and with other filaments.
We have thus implemented filament-to-filament and filament-to-wall collision models to prevent the stems from crossing each other or the wall while deforming \citep{snook-guazzelli-butler-2012}.
Nevertheless, after the extensive testing of different collision models and of their calibration parameters conducted in previous investigations \citep{monti-olivieri-rosti-2023}, the influence of the filament-to-filament collision term on both the filament and the fluid dynamics was found to be very weak, whereas the filament-to-wall interaction model turned out to be necessary only to correctly describe the dynamics of the most flexible filaments, at large values of $Ca$.
We thus resort to an inelastic collision model, applying a repulsive force to all the filament points approaching the wall within a range of four grid points.

The coupling between the fluid and the structure is attained spreading over the Eulerian grid points the force distribution computed by means of the Lagrangian IBM, ensuring the no-slip condition $\partial\mathbf{X}/\partial t=\mathbf{u}[\mathbf{X}(s,t),t]$ at the Lagrangian points representing the filaments.
The intensity of the force $\mathbf{F}$ exerted by the fluid on the structure is proportional to the difference among the velocity of the structure and that of the fluid interpolated at the structure points, $\mathbf{u}_{\mathbf{\rm{IBM}}}$. We therefore write $\mathbf{F}= \beta \left(\mathbf{u}_{\mathbf{\rm{IBM}}} - {\partial \mathbf{X}}/{\partial t}\right)$, where $\beta$ is a properly tuned coefficient here set equal to $10$.
Finally, $\mathbf{F}$ is spread to the nearby grid points in order to compute the back-reaction on the fluid, $\mathbf{f}_{\mathbf{\rm{fib}}}=\displaystyle\int_\Gamma \mathbf{F}_{\mathbf{\rm{IBM}}}(s,t)\delta(\mathbf{x}-\mathbf{X}(s,t))ds$, with $\Gamma$ the support of the IBM. 
The interface between the fluid and the filaments is therefore not sharply captured, but spread over the support of the IBM through the action of a window function, which determines the diameter of the filaments.

The ability of our numerical setup and methods to correctly describe the dynamics of a whole submerged canopy made of flexible filaments, without introducing spurious effects due to the finite size of the domain or the grid, is extensively assessed in \cite{monti-olivieri-rosti-2023}. The length of the domain along the homogeneous directions is sufficient to contain the largest turbulent flow structures \citep{bailey-stoll-2013}. The vertical size of the domain, instead, is most likely affecting the results.
Our simulations, in facts, aim at investigating a submerged canopy rather than a canopy exposed to a boundary layer, for which a significantly higher domain would be needed.
In the numerical simulation of submerged canopies, instead,  it is  customary not to simulate the fluid interface far above their tip, but rather to approximate it as a free slip surface \citep{sharma-garciamayoral-2018,tschingale-etal-2021, wang-etal-2022-1,he-liu-shen-2022,lohrer-frohlich-2023}; its distance from the bottom wall thus becomes a parameter of the simulations. 
In this work, we follow such practice.

\begin{figure}
\centering
\includegraphics[width=.95\textwidth]{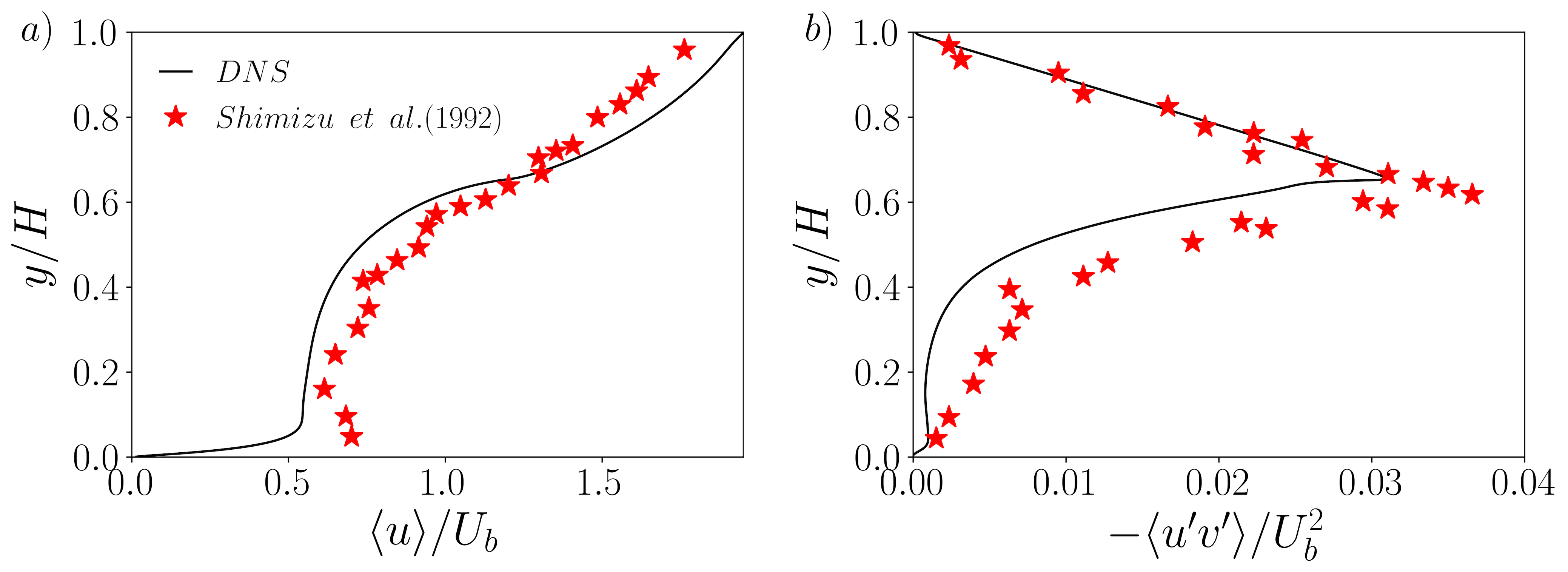}
\caption{Mean streamwise velocity profile (panel $a$) and Reynolds shear stress (panel $b$) in and above a rigid canopy with $h=0.65H$ and $\lambda=0.41$, at $Re_b=7070$. Red stars denote the experimental measurements of \cite{shimizu-etal-1992}, while black lines are the outcome of a direct numerical simulation matching the experimental parameters, performed with our code.} 
\label{fig:expProf}
\end{figure}

Canopy flows have been investigated experimentally since the origins of the field, and a broad range of measurements is now available \citep[e.g.,][]{gao-shaw-paw-1989,okamoto-nezu-2009,nicolai-etal-2020}.
Nevertheless, the comparison to simulation results is often hindered by the inevitable differences between numerical and experimental setups, especially in the canopy arrangement and flow parameters.
Experiments are often performed at significantly higher Reynolds numbers than simulations, and they are frequently characterised by the presence of secondary flows. 
Despite these difficulties, we found that our setup compares well to that adopted by \cite{shimizu-etal-1992}, where a rigid canopy of height $h=0.65H$ and solidity $\lambda=0.41$ is exposed to a turbulent channel flow at $Re_b=7070$.
Therefore, after purposely simulating the flow within and over a rigid canopy matching the same parameters \citep[further details in the supplementary information of][]{monti-olivieri-rosti-2023}, we contrast the computed mean flow profile and Reynolds shear stress with their measurements in figure \ref{fig:expProf}.
This comparison confirms the ability of our code to correctly describe the back-reaction of the structure on the fluid.
\section{Dynamics of the fluid}
\label{sec:fluiddyn}

\subsection{Mean flow quantities}
\label{sec:meanflow}

To investigate the flow above and within the canopy for different values of $Ca$, we start looking at the mean profiles of the velocity and of the Reynolds stresses, along with relevant derived quantities.
In the following, averaging in time and along the homogeneous directions is denoted with angle brackets, while fluctuations with respect to such mean are marked with an apostrophe.
Our analysis is based on $100$ flow fields for each value of $Ca$, regularly collected over $50$ bulk time units. The flow statistics do not show any appreciable variation upon computation with half of the fields, thus confirming that they are converged.

\begin{figure}
\includegraphics[width=0.48\textwidth]{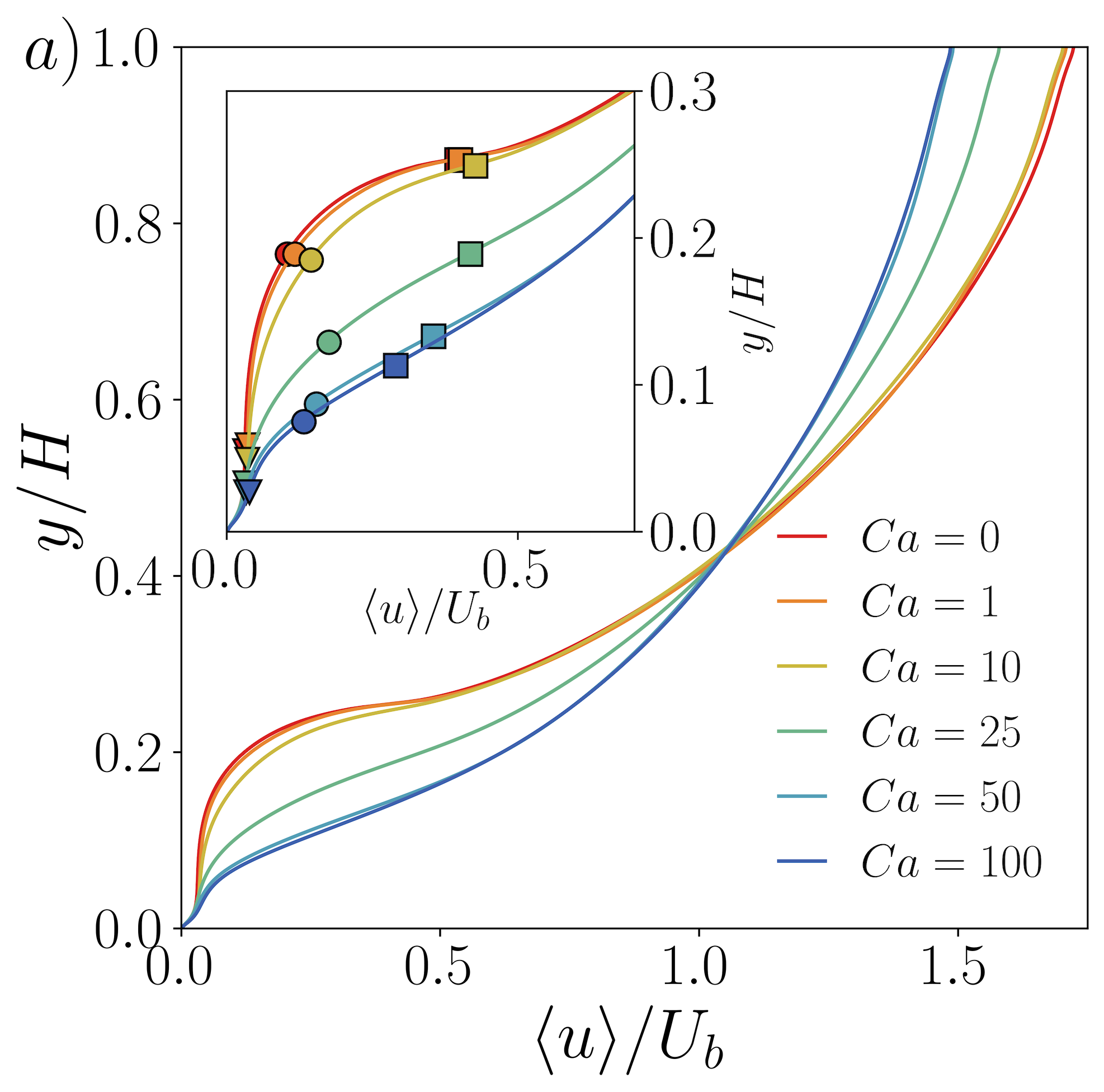}
\hfill
\includegraphics[width=0.48\textwidth]{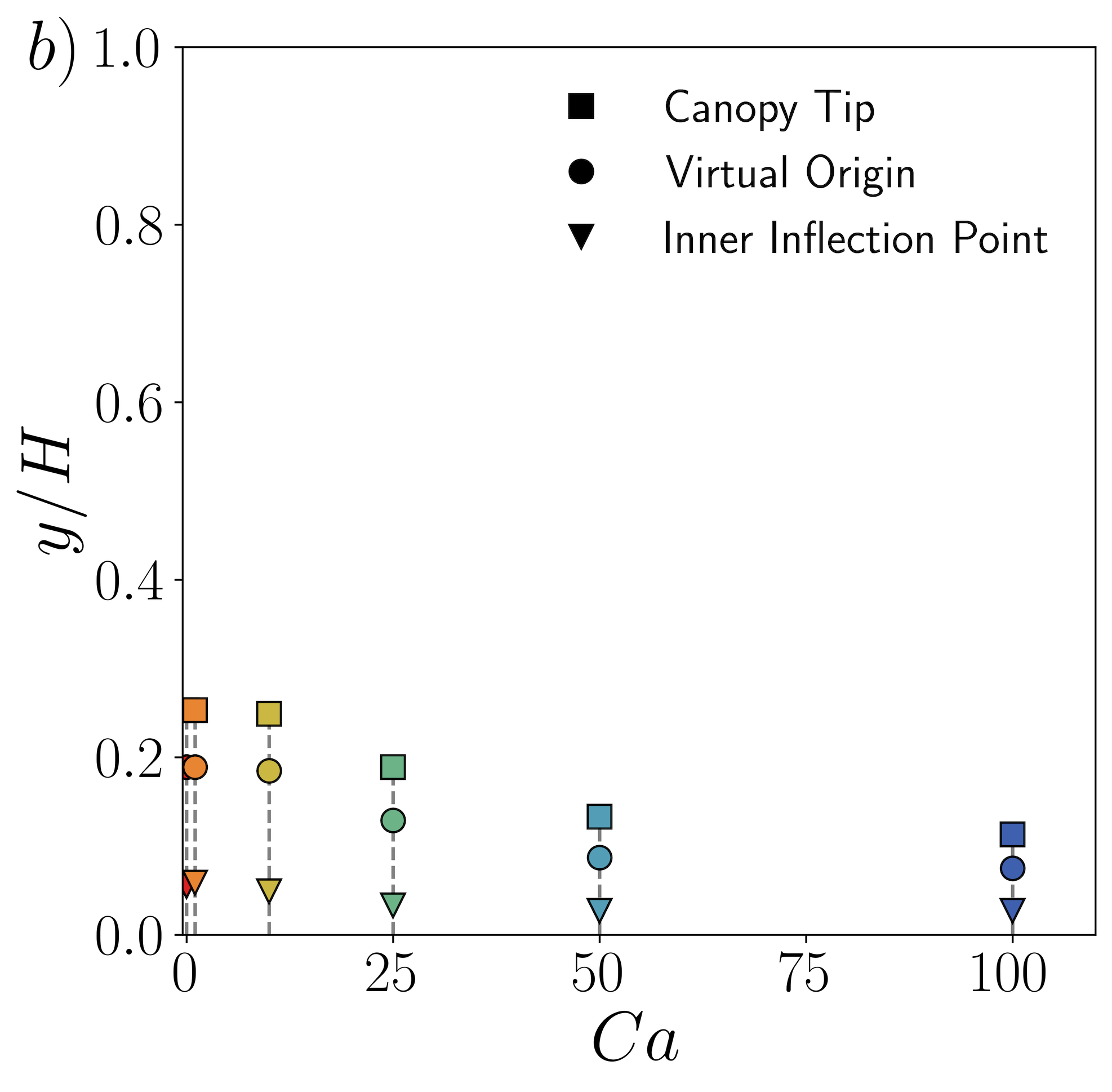}
\caption{Mean profiles of the streamwise velocity (panel $a$) for different values of $Ca$ and associated relevant points (inset and panel $b$). In panel $b$ we show the position of the relevant points for different values of $Ca$, maintaining the vertical scale unchanged with respect to that of panel $a$.}
\label{fig:meanVelProf}
\end{figure}

The mean profiles of the streamwise velocity $u$, shown in panel $a$ of figure \ref{fig:meanVelProf}, exhibit two distinct inflection points: an \textit{outer} inflection point ($y_{out}$, square symbol) generated by the drag discontinuity at the average canopy tip position and an \textit{inner} inflection point ($y_{in}$, triangle symbol) closer to the wall, where the inflected velocity profile connects to the wall boundary layer. 
Below the outer inflection point the \textit{inner} flow is reminiscent of that attained in an anisotropic porous medium \citep{rosti-brandt-pinelli-2018}, while immediately above that the \textit{outer} flow is similar to a turbulent mixing layer \citep{raupach-finnigan-brunei-1996}.
Yet, canopy flows can also be considered instances of obstructing substrates for the assessment of outer-layer similarity \citep{chen-garciamayoral-2023}. 
Fitting a logarithmic profile to the mean velocity well above the canopy is in facts a convenient numerical expedient to simplify its parametrisation for modelling purposes, even though it does not satisfactorily represent the physical features of the flow. 
We therefore compute the virtual origin of the outer flow, $y_{vo}$, imposing the matching with a canonical logarithmic profile and notice that, for all the cases of interest, it lays well between the two inflection points (panel $b$ of figure \ref{fig:meanVelProf}), thus confirming that we are in a \textit{dense} canopy regime \citep{monti-etal-2020,monti-etal-2022}.
The canopy becomes more compliant to the flow increasing $Ca$, hence the mean position of its tip as well as all the other relevant points are monotonously shifted downward, maintaining their relative order; the mean streamwise velocity at those points, instead, exhibits a non monotonous trend. 
At the canopy tip, in particular, it first undergoes a slight increase due to the reduction of the filaments drag, later decreasing again as the effect of the downward shift becomes dominant.
The positions of all points reach a plateau for high values of $Ca$, where the vertical stacking of the deflected filaments poses a lower bound to the thickness of the inner flow region.

\begin{figure}
\includegraphics[width=0.49\textwidth]{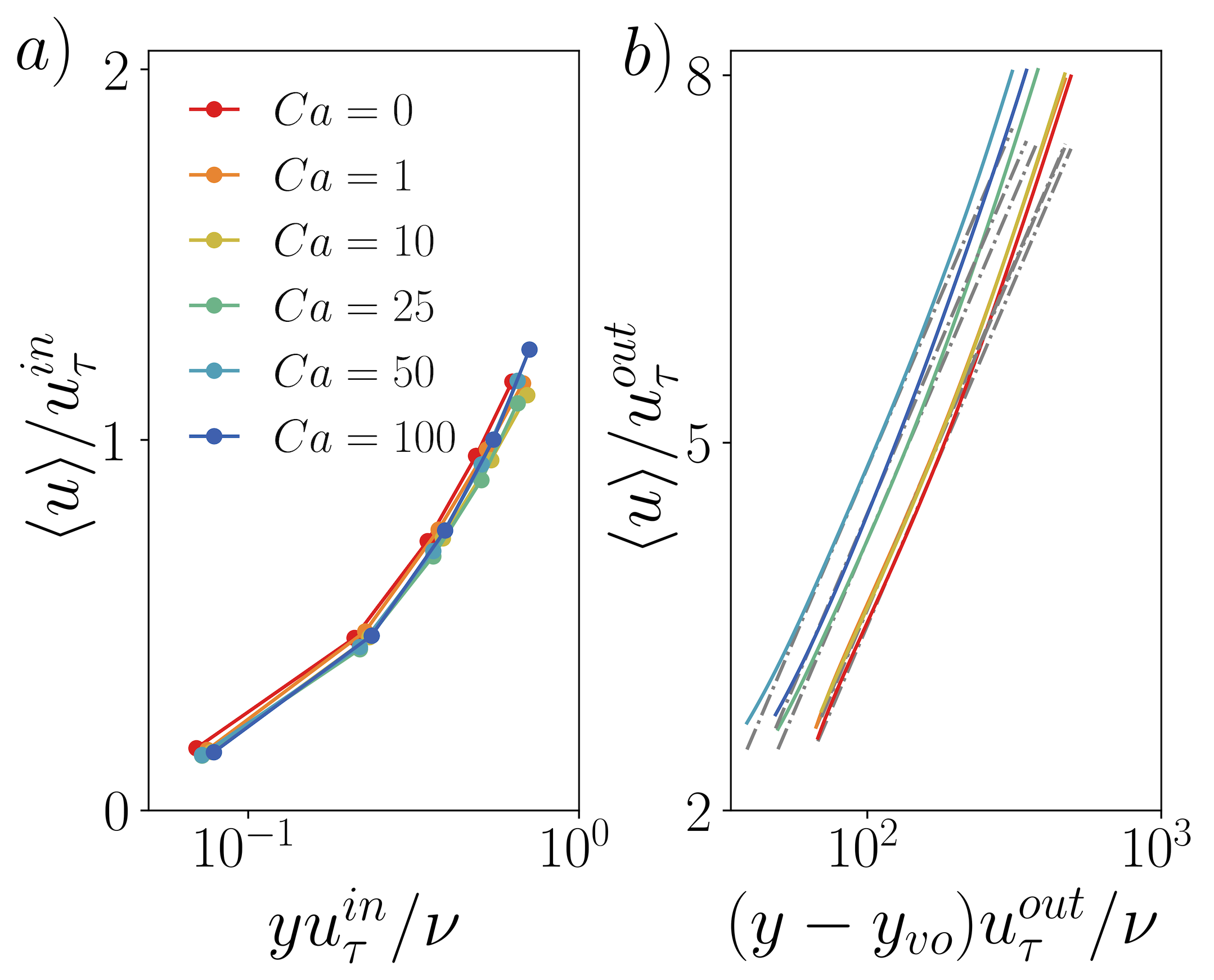}
\hfill
\includegraphics[width=0.47\textwidth]{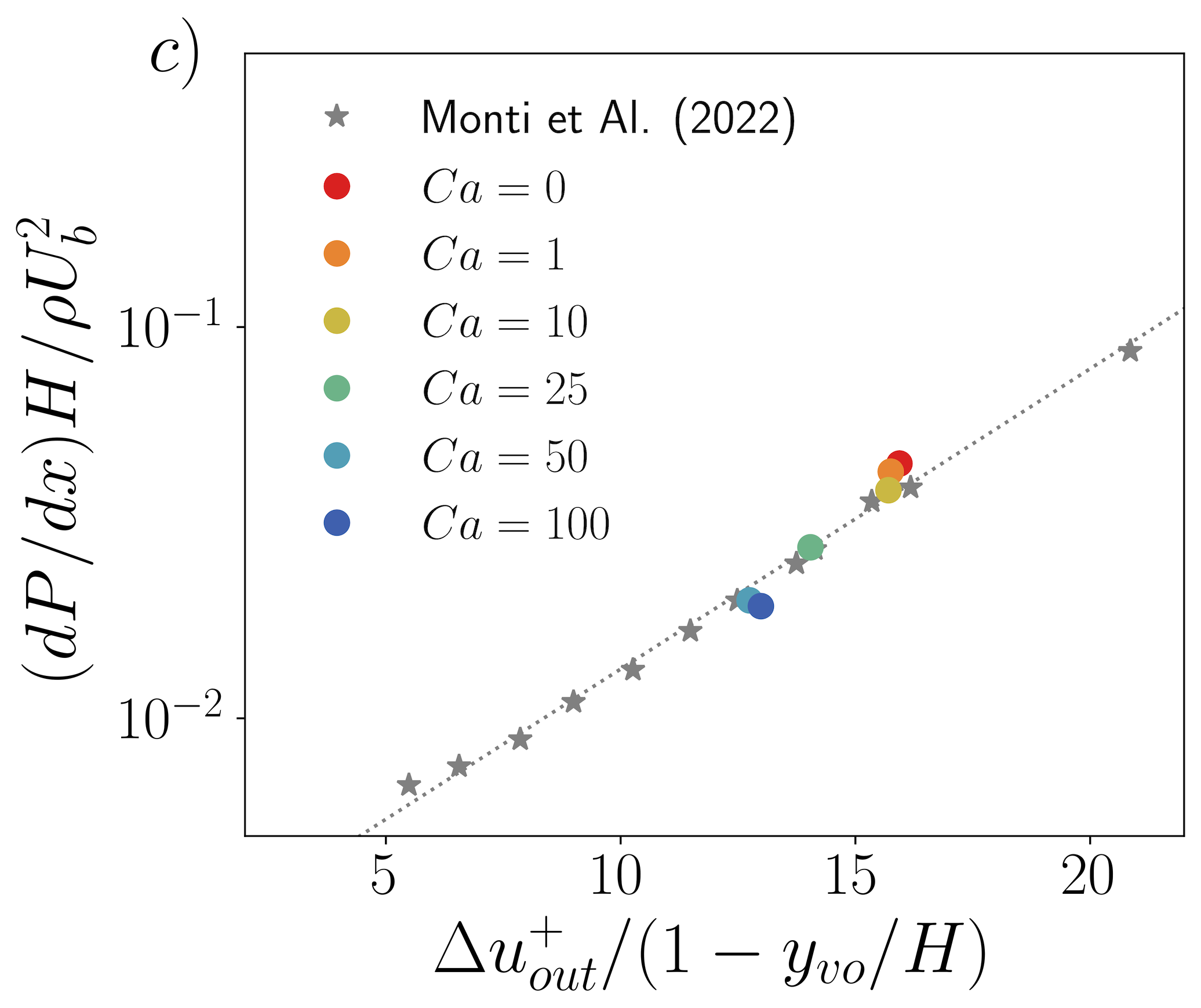}
\caption{Panel $a$ reports the first five computed points of the mean velocity profiles, made dimensionless upon the friction velocity at the wall, $u_{\tau}^{in}$, against the wall distance. Panel $b$, instead, reports the mean velocity profiles made dimensionless upon the friction velocity at the virtual origin, $u_{\tau}^{out}$, against the wall distance shifted by $y_{vo}$, within the region where the scaling holds. In panel $b$, for each case, we also report the logarithmic profile computed with equation \ref{eq:logLaw} as a dash-dotted gray line. The inner scaling yields good overlapping of the different profiles with a quasi-linear trend over the first grid points off the wall while, with the outer scaling, the profiles collapse on the analytical predictions. Finally, in panel $c$, we show the exponential trend of the friction function $\Delta u^{+}_{out}$ appearing in equation \ref{eq:logLaw} with respect to the driving pressure gradient, $\mathrm{d}P/\mathrm{d}x$, and compare it with the numerical data of \cite{monti-etal-2022}, who studied rigid canopies with different inclinations.}
\label{fig:velScaling}
\end{figure}

As expected, on scaling the velocity profile of the inner flow with the friction velocity computed at the wall, $u_{\tau}^{in}=\sqrt{\tau_w/\rho_f}=\sqrt{\nu \partial \langle u\rangle / \partial y \vert_0}$, the typical trend ${\langle u\rangle}/{u_{\tau}^{in}} = y u_{\tau}^{in} / \nu$ is recovered close to the bottom wall (panel $a$ of figure \ref{fig:velScaling}). 
Instead, on scaling the velocity profile of the outer flow with the friction velocity computed at the virtual origin, $u_{\tau}^{out}=\sqrt{\nu \partial \langle u\rangle / \partial y \vert_{y_{vo}}- \langle u'v' \rangle \vert_{y_{vo}}}$, we confirm good agreement with a logarithmic profile (panel $b$ of figure \ref{fig:velScaling}) of the form 
\begin{equation}
	\frac{\langle u\rangle}{u_{\tau}^{out}} = \frac{1}{\kappa}log\left( \frac{(y-y_{vo})u_{\tau}^{out}}{\nu}\right)+B-\Delta u^{+}_{out}
\label{eq:logLaw}
\end{equation}
where $\kappa=0.41$ and $B=5.2$, while $\Delta u^{+}_{out}$ denotes the friction function accounting for the mean velocity shift in the outer flow due to the presence of the canopy \citep[or wall roughness, as in][]{jimenez-2004}.
As noted by \cite{monti-etal-2022},  to whom we compare our results, $\Delta u^{+}_{out}$ exhibits an exponential trend with the driving pressure gradient $\mathrm{d}P/\mathrm{d}x$ (panel $c$ of figure  \ref{fig:velScaling}), once made dimensionless upon the height of the channel above the virtual origin. We acknowledge a minor deviation from their data, that we impute to the different shape of our filaments and to their movement.
Differently from our, in facts, the study of  \cite{monti-etal-2022} only concerned rigid canopies: 
in particular, at a chosen value of the Reynolds number ($Re_b=6000$, higher than ours), they varied the canopy solidity $\lambda$ by changing the inclination of the filaments.


\begin{figure}
\includegraphics[width=0.48\textwidth]{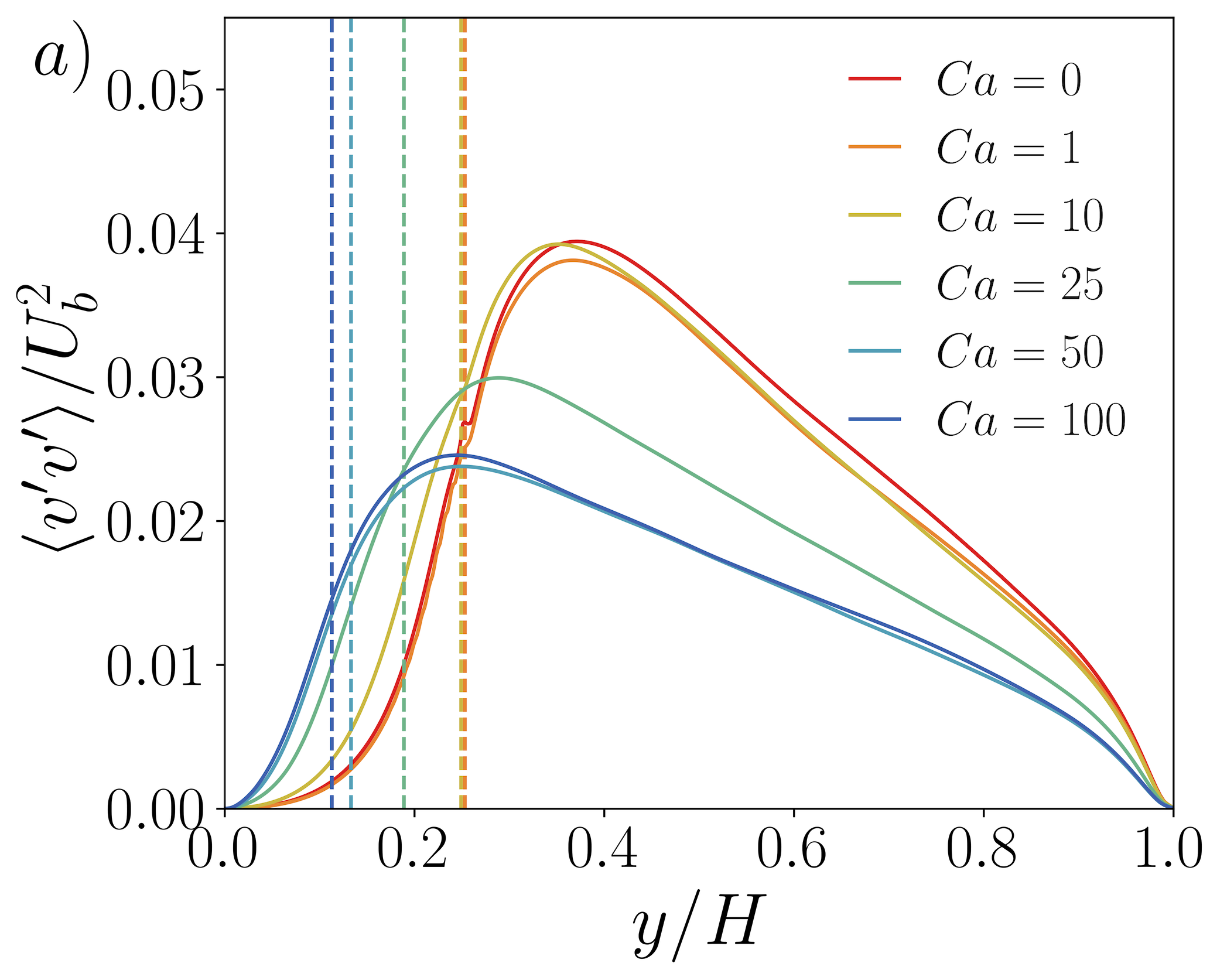}
\hfill
\includegraphics[width=0.48\textwidth]{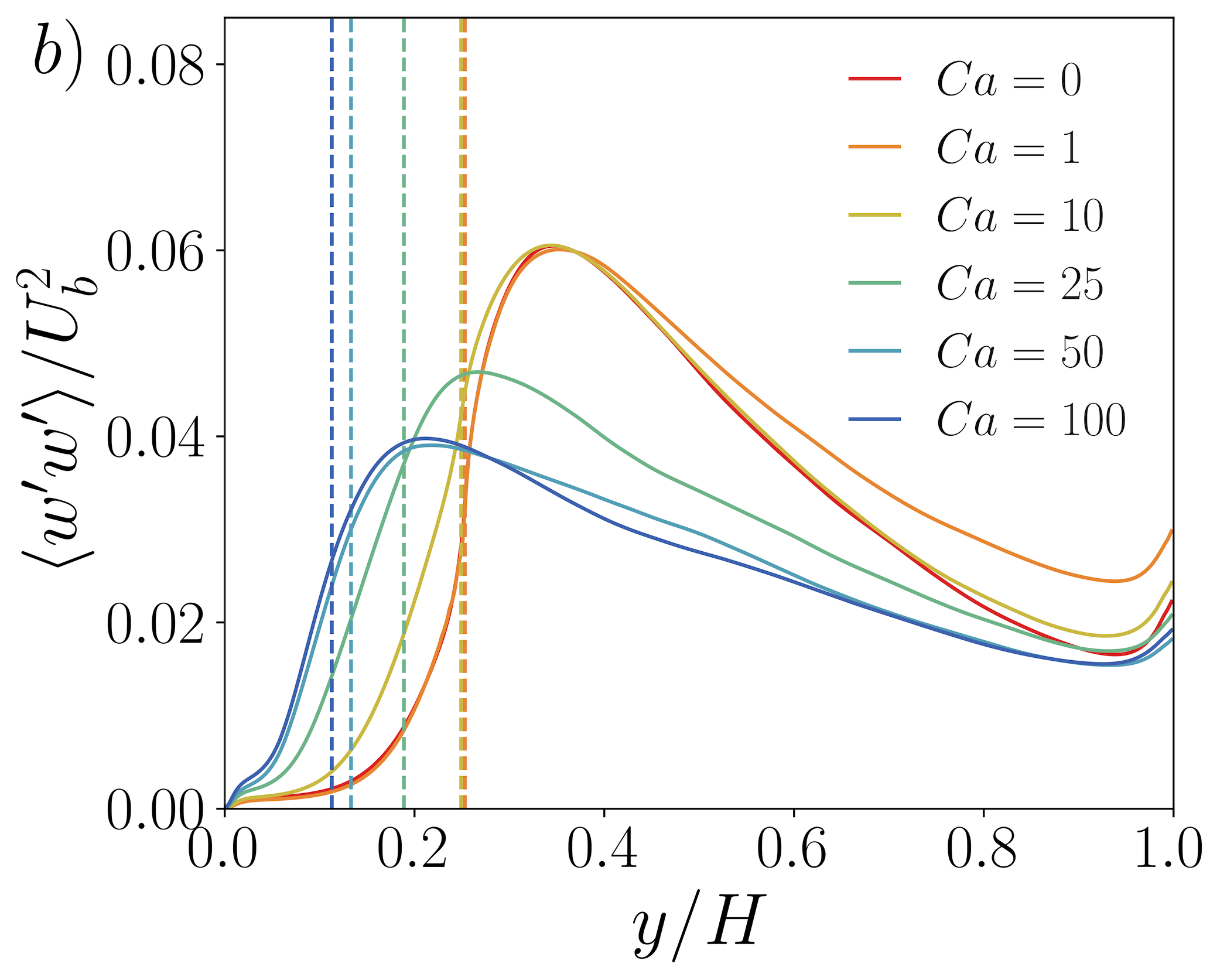}
\caption{Fluctuations of the wall normal (panel $a$) and spanwise (panel $b$) velocity components for different values of $Ca$. The positions of the canopy tip (identified with the outer inflection point) are denoted by vertical dashed lines.}
\label{fig:vv&ww}
\end{figure}

\begin{figure}
\includegraphics[width=0.48\textwidth]{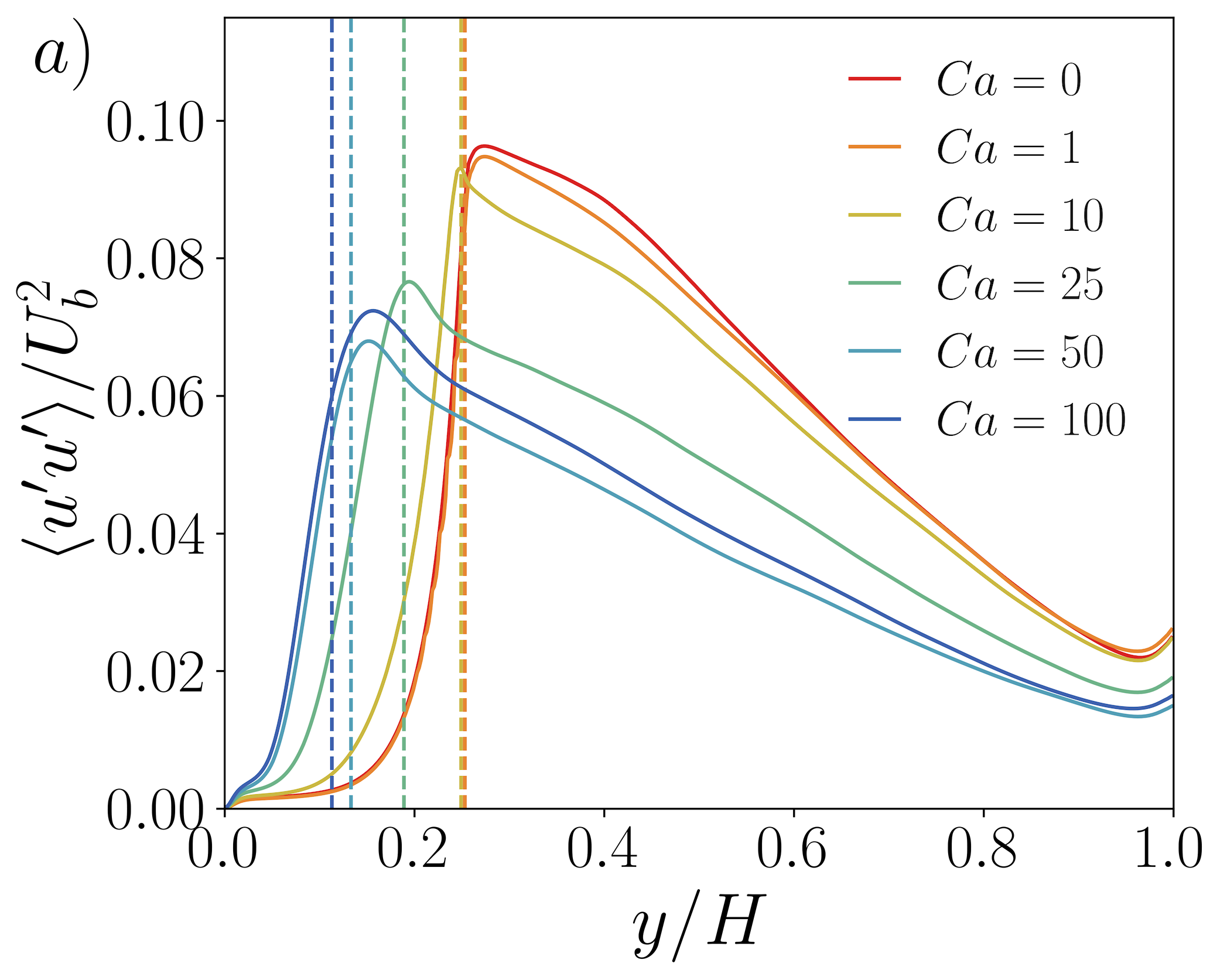}
\hfill
\includegraphics[width=0.48\textwidth]{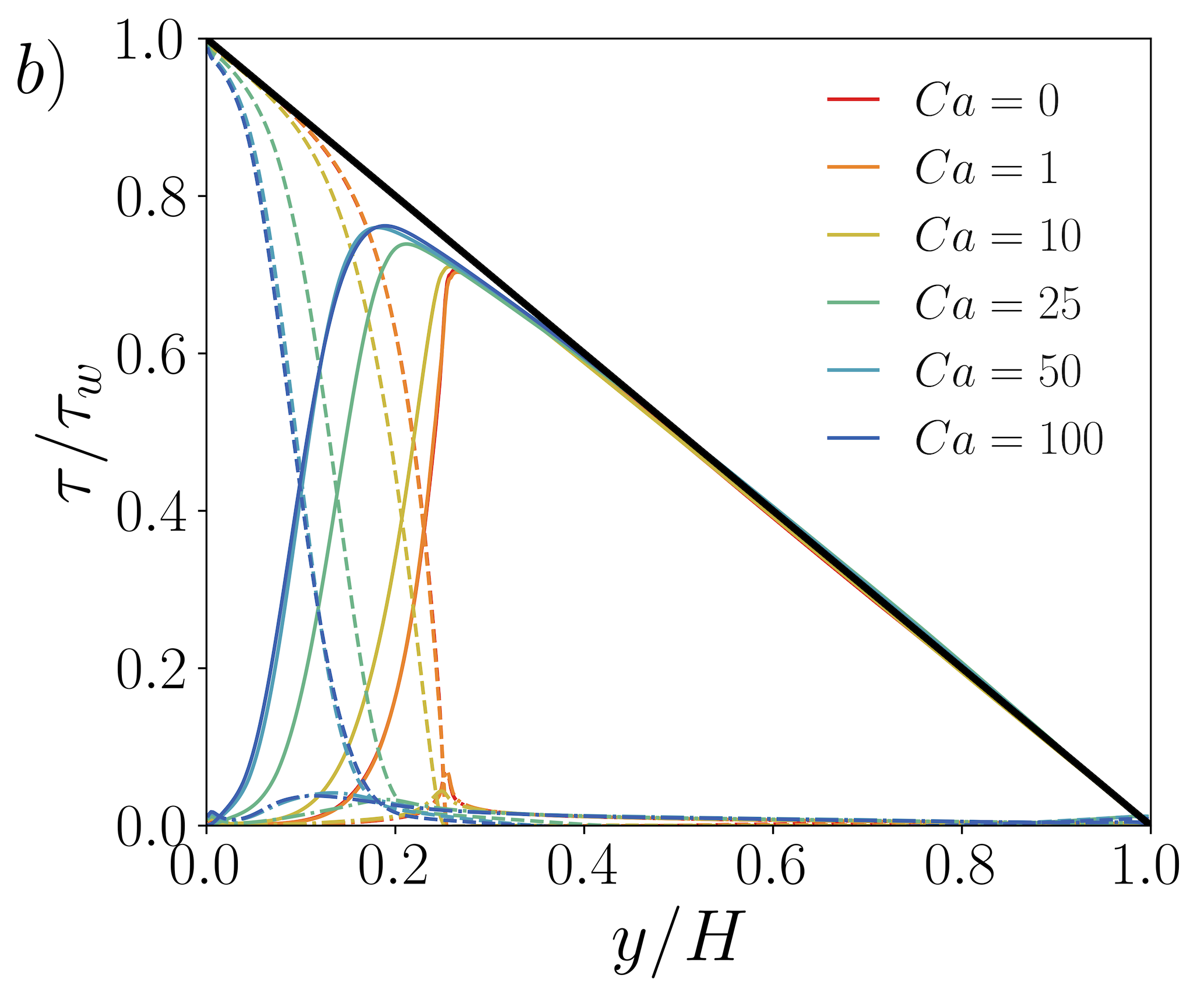}
\caption{Fluctuations of the streamwise velocity component (panel $a$) and shear stress balance (panel $b$) for different values of $Ca$. In panel $a$, the positions of the canopy tip (identified with the outer inflection point) are denoted by vertical dashed lines. In panel $b$, the total shear stress (black line) normalised by the wall shear stress is given by the sum of the turbulent shear stress (continuous lines), the viscous shear stress (dash-dotted lines) and the canopy drag (dashed lines), as described in the main text.}
\label{fig:uu&tau}
\end{figure}

The diagonal components of the Reynolds stress tensor (in figure \ref{fig:vv&ww} and \ref{fig:uu&tau}$a$) increase monotonously moving away from the wall in the canopy region, and peak at a position variable with $Ca$.
They therefore decrease towards the centre of the channel until the no-penetration condition becomes relevant and damps the fluctuations of the wall normal velocity,  enhancing those of the wall parallel components because of continuity.
All the peaks move towards the wall as $Ca$ increases and the filaments get more deflected by the action of the fluid, but the effect appears to saturate for the highest values of $Ca$.
The fluctuations of the wall normal and spanwise velocity components, shown in figure \ref{fig:vv&ww}, always reach their peak value above the canopy, highlighting the intense turbulent activity caused by the unstable shear layer at the drag discontinuity.
The maximum in the fluctuations of the streamwise velocity (reported in panel $a$ of figure \ref{fig:uu&tau}), instead, lays close to the canopy tip for the lowest values of $Ca$ and moves slightly above that for the highest ones. This picture is compatible with the existence of high and low streamwise velocity regions at the canopy tip, induced by the overlying Kelvin-Helmholtz like instability.
Those velocity structures alternatively deflect the filaments and penetrate the upper region of the canopy; nevertheless, for the most flexible cases, the filaments are significantly bent forward and therefore shield the inner flow, causing a slight shift up of those structures with respect to the canopy tip.

\begin{figure}
\includegraphics[width=\textwidth]{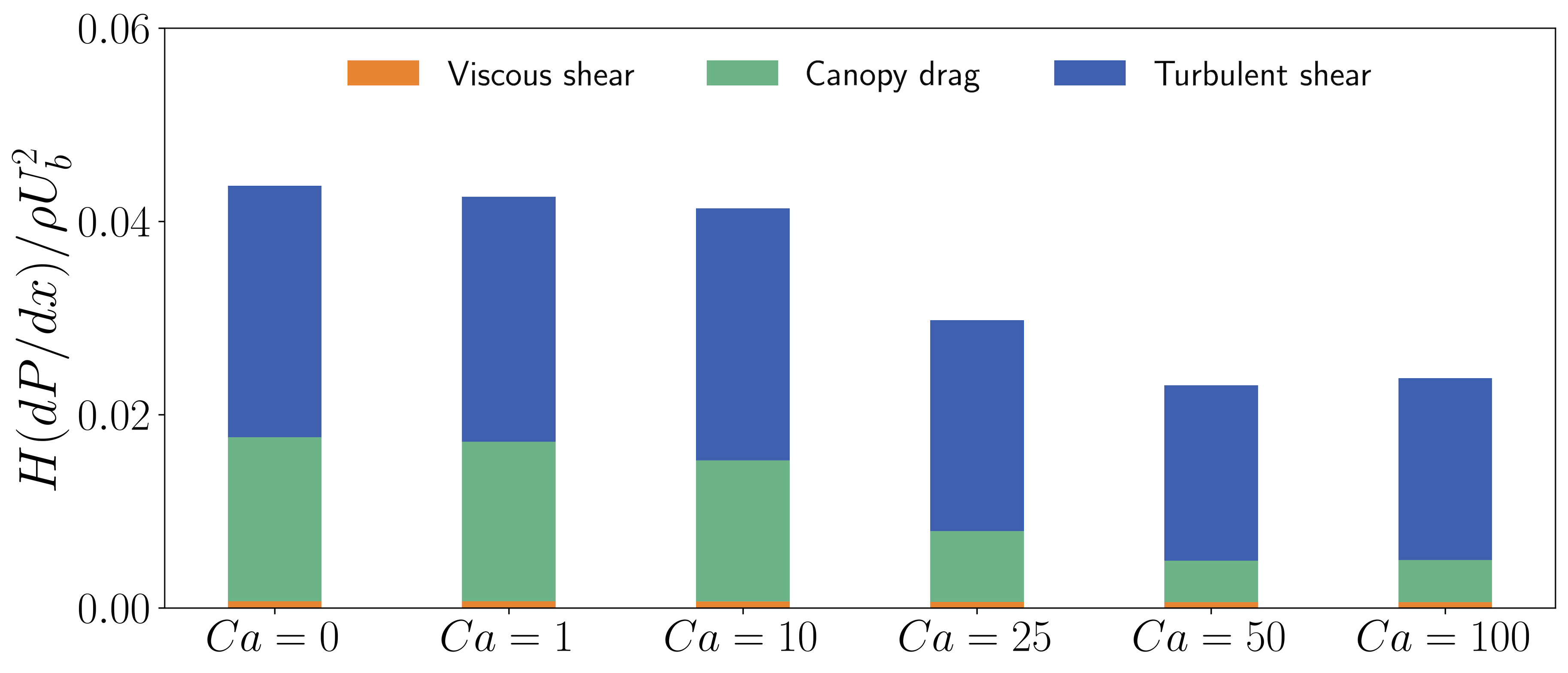}
\caption{Decomposition of the driving streamwise pressure gradient $\mathrm{d}P/\mathrm{d}x$ into the contributions of the viscous and turbulent shear stresses along with the canopy drag, integrated across the wall normal direction, for different values of $Ca$.}  
\label{fig:tauHist}
\end{figure}
The streamwise mean momentum equation imposes 
\begin{equation}
	\mathrm{d}\tau/\mathrm{d}y = \mathrm{d}P/\mathrm{d}x
\label{eq:shearBalance}
\end{equation}
where the total shear stress writes $\tau=\rho\nu \mathrm{d}\langle u\rangle /\mathrm{d}y - \rho \langle u'v'\rangle + D_c$, where $D_c$ is the canopy drag.
Denoting with $\tau_w$ the total shear stress at the wall, the sum of the three components constituting $\tau$ is therefore constrained by $\tau(y)=\tau_w(1-y/H)$.
Observing the separate contributions to $\tau$ normalised by $\tau_w$, in panel $b$ of figure \ref{fig:uu&tau}, we notice that the viscous shear stress ($\rho\nu \mathrm{d}\langle u\rangle /\mathrm{d}y$, plotted with dash-dotted lines) has two local maxima, one at the wall and one at the canopy tip, while it is negligible elsewhere. Inside the canopy, the stress is dominated by the drag contribution $D_c$ (plotted with dashed lines), which vanishes when moving out of it, giving way to the turbulent shear stress ($- \rho \langle u'v'\rangle$, plotted with continuous lines), which dominates the outer region as in conventional turbulent channel flows.
We once again notice how the transition from the inner to the outer regions occurs closer to the wall for growing values of $Ca$ due to the deflection of the filaments; nevertheless, the variation of their flexibility affects also the intensity of the shear layer above them. 
The viscous shear stress exhibits a sharp peak at the canopy tip for the lowest values of $Ca$, while for the highest ones the peak is less pronounced and spans a wider vertical span.
Indeed in this case, the shear layer is less definite and penetrates more into the canopy due to the filaments motion. 
Integrating equation \ref{eq:shearBalance} in the wall normal direction, we notice that the sum of the different integral components of the shear stress decreases as $Ca$ increases and plateaus for the most flexible cases (as in figure \ref{fig:tauHist}), always matching the streamwise pressure gradient needed to drive the flow at a constant value of $Re_b=5000$. 
The component coming from the viscous shear remains small and practically constant across all the cases, while those induced by the turbulent shear and by the canopy drag significantly decrease.
We impute the depletion of the former to a lower level of turbulent activity, associated to a weaker shear layer, while the latter is mainly reduced by the compliant nature of the filaments and the consequent reduction of the frontal canopy area.

\subsection{Canopy drag}
\label{sec:experiments}

\begin{figure}
\centering
\includegraphics[width=.95\textwidth]{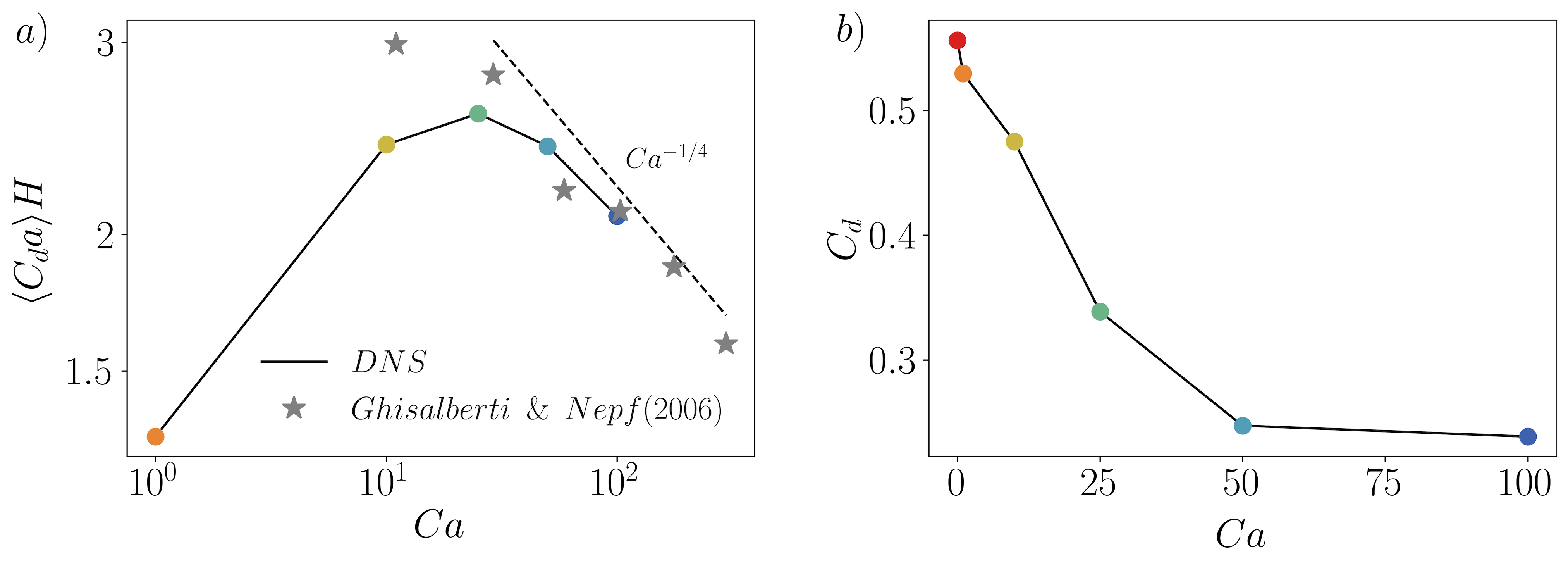}
\caption{Canopy drag measurements. We show (panel $a$) the mean value of $C_d a$ from our simulations, compared to that measured experimentally by \cite{ghisalberti-nepf-2006}. We also report (panel $b$) the value of the canopy drag, $C_d$, throughout our simulations.} 
\label{fig:expCd}
\end{figure}

Here we focus on the measurement of the canopy drag coefficient, $C_d$. 
First, in close analogy to \cite{ghisalberti-nepf-2006}, we define 
\begin{equation}
	C_d a (y)= \displaystyle \frac{\displaystyle \frac{\partial \langle u' v' \rangle}{\partial y}\bigg\rvert_{y_{out}<y<H} - \frac{\partial \langle u' v' \rangle}{\partial y}(y)}{\displaystyle \frac{1}{2}\langle u \rangle ^2 (y)}
\end{equation}
where $a$ is the frontal canopy area per unit volume, and the first term in the numerator denotes the mean vertical gradient of the Reynolds' shear stress above the canopy tip, $y_{out}$.
We thus compute the mean value of $C_d a$, denoted with angle brackets, between the inner and the outer inflection points for different values of $Ca$. 
This approach circumvents the complexity of the computation of $a$ \citep{ghisalberti-nepf-2006}, but yields a dimensional quantity which we thus make dimensionless upon multiplication with the channel height. 
The outcome is compared with the experimental measurements of \cite{ghisalberti-nepf-2006} in panel $a$ of figure \ref{fig:expCd}, highlighting a good correspondence between the two for the most flexible cases considered in our investigation. The data appear to tend towards $\langle C_d a \rangle H \sim Ca^{-1/4}$ at large values of $Ca$.

In order to decouple the value of $C_d$ from $a$, we consider the integral form of the streamwise momentum balance 
\begin{equation}
	-H\displaystyle\frac{dP}{dx}=\tau_w + \int_0^{y_{out}}D_c(y) dy
\end{equation}
where the streamwise pressure gradient $\displaystyle \frac{dP}{dx}$ equates the sum of the mean shear stress at the wall, $\tau_w$, and of the canopy drag, $D_c(y)$, integrated across the canopy height.  
$C_d$ can thus be computed through such balance, referring it to the frontal canopy area,
\begin{equation}
	C_d=\displaystyle \frac{L_x L_z}{H L_z} \displaystyle \frac{\int_0^{y_{out}}D_c(y) dy}{\frac{1}{2}\rho U_b^2} = - \frac{L_x}{H} \displaystyle \frac{H\displaystyle\frac{dP}{dx}+\tau_w}{\frac{1}{2}\rho U_b^2} 
\end{equation}
The outcome of this approach in our simulations is reported in panel $b$ of figure \ref{fig:expCd}.
The value of $C_d$ in the rigid case is compatible with what reported in literature for canopies with similar properties \citep{raupach-thom-1981,shimizu-etal-1992,finnigan-2000,ghisalberti-nepf-2006,nepf-2012-1}, where $C_d \approx 0.5-1.5$, and quickly decreases reducing the rigidity of the filaments, reaching a plateau for $Ca \gtrapprox 50$. 
Overall, the drag measurements reported here are in good agreement with the experimental measurements and shed light on the trend of $C_d$ with $Ca$.

\subsection{Energy spectra} 
\label{sec:spectra}

The amount of kinetic energy retained by the turbulent fluctuations of the flow is quantified by the mean turbulent kinetic energy (TKE), defined as $K(y)=0.5(\langle u' u' \rangle + \langle v' v' \rangle +\langle w' w' \rangle)$. 
Nevertheless, the space averaging operation cancels any information about the distribution of the kinetic energy across the different scales of motion. 
To this purpose we therefore resort to the spatial spectrum of the TKE, $E(k_x,k_z;y)=\overline{\mathcal{F}_x(\mathcal{F}_z(0.5(u' u'+v' v'+w' w')))}$, where $\mathcal{F}$ denotes the Fourier transform operator across either of the homogeneous directions and the over bar indicates averaging in time.
Integration along the spanwise wave numbers $k_z=2\pi i/L_z$ (with $i \in \mathbb{N}; i = 1, ..., n_z$) yields the streamwise spectrum $E_x(k_x;y)$ as a function of the streamwise wave number $k_x$ and the wall normal coordinate; with a similar procedure we also attain the spanwise spectrum $E_z(k_z;y)$.
\begin{figure}
\includegraphics[width=\textwidth]{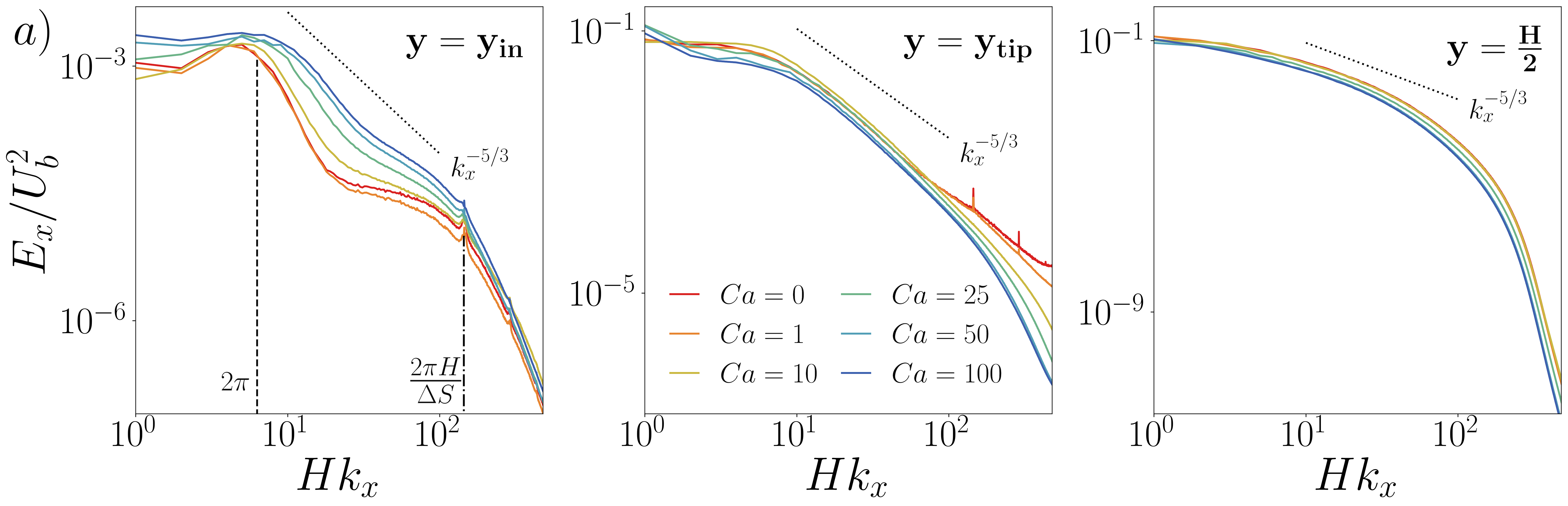}
\hfill
\includegraphics[width=\textwidth]{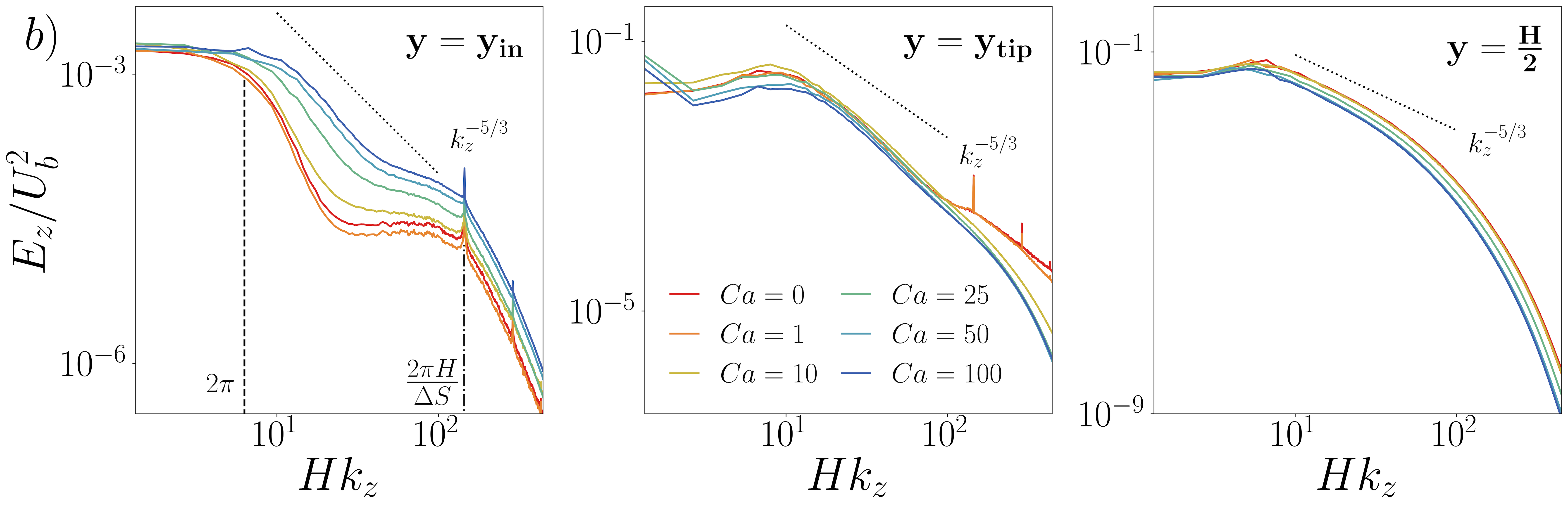}
\caption{(panel $a$) Streamwise and (panel $b$) spanwise 1D spectra of the turbulent kinetic energy, integrated along the remaining homogeneous direction, for different values of $Ca$. The spectra are sampled (left) at the inner inflection point, (centre) at the canopy tip and (right) in the outer flow.}  
\label{fig:spectra}
\end{figure}
In panels $a$ and $b$ of figure \ref{fig:spectra} we inspect the profiles of $E_x$ and $E_z$, respectively, for different values of $Ca$ at the inner inflection point, $y=y_{in}$, at the outer inflection point, $y=y_{tip}$, and above the canopy, at $y=H/2$.
We also report the typical $-5/3$ slope for comparison.
Within the canopy, both spectra exhibit a non-monotonous behaviour characterised by a first peak at the wave number associated to the channel height and a second, sharper one, associated to the mean separation of the filaments. 
In between the two, the spectra approach a $-5/3$ slope when $Ca$ increases.
The filaments absorb energy in the shear production range, close to the largest scales of motion, and re-inject it in the flow through their wakes and their waving motion, in close agreement to the spectral short cut process described by \cite{finnigan-2000} and \cite{olivieri-etal-2020-2}.
A variation in the $Ca$ is observed to affect both the spectral short cut and the amplitude of the spectra.
In facts, the spectral short cut is more intense in the case of rigid filaments, where the spectra almost approach a plateau before their second peak, while it appears less accentuated in the cases at higher $Ca$, consistently with the observations of \cite{olivieri-cannon-rosti-2022}.
A more regular decay of the spectra is recovered moving up to the drag discontinuity at the canopy tip.
In particular, while $E_x$ decreases monotonously from the largest scales of motion, $E_z$ peaks at $k_z\sim\mathcal{O}(10)$: this behaviour, which is also found in the outer flow at $y=H/2$, appears compatible with the presence of large structures dominating the outer flow above the canopy \citep{monti-olivieri-rosti-2023}.
Conventional full spectra are observed in the outer flow, denoting the persistence of a fully-developed turbulent state there.
Interestingly, the amplitude of the spectra within the canopy appears to increase with $Ca$, while an opposite trend is observed outside.
An increase in the flexibility of the filaments is in facts associated to more intense velocity fluctuations at the inner inflection point, located deep in the canopy and marginally affected from the outer flow turbulence, due to the increased motion of the filaments. 
At and above the canopy tip, instead, more flexible filaments are associated to the generation of a weaker shear layer with respect to the rigid case due to the lower drag discontinuity at their tip.

\subsection{Lumley triangle}
\label{sec:lumley}

The direct numerical simulation of turbulent canopy flows is a challenging task that might provide an excessive level of detail for most engineering applications, where an accurate model of the principal phenomena governing the fluid motion can instead prove sufficient.
Turbulence models often rely on a detailed description of the Reynolds stress anisotropy tensor $\mathbf{b}$, which we therefore thoroughly characterise in this section resorting to the Lumley triangle formalism. 
As outlined by \cite{pope-2000}, the turbulent stress state can be described with only two scalar variables, $\xi$ and $\eta$, with the latter quantifying the anisotropy of the stress state.
Adopting the index notation and implying Einstein's summation over repeated indices ($i,j \in \mathbb{N}; i,j=1,2,3$) we can write
\begin{equation}
	b_{ij}=\frac{\langle u'_i u'_j \rangle}{\langle u'_k u'_k \rangle}-\frac{1}{3}\delta_{ij},
\label{eq:anisotropy}
\end{equation}
where $\delta_{ij}$ represents the second order tensor identity. $\xi$ and $\eta$ are therefore defined as 
\begin{equation}
	6\eta^2=b_{ij}b_{ji}, \enspace 6\xi^3=b_{ij}b_{jk}b_{ki}.
\label{eq:xiEta}
\end{equation}
The so-called Lumley triangle is attained delimiting in the $\xi-\eta$ plane the set of realisable states of the Reynolds stress tensor, hence those associated to real and positive eigenvalues. 

\begin{figure}
\includegraphics[width=\textwidth]{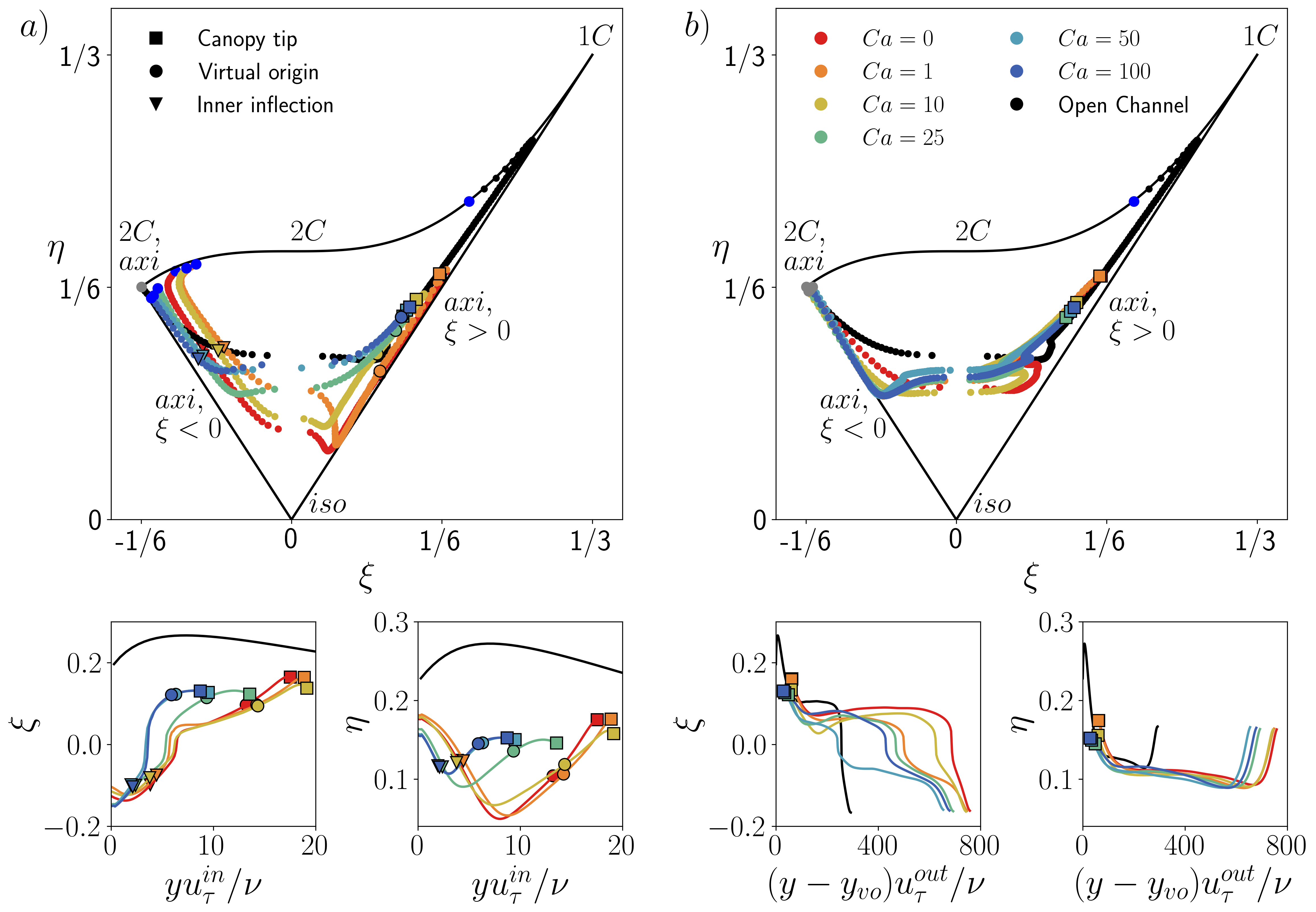}
\caption{Characterisation of the turbulence state (panel $a$) inside and (panel $b$) outside the canopy for different values of $Ca$, according to the Lumley triangle formalism. Data for the reference open channel case are reported across the whole channel height in both panels and relevant points within the canopy are denoted with distinctive symbols. 
In the two main plots, each dot corresponds to a grid point along the wall normal direction, with the first one close to the wall coloured in blue and the last one at the centreline coloured in grey; note how the first grid point of the open channel lays in the right half of the triangle, while those of all the canopy cases lay in the left one.
The smaller plots at the bottom report the trends of $\xi$ and $\eta$ (panel $a$) inside / (panel $b$) outside the canopy, scaled in the viscous units of the inner / outer flow introduced in \S\ref{sec:meanflow}, respectively. }  
\label{fig:lmt}
\end{figure}

To set a starting reference, we first report in all panels of figure \ref{fig:lmt} the states attained at each wall normal grid point of a turbulent open channel flow simulation carried out in the same setup detailed in \S\ref{sec:setupnmethods} at $Re_b=5000$, without the filaments.
The first grid point close to the wall is marked in blue, while the one at the centreline is coloured in grey.
Turbulence at the wall is two component (2C) due to the no penetration condition and the anisotropy peaks moving upward, at about $y u_\tau^{in}/\nu\approx7$. 
An axisymmetric state with $\xi>0$ is achieved towards the log layer (the Reynolds stress ellipsoid therefore resembles a prolate spheroid); such condition is nevertheless lost approaching the centreline, in contrast to what observed for a full channel, as the no penetration condition forces again a 2C state, which this time is approached for $\xi<0$ (the Reynolds stress ellipsoid therefore resembles an oblate spheroid).
We impute the difference in the sign of $\xi$ attained at the wall and at the free-slip surface to the effect of the no-slip condition. 
Next, we plot the trends of $\xi$ and $\eta$ for the different canopy cases: panel $a$ refers to the flow within the canopy and the evolution of the variables is therefore truncated at the canopy tip, highlighting the relevant points with distinctive symbols.
Panel $b$, instead, refers to the outer flow and reports the evolution of the variables from the canopy tip upward. 
Since any information concerning the wall normal coordinate is lost in the triangles, we complement them with two additional panels each, showing the trends of $\xi$ and $\eta$ with $y$ scaled in inner (panel $a$) and outer (panel $b$) viscous units, as introduced in \S\ref{sec:meanflow}.
As a matter of consistency, for the open channel case, we set $u_\tau^{in} = u_\tau^{out} = \sqrt{\tau_w/\rho_f}$ and $y_{vo}=0$.

Turbulence close to the bottom of the canopy lays in a $2C$ axisymmetric state with $\xi<0$ (opposite to the 2C state at  $\xi>0$ found in the open channel), arguably, due to the shielding effect of the filaments on the wall and the consequent attenuation of sweep events ($u'>0,v'<0$) reaching the bottom layer.
The anisotropy reduces moving upward and reaches a minimum between the inner inflection point and the virtual origin, where turbulence becomes almost isotropic for the most rigid canopy cases. At the same time, starting close to the inner inflection point, $\xi$ begins to grow more rapidly and shifts to positive values, leading to an axisymmetric state with  $\xi>0$ characteristic of both the virtual origin and the outer inflection point, where $\eta$ has a local maximum.
Moving out of the canopy, above the shear layer at the outer inflection point, the anisotropy is depleted and approaches an almost linear trend which is maintained far into the outer flow for all cases.  
For the most rigid canopies, $\xi$ settles on a constant value and the stress state is reminiscent of that attained in the log-law region of the open channel. Nevertheless, for the most flexible canopies, $\xi$ exhibits an irregular decrease and soon switches sign again.
Finally, approaching the free-slip wall, a 2C axisymmetric state with $\xi<0$ is forced by the boundary condition; $\xi$ therefore undergoes a sudden fall, as $\eta$ sharply increases.

Our analysis supports a multi-layer approach to the modelling of canopy flow turbulence like in \cite{poggi-etal-2004}, based on the collection of different turbulence states along the wall normal direction.
Turbulence in the inner flow exhibits a peculiar structure close to the bottom wall, tending towards a more isotropic condition immediately above and further transitioning smoothly towards a state similar to that attained in the log-law region of turbulent wall flows. 
This picture is also supported by the visualisations of the instantaneous, local anisotropy of the flow reported in appendix \ref{app:flowVis}.
The profiles of $\xi$ and $\eta$  in figure \ref{fig:lmt} show a reasonable collapse in both the most rigid ($Ca=0,1,10$) and most flexible cases ($Ca=50,100$), once scaled in inner viscous units, with the case at $Ca=25$ laying in between the two.
Turbulence immediately above the canopy is more isotropic than over the no-slip wall of the open channel ($\eta$ is smaller), consistently with the observations of \cite{kuwata-suga-2016} for an isotropic porous medium.
Instead, the portion of the outer flow from the log-law region up to where the upper free-slip wall is felt appears analogous to a turbulent wall flow, thus suggesting outer similarity arguments.
There, the anisotropy settles on an almost linear trend common to all $Ca$.

\subsection{Inner / outer flow interaction}
\label{sec:quadrant}

\begin{figure}
\centering

\includegraphics[width=.3\textwidth]{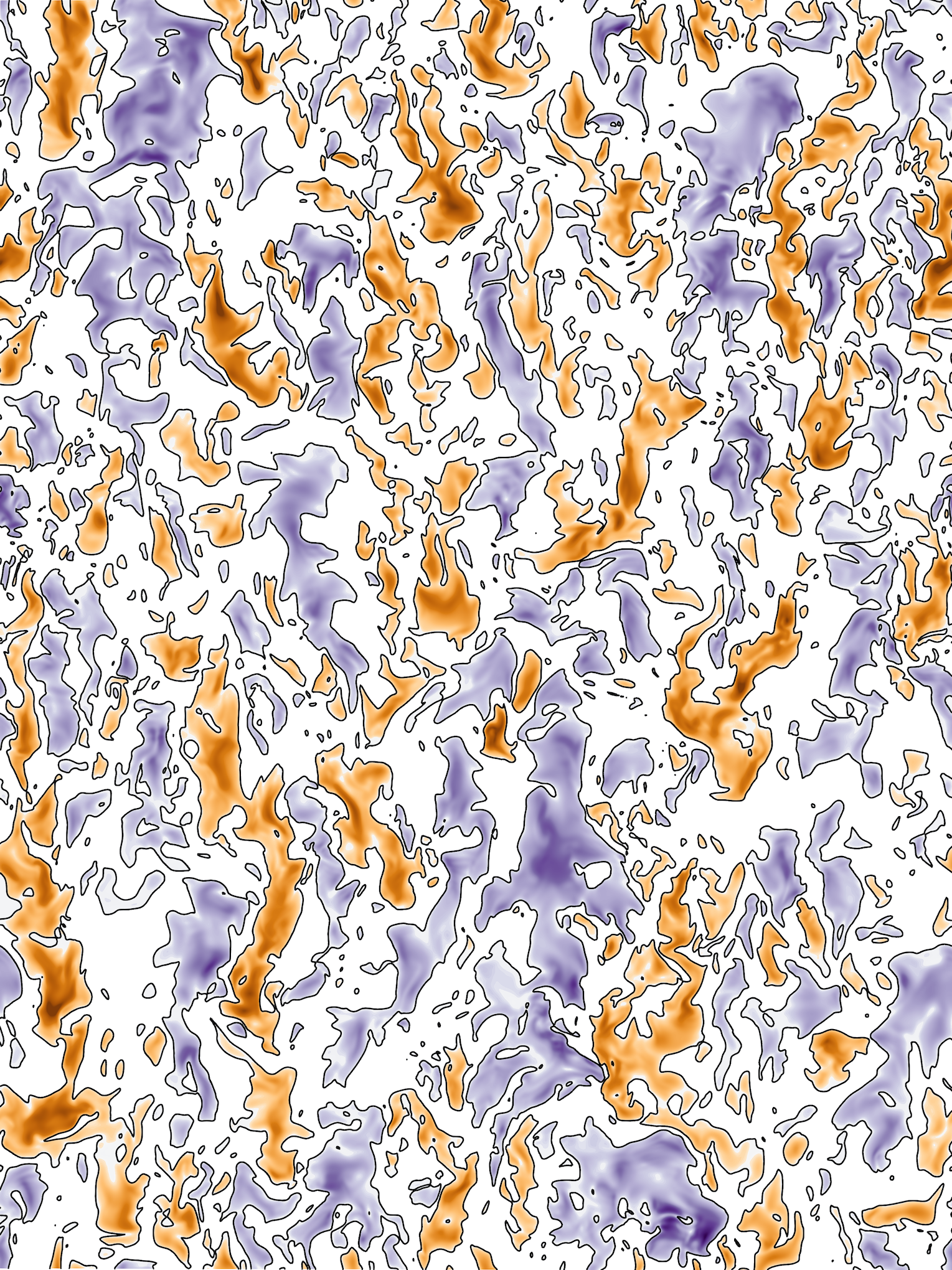}
\put(-110,140){\colorbox{white!30}{a)}}
\hspace{0.01cm}
\includegraphics[width=.3\textwidth]{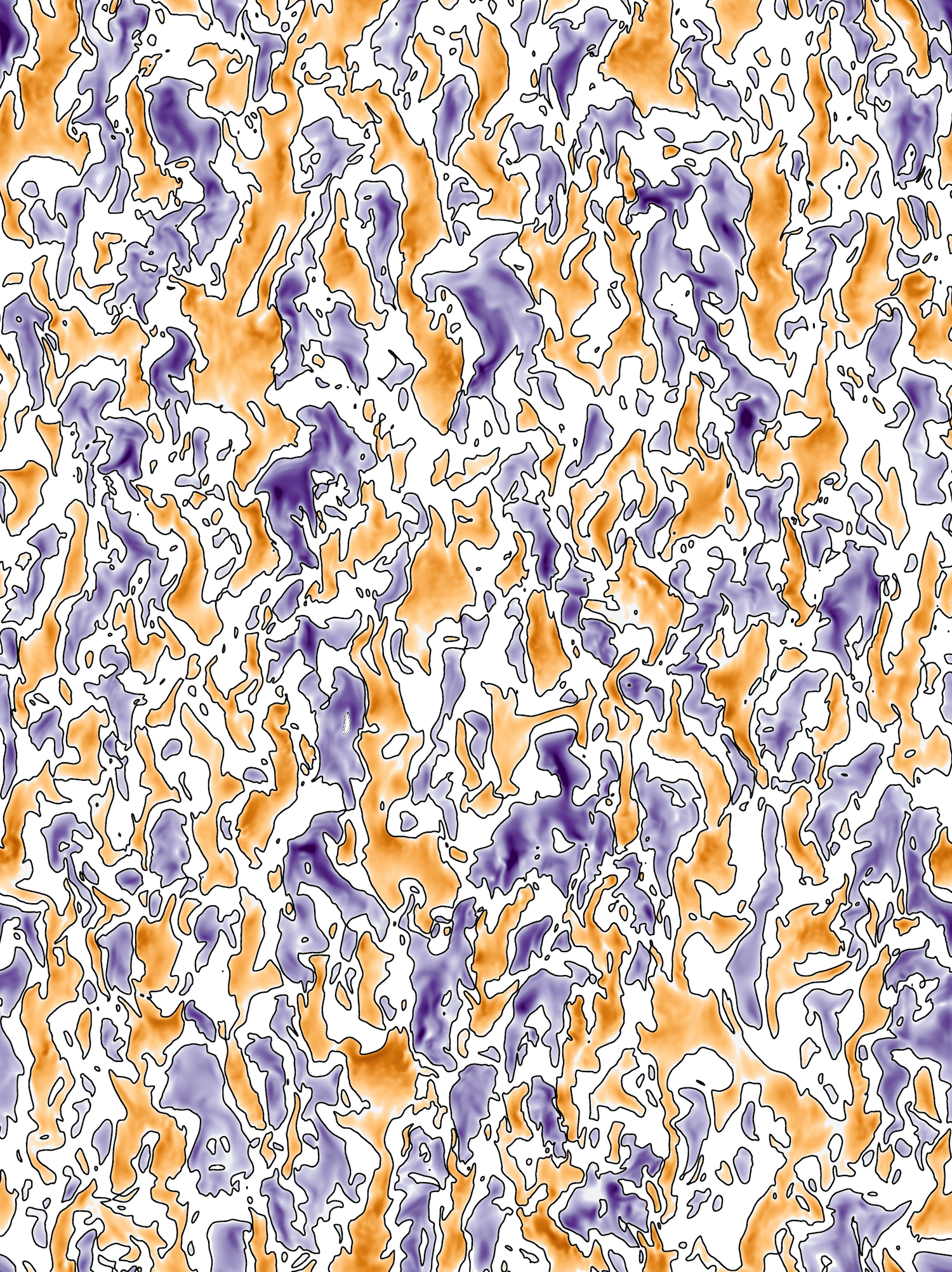}
\put(-110,140){\colorbox{white!30}{b)}}
\hspace{0.01cm}
\includegraphics[width=.3\textwidth]{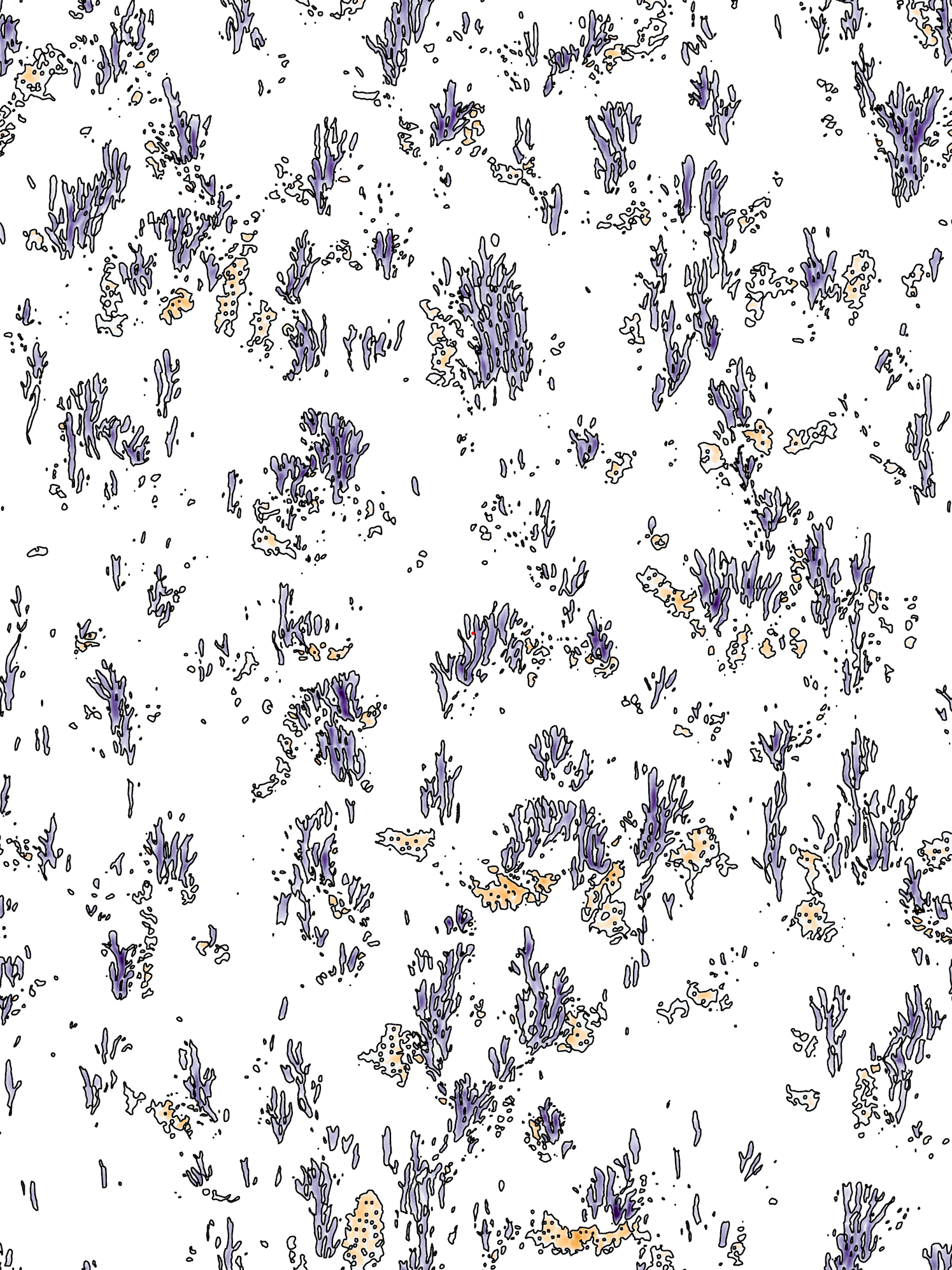}
\put(-110,140){\colorbox{white!30}{c)}}

\vspace{.03cm}

\includegraphics[width=.3\textwidth]{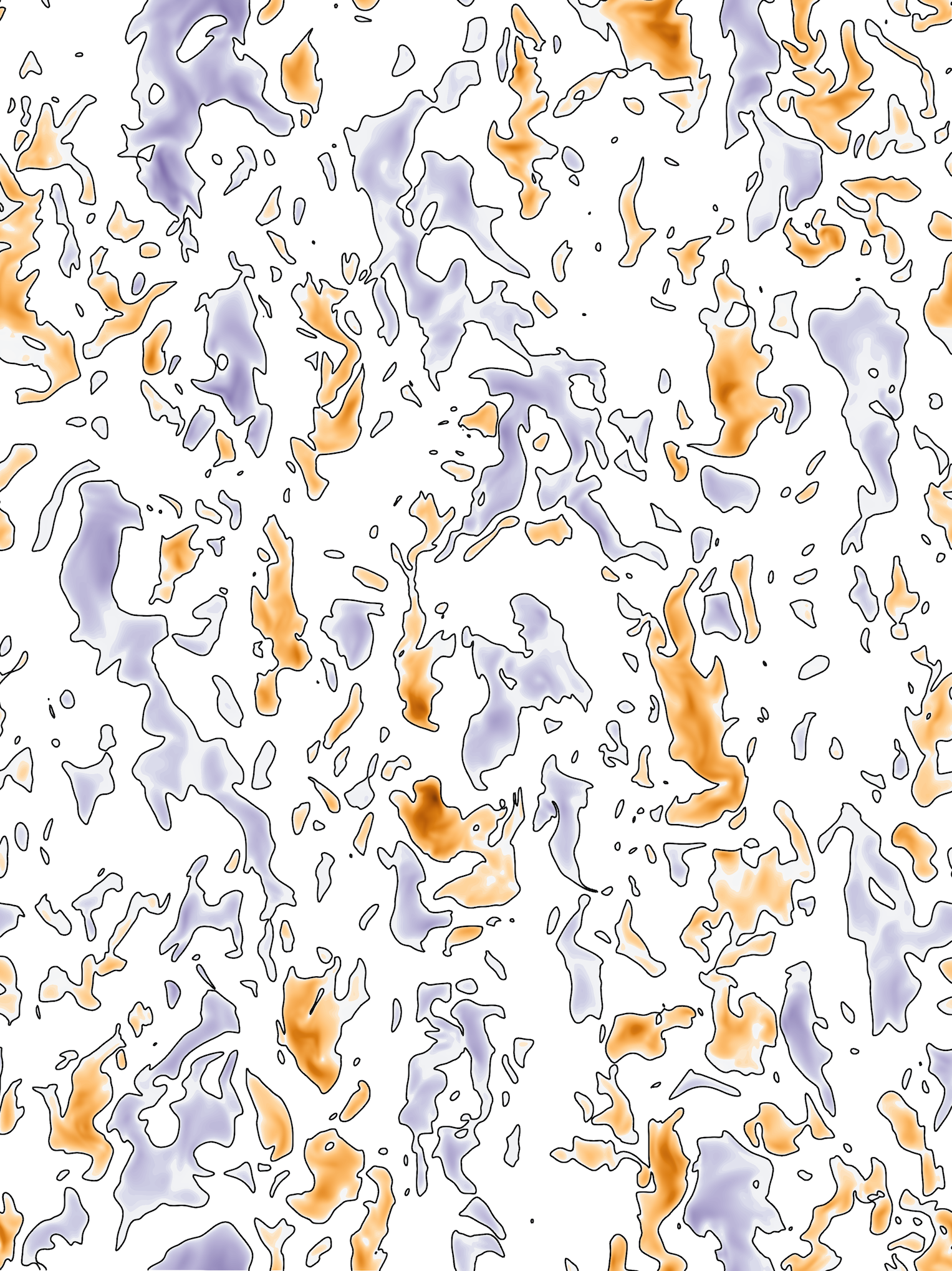}
\put(-110,140){\colorbox{white!30}{d)}}
\hspace{0.01cm}
\includegraphics[width=.3\textwidth]{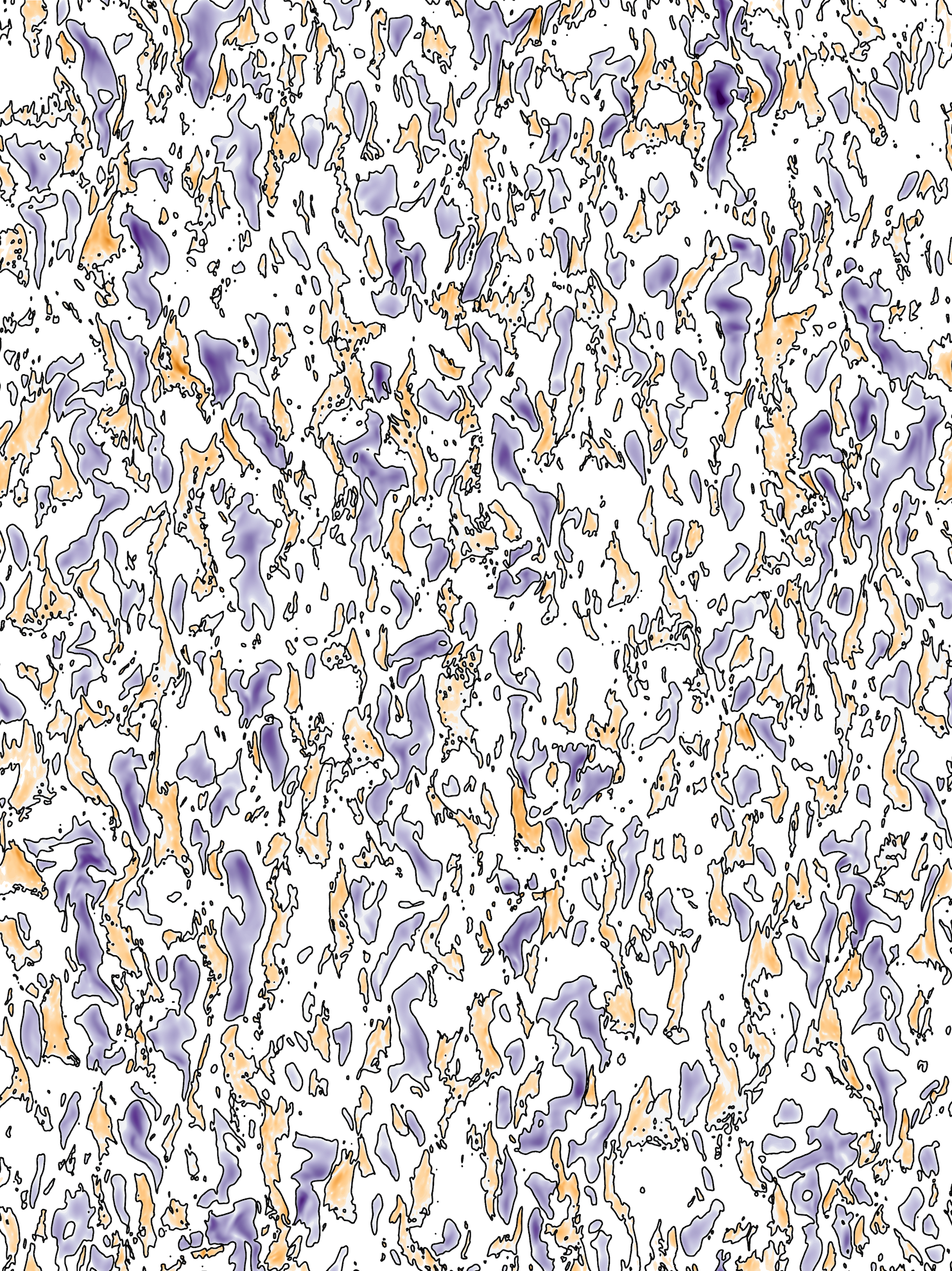}
\put(-110,140){\colorbox{white!30}{e)}}
\hspace{0.01cm}
\includegraphics[width=.3\textwidth]{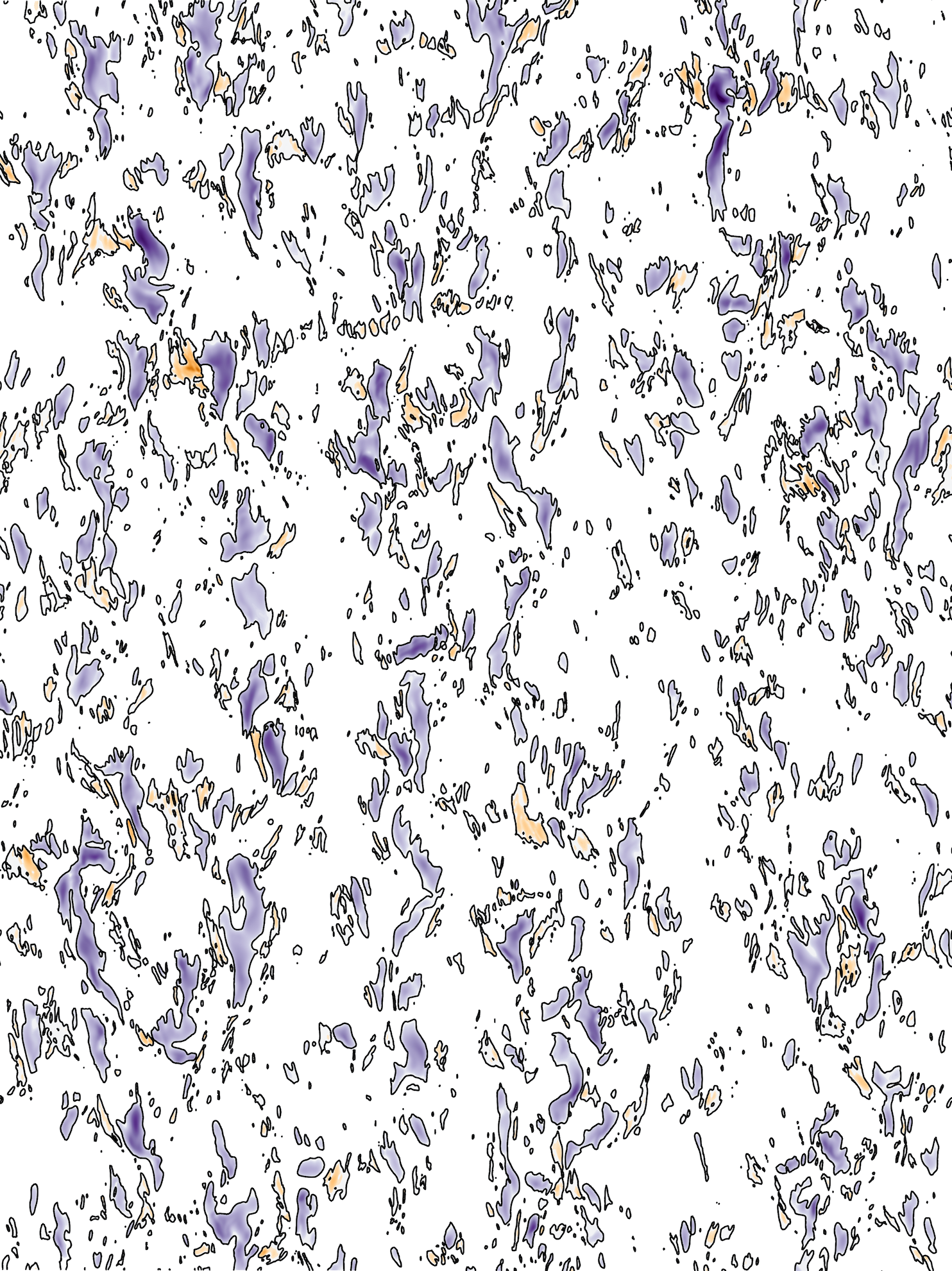}
\put(-110,140){\colorbox{white!30}{f)}}

\caption{Instantaneous sweep and ejection events in a rigid (panels $a,b,c$) and a flexible (panels $d,e,f$) canopy flow with $Ca=100$, at $y=H/2$ (panels $a,d$), at the canopy tip (panels $b,e$) and at the virtual origin (panels $c,f$). The flow is sampled on wall-parallel planes with the mean velocity aligned to the vertical direction, going from bottom to top. Regions where the events are occurring are delimited with black lines, while their magnitude is quantified as $|uw|/U_b^2$ and visualised with a linear colormap ranging from white to orange (ejections) or violet (sweeps) in [0,0.4].}
\label{fig:events}
\end{figure}

Multiple authors have proposed layered descriptions of canopy flows, identifying regions with specific dynamical properties while moving along the wall-normal direction. 
Despite the differences between the various approaches proposed \citep{belcher-jerram-hunt-2003,poggi-etal-2004,okamoto-nezu-2009}, there is consensus in separating the flow in the canopy from that outside, and in isolating the thin shear layer region at the canopy tip. 
In the previous subsection we characterised the inner and the outer flow separately; here, instead, we aim at elucidating how they interact with each other.  
As extensively discussed in literature, the flows over plant canopies \citep{finnigan-2000} along with those above permeable \citep{breugem-boersma-2005} and elastic \citep{rosti-brandt-2017} walls are often characterised by the presence of large scale vortical structures elongated in the spanwise direction (\textit{rollers}), generated by a Kelvin-Helmholtz like instability.
We provide evidence of the presence of these structures in our setup in appendix \ref{app:flowVis}.
These structures are likely to dominate the momentum exchange between the inner and the outer flow, but they are not the only process active close to the shear layer: also high/low speed streaks \citep{monti-olivieri-rosti-2023} and quasi-streamwise vortices might affect the interaction.
The flow dynamics close to the canopy tip is in facts a highly non-trivial phenomenon, which can not be exhaustively described only accounting for the presence of the rollers.
Furthermore, as noted by \cite{nicholas-etal-2023}, drag model-based simulations provide reasonable results in capturing the flow behaviour near the canopy, but they fall short in providing a detailed characterisation of the flow within the canopy layer and at the interface with the outer flow. 
A better understanding is therefore crucial for the development of more accurate models. 

Thus, in figure \ref{fig:events} we observe the sweep ($u'>0,v'<0$) and ejection ($u'<0,v'>0$) events taking place at three locations distinctive of the three regions mentioned above\footnote{The continuous evolution of sweeps and ejections along the wall-normal direction can be found in \cite{poggi-etal-2004}, for a rigid canopy.}: at $y=H/2$, at the canopy tip, corresponding to the outer inflection point $y_{out}$, and at the virtual origin $y_{vo}$. We consider the rigid canopy case (panels $a,b,c$) as well as a flexible one ($Ca=100$, panels $d,e,f$).
The most intense sweeps and ejections occur at the canopy tip (panels $b,e$), while they attenuate and become more coherent moving in the outer flow (panels $a,d$).
Consistently with the observations of \cite{gao-shaw-paw-1989}, sweep and ejections are of about equal strength at approximatively twice the canopy height. 
At the virtual origin (panels $c,f$), instead, only the most intense sweep events are able to penetrate and they occupy small, well defined regions; ejections there are less frequent.
Overall, the turbulent state is more strongly driven in the rigid canopy case, where it is characterised by more coherent events compared to the flexible one.
Sweep events reach the virtual origin of the flexible canopy flow more frequently, but with diminished intensity.

\begin{figure}
\includegraphics[width=\textwidth]{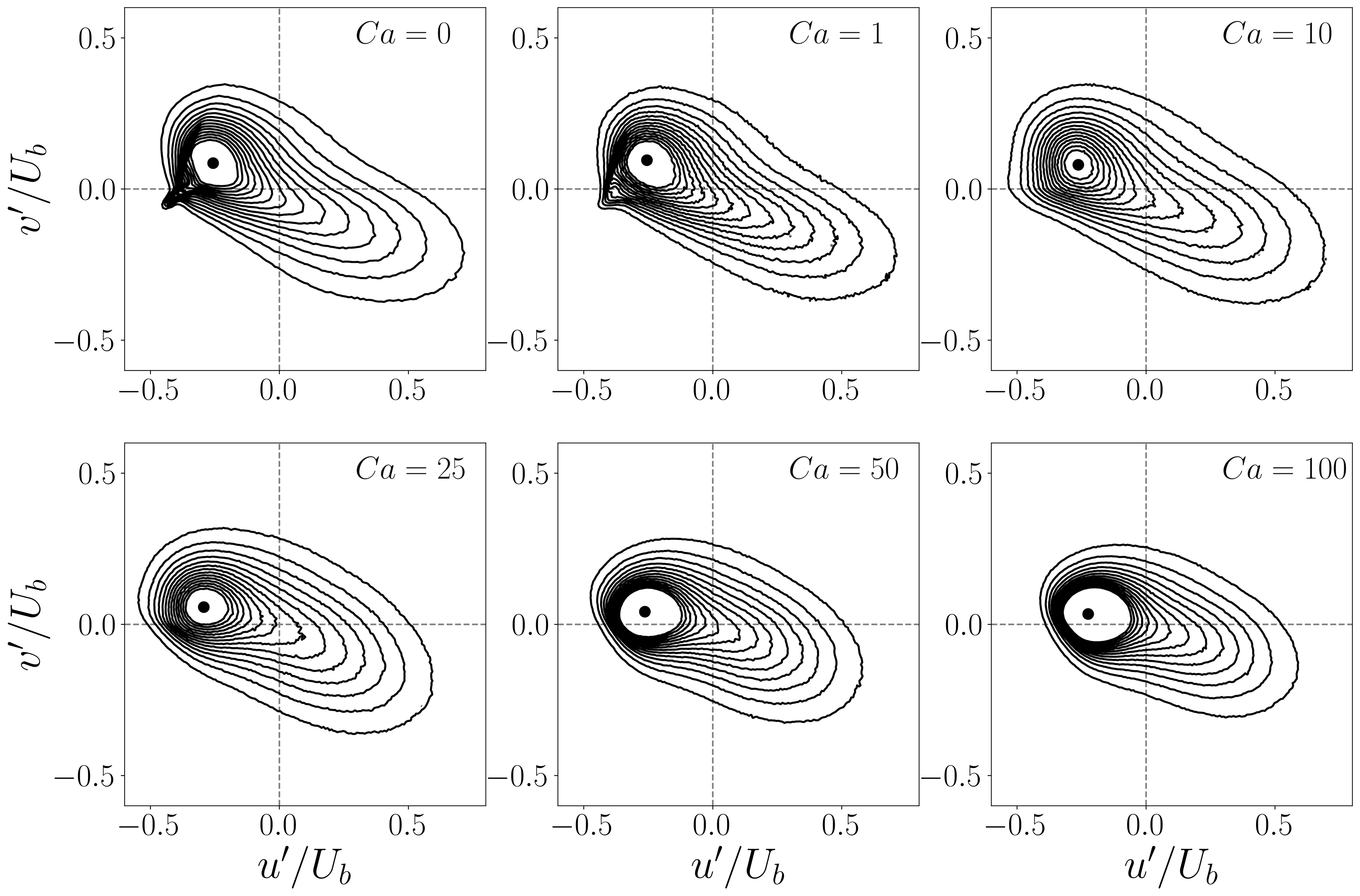}
\caption{Isolines of the J-PDF (normalised to a unitary integral over the domain) associated to the streamwise and wall normal velocity fluctuations at the mean position of the canopy tip, for different values of $Ca$. Levels are evenly distributed between $0.4$ and $6$ with $0.4$ increments, while the locations of the peaks are denoted by black dots.}
\label{fig:quadrant}
\end{figure}

In order to provide a more quantitative description of the momentum transfer between the inner and the outer flows, we compute the joint probability density function (J-PDF) of the streamwise and wall-normal velocity fluctuations at the canopy tip for different values of $Ca$.
As visible in figure \ref{fig:quadrant}, the peaks always lay in the second quadrant ($u'<0,v'>0$) denoting the dominance of ejections over sweeps, as well documented in literature \citep{gao-shaw-paw-1989,finnigan-2000}. 
The downward motion of high-speed fluid is mitigated by the presence of the canopy, which instead opposes little resistance to the uplift of low-speed fluid from its interior.
Such observation is consistent with the experiment of \cite{chowdhuri-ghannam-banerjee-2022}, who noted that long-lasting ejection events are more relevant than short-lived sweep events close to the canopy tip. 
Furthermore, the shape of the J-PDFs closely resembles that reported by \cite{manes-poggi-ridolfi-2011} in the case of a highly-permeable porous medium.
Notwithstanding this similarity, our J-PDFs differ from one another due to the effect of canopy flexibility.
In the most rigid cases the J-PDFs elongate in the fourth quadrant ($u'>0,v'<0$), remarking the occurrence of rare and intense sweep events which are attenuated increasing the flexibility. 
In the most flexible cases, in facts, the filaments are deflected by the mean flow and shield the inner region of the canopy from downward interactions.
Furthermore, at $Ca=0$ and $Ca=1$, the J-PDFs exhibit a secondary peak in the third quadrant ($u'<0,v'<0$) associated with the deflection of the streamlines around the filament tip.
As seen in \cite{monti-etal-2020} and in figure 16 of \cite{nicholas-etal-2023}, immediately after the tip the flow decelerates and plunges.

Our analysis appears to support a scenario where turbulence inside the canopy is sustained by intense events induced by the outer flow. Nevertheless, while the sweeps are hindered by the presence of the filaments, the canopy opposes little resistance to ejections. The filaments, in facts, regulate the vertical exchange of streamwise momentum allowing for the transpiration of the canopy and shielding the inner flow from all but the most intense downward interactions \citep[likely associated to the secondary instability of the rollers,][]{nepf-2012-1}.
Such intense sweeps induce a significant local deflection of the filaments and a large-amplitude flapping motion. 
Turbulence inside the canopy is therefore driven by frequent ejections and rare intense sweep events \citep{nepf-2012-1}, along with the flapping motion of the filaments in the flexible cases.

\subsection{Filament density variation}
\label{sec:drho}

In order to investigate the effects of a variation in the inertial properties of the filaments, we consider two additional values of the density ratio $\{ 1.0+1.46\cdot 10^{-2}, 1.0+1.46\cdot 10^{-1}\}$ starting from the case at $Ca=25$ and $\rho_s/\rho_f=1.0+1.46\cdot 10^{-3}$. 

\begin{figure}
\includegraphics[width=0.48\textwidth]{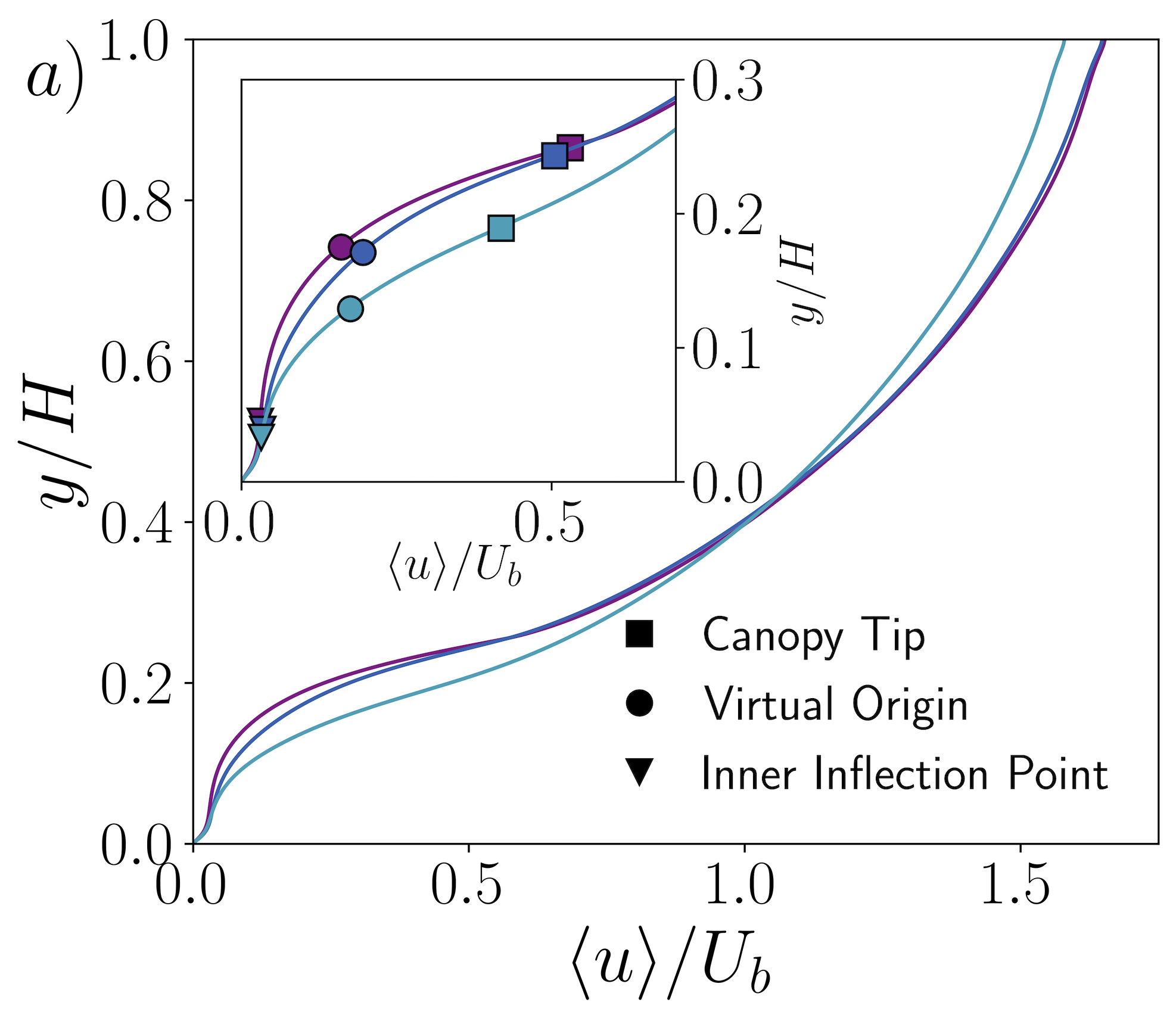}
\hfill
\includegraphics[width=0.48\textwidth]{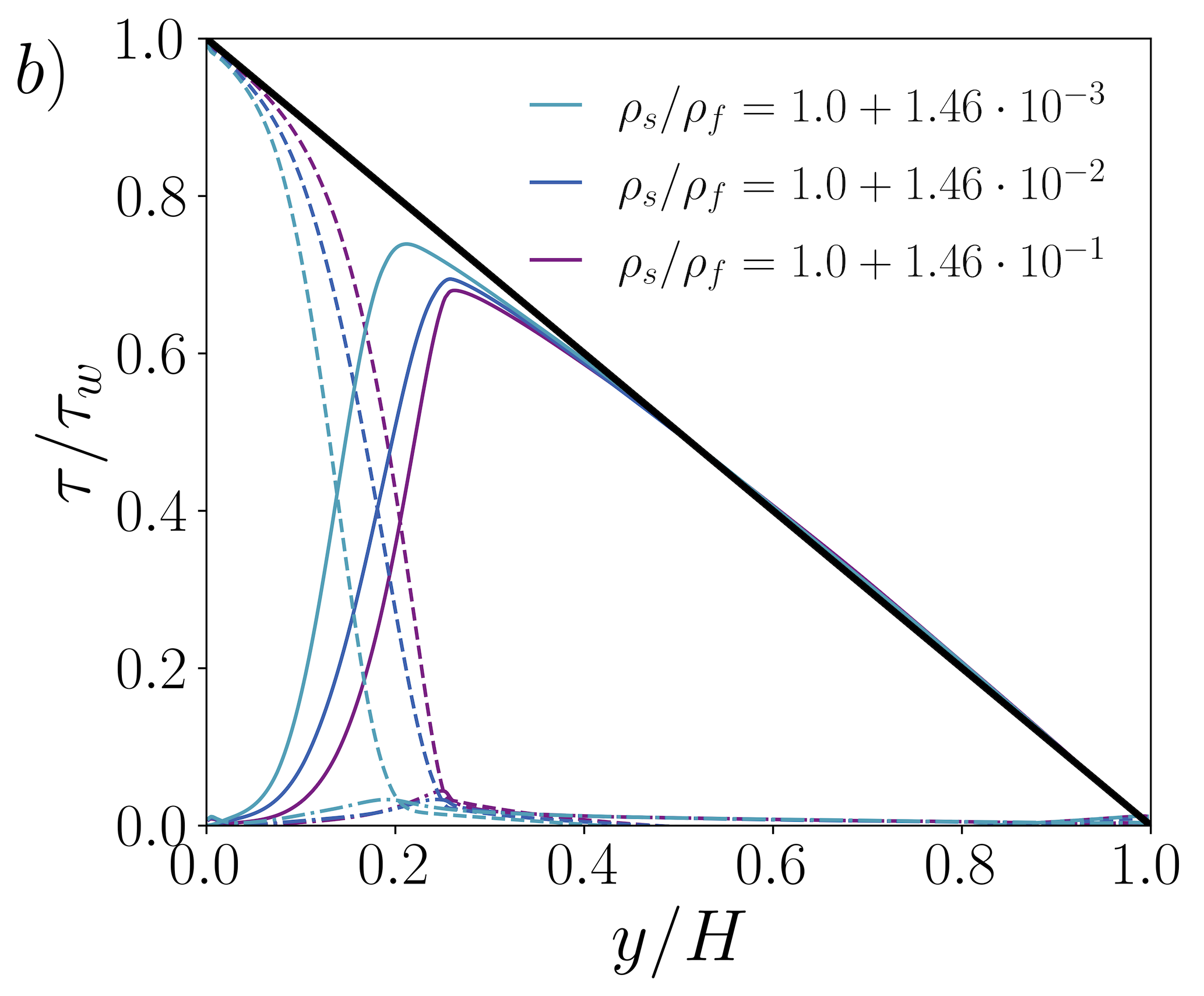}
\caption{(panel $a$) Mean profiles of the streamwise velocity for different values of $\rho_s/\rho_f$ at $Ca=25$ and associated relevant points (inset). In panel $b$, the total shear stress (black line) for the same cases, normalised by the wall shear stress, is given by the sum of the turbulent shear stress (continuous lines), the viscous shear stress (dash-dotted lines) and the canopy drag (dashed lines). }
\label{fig:dRho_profiles}
\end{figure}

\begin{figure}
\includegraphics[width=\textwidth]{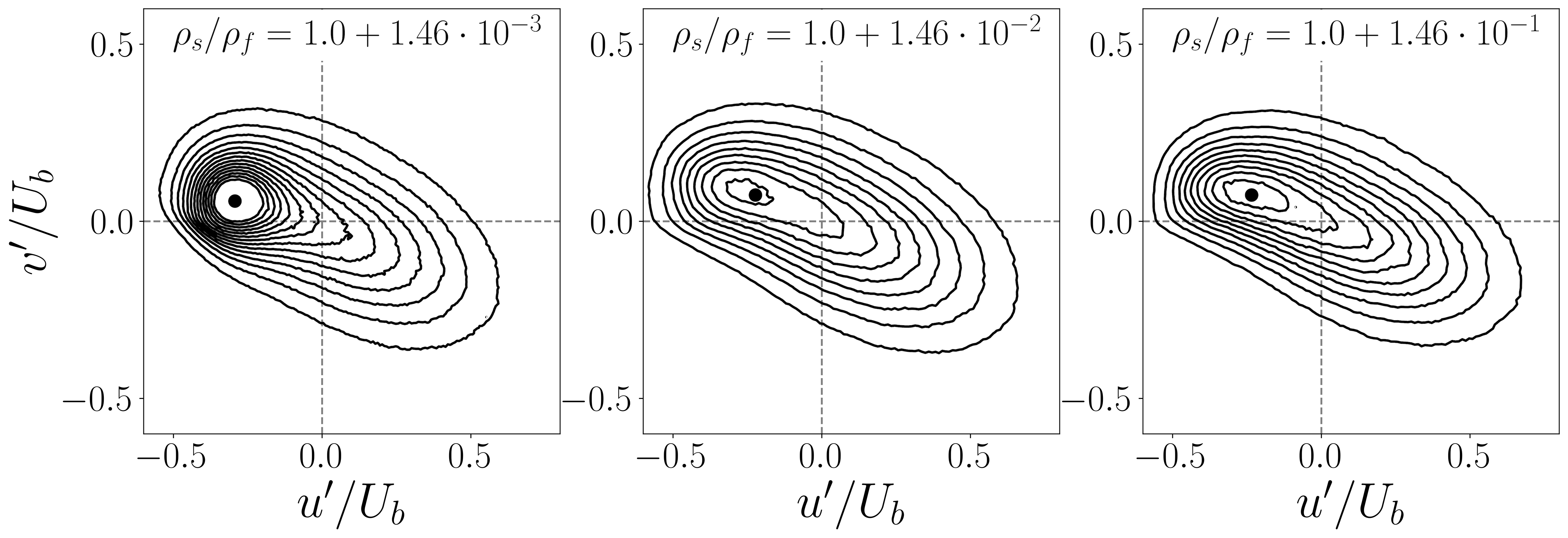}
\caption{Isolines of the $u'/v'$ J-PDF at the canopy tip, for different values of $\rho_s/\rho_f$ at $Ca=25$. Levels are distributed between $0.4$ and $6$ with $0.4$ increments, while peaks are denoted by black dots.}
\label{fig:dRho_quadrant}
\end{figure}

The first, macroscopic effect of an increase in the density of the filaments is their reduced compliance to the flow and the consequent upshift of the canopy tip. 
A taller canopy is associated to a higher blockage effect: the streamwise velocity profiles close to the bottom wall are therefore depleted (panel $a$ of figure \ref{fig:dRho_profiles}), while higher velocities are attained far above the canopy tip due to the imposition of a constant flow rate.
The positions of the inner inflection point and of the virtual origin remain essentially unchanged (inset of panel $a$, figure \ref{fig:dRho_profiles}), while the upward shift of the canopy tip is associated to an increase in the mean streamwise velocity there. 
Moving to the shear stress balance reported in panel $b$ of figure \ref{fig:dRho_profiles}, we observe how the peak in the viscous shear stress becomes more definite in the case of denser filaments, denoting the presence of a sharp shear layer above their tip. Nevertheless, their limited flapping motion yields less intense turbulent fluctuations there. 
Equivalently, the shear layer is blurred for lower density ratios and the filaments undergo a more pronounced motion, responsible for the higher Reynolds shear stress.
As the most significant effects of a variation in the filament density are appreciated at the canopy tip, we are motivated to better investigate the events taking place there observing the J-PDFs of the streamwise and wall normal velocity fluctuations reported in figure \ref{fig:dRho_quadrant}.
While a clear peak is always observed in correspondence of ejection events, sweep events penetrating the canopy become more frequent for higher density ratios due to the diminished shielding of the inner flow exerted by the filaments.
The J-PDFs therefore elongate in the fourth quadrant, as the ejection peak becomes less pronounced.
A deviation towards negative values of $u'$ close to $v'=0$ can also be appreciated, due to the augmented inertia of the filaments and the consequently increased intensity of impact interactions with high speed fluid.

For the dynamics of the fluid considered so far, an increase in the density of the filaments yields effects similar to those of an increase in their rigidity from one of the most flexible cases. 
Such effects can be appreciated comparing the case with lowest density ratio (belonging to the bulk of our investigation) to the others. 
Nevertheless, the differences between the two cases at higher $\rho_s/\rho_f$ are marginal and denote the attainment of a saturation for $\rho_s/\rho_f-1=\mathcal{O}(10^{-1})$, from which most of our investigation remains far.

\subsection{The effect of the filaments' flexibility}
\label{sec:freeze}

In a flexible canopy, the motion of the filaments is tightly coupled to the large scale coherent fluctuations of the turbulent flow \citep{monti-olivieri-rosti-2023}; in turn, the flow is affected by the flapping motion of the filaments along with their wakes at significantly smaller scales, due to the \textit{spectral short-cut} mechanism highlighted in \S\ref{sec:spectra}.
To assess the consequences of this complex fluid-structure interaction we inhibit it, by ``freezing'' the flexible canopy in one of its instantaneous configurations, and we compare the flow attained above such peculiar rigid canopy to the flow developed above its flexible counterpart. 
We repeat the comparison for all the non-zero values of $Ca$ in this study. 
The independence of our observations from the specific canopy configuration considered is assessed comparing the pressure gradient needed to impose the same flow rate above two different frozen canopies originated from the same simulation at $Ca=100$: no significant variation is observed between the two, thus confirming the adequacy of our numerical test. 
In particular, the canopy proves large enough to accommodate multiple configurations of the filaments in spite of their large scale coherent motion and therefore retains almost constant space-averaged characteristics across time. 

\begin{figure}
\includegraphics[width=\textwidth]{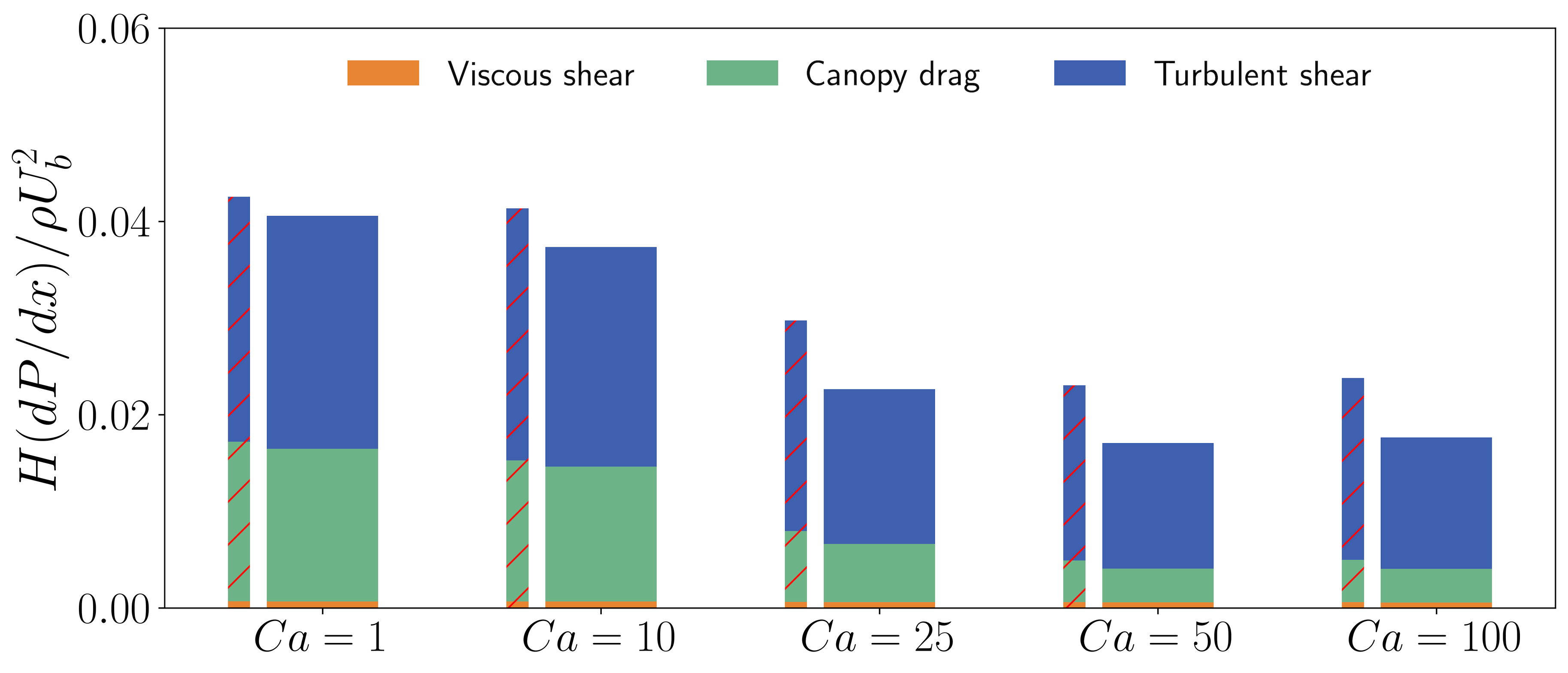}
\caption{Different contributions to the shear stress balance, integrated across the wall normal direction, for the frozen canopy cases at different initial values of $Ca$. Results for the corresponding flexible cases are reported, for reference, as thinner bars with a red hatched fill.}
\label{fig:tauHist_frozen}
\end{figure}

\begin{figure}
\centering
\includegraphics[width=0.48\textwidth]{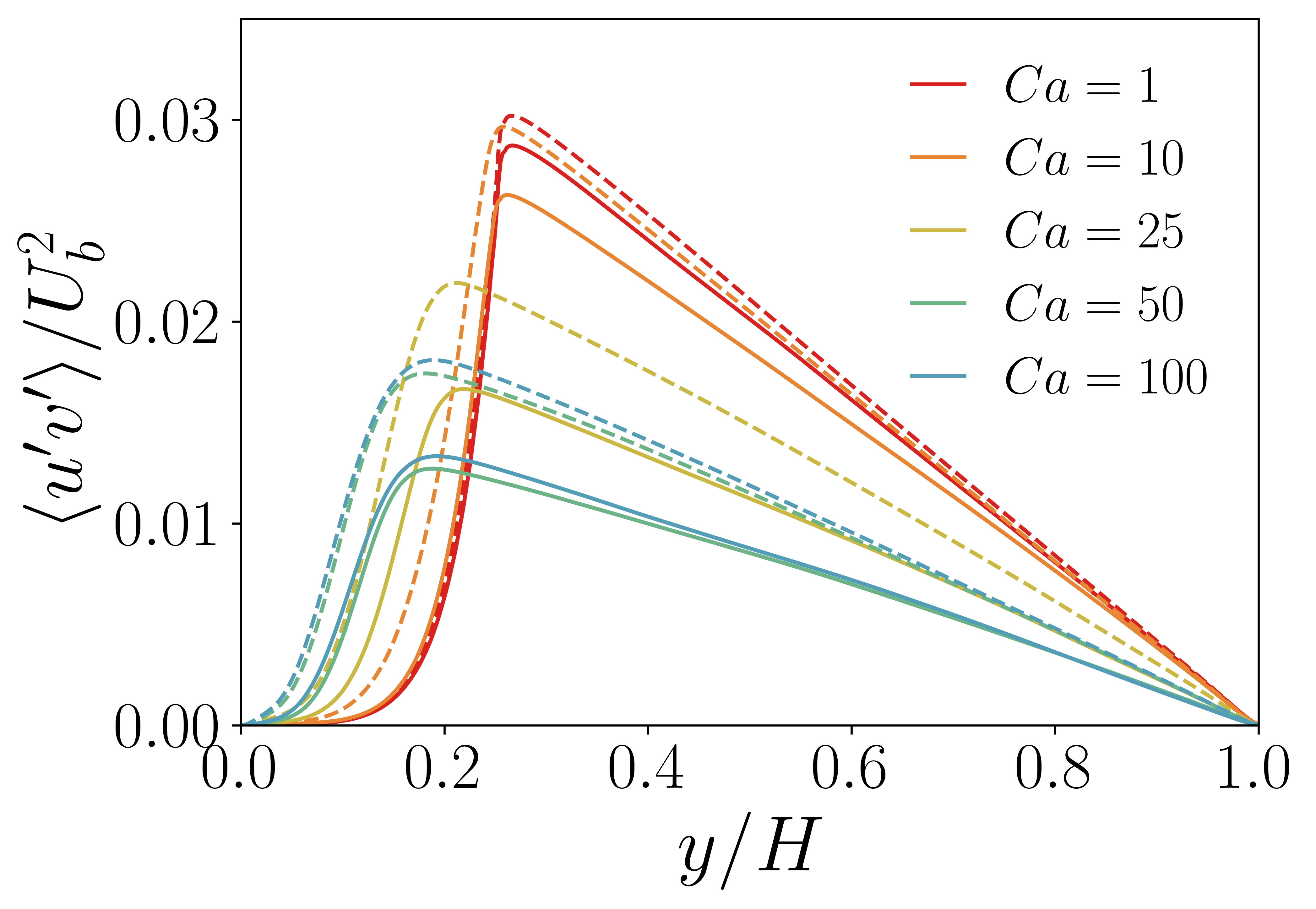}
\caption{Shearing component of the Reynolds stress tensor for different values of $Ca$. Data from the frozen canopy cases are denoted with continuous lines and data from the corresponding flexible cases are shown with dashed lines.}
\label{fig:turb_frozen}
\end{figure}

\begin{figure}
\includegraphics[width=\textwidth]{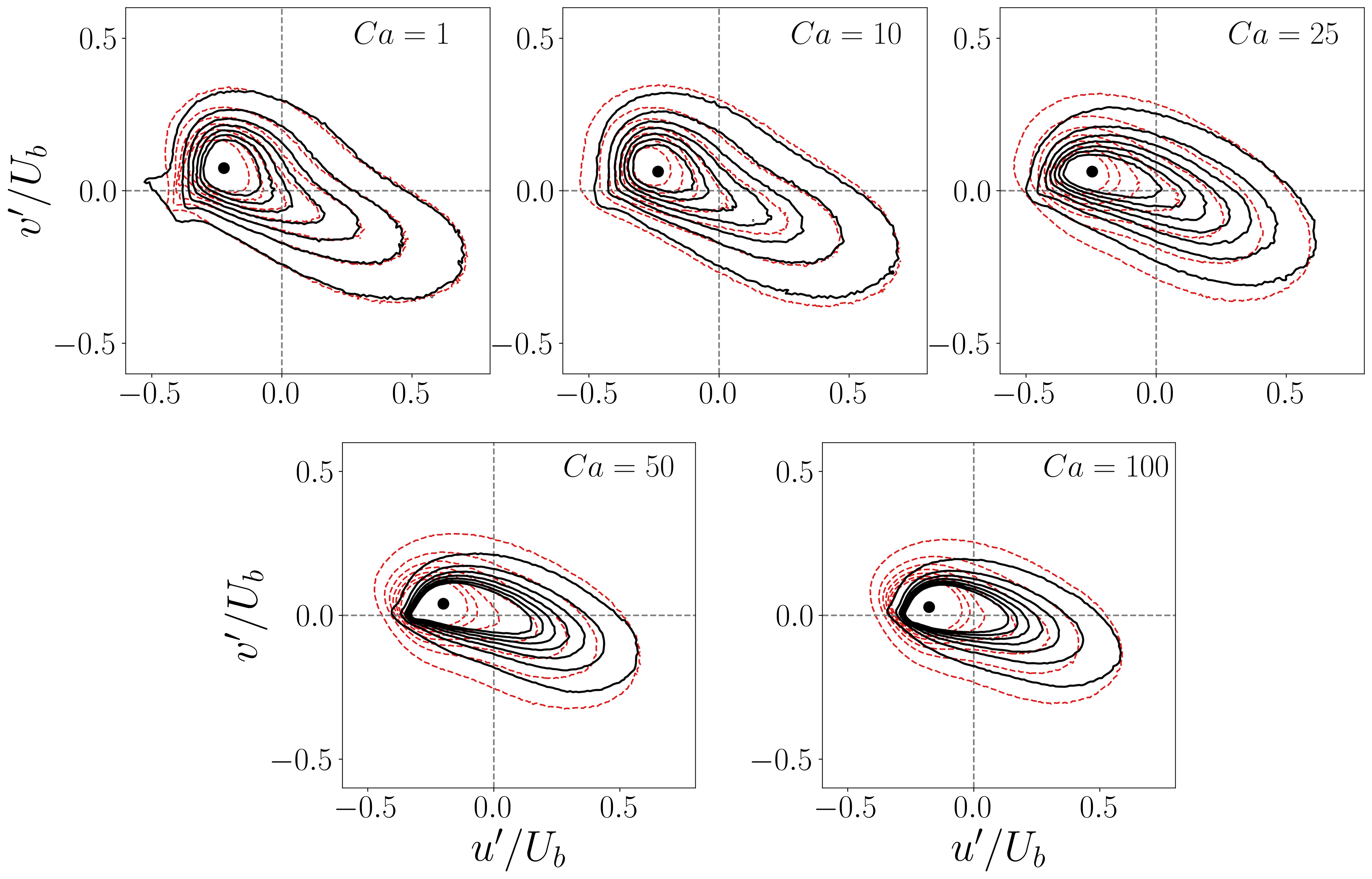}
\caption{Isolines of the $u'/v'$ J-PDF at the tip of the frozen canopies for different initial values of $Ca$. Levels are distributed between $0.4$ and $6$ with $0.8$ increments, while peaks are denoted by black dots. The same isolines from the corresponding flexible cases are reported, for reference, as dashed red lines.}
\label{fig:quadrant_rigid}
\end{figure}

The first macroscopic effect of freezing the flexible canopy in its instantaneous configuration, despite the increased average velocity difference between the filaments and the fluid, is a reduction of its total drag (which nevertheless remains higher then in an open channel at $Re_b=5000$ without the canopy). 
As visible in figure \ref{fig:tauHist_frozen}, the driving pressure gradient $\mathrm{d}P/\mathrm{d}x$ is reduced with respect to all the flexible canopy cases, reported for reference as thinner vertical bars highlighted with a red hatched fill.
The viscous shear contribution remains small and constant, while the canopy drag is slightly depleted. 
Nevertheless, the most significant reduction is undergone by the turbulent shear, which is therefore almost completely responsible for the drag reduction effect. 
Indeed, the depletion of the driving pressure gradient is accompanied by a reduction of all the components in the Reynolds stress tensor across the whole half-channel height.
Such effect is shown for the shearing contribution only in figure \ref{fig:turb_frozen}, where data from the frozen canopy cases are denoted with continuous lines and data from the flexible ones are shown with dashed lines. 
No significant variation can be appreciated in the location of the maxima of $\langle u' v' \rangle$ between the flexible and the frozen canopy cases, suggesting that the position of the shear layer above the canopy remains essentially unchanged.
To clarify how the nature of the turbulent fluctuations relates to the (now prevented) motion of the filaments, we once again observe the J-PDFs of the streamwise and wall normal velocity fluctuations at the canopy tip.
In figure \ref{fig:quadrant_rigid} the same levels are shown for both the frozen and the flexible canopy cases, respectively denoted with red dashed lines and black continuous lines, for different values of $Ca$.
The deviation from the flexible cases is understandably enhanced increasing the initial $Ca$, as we compare canopies of increasing flexibility to rigid ones. 
For the frozen cases, intense events of the wall normal velocity become less relevant in both the positive and negative quadrants, leading to sharper peaks in the second quadrant ($u'>0, v'<0$). Intense sweep and ejection events are therefore diminished: the frozen canopy behaves as a less permeable porous medium compared to the flexible one, thus justifying its decreased drag \citep{manes-etal-2009}.

The drag reduction observed freezing a flexible canopy in its instantaneous configuration can therefore be imputed to an overall reduction of the turbulent activity with respect to the flexible case. 
We identify two phenomena contributing to such effect: first, upon freezing the filaments, no more energy is pumped into the small scale turbulent fluctuations of the flow by their flapping motion. 
This is consistent with the maximum drag reduction being attained in the intermediate regime ($Ca=25\sim50$), where the large amplitude flapping of the filaments is most regular. As further discussed in the following section, in facts, at intermediate values of $Ca$ the natural response of the filaments resonates with a characteristic frequency of the turbulent forcing.
Second, the frozen canopies are more effective then the flexible ones at preventing intense vertical interactions between the inner and the outer flows, thus yielding a weakly driven turbulent state within the canopy and depleting the turbulent fluctuations close to the shear layer.

\section{Dynamics of the filaments}
\label{sec:fildyn}

\subsection{Characterisation of motion}
\label{sec:filmotion}

\begin{figure}
   \includegraphics[width=0.48\textwidth]{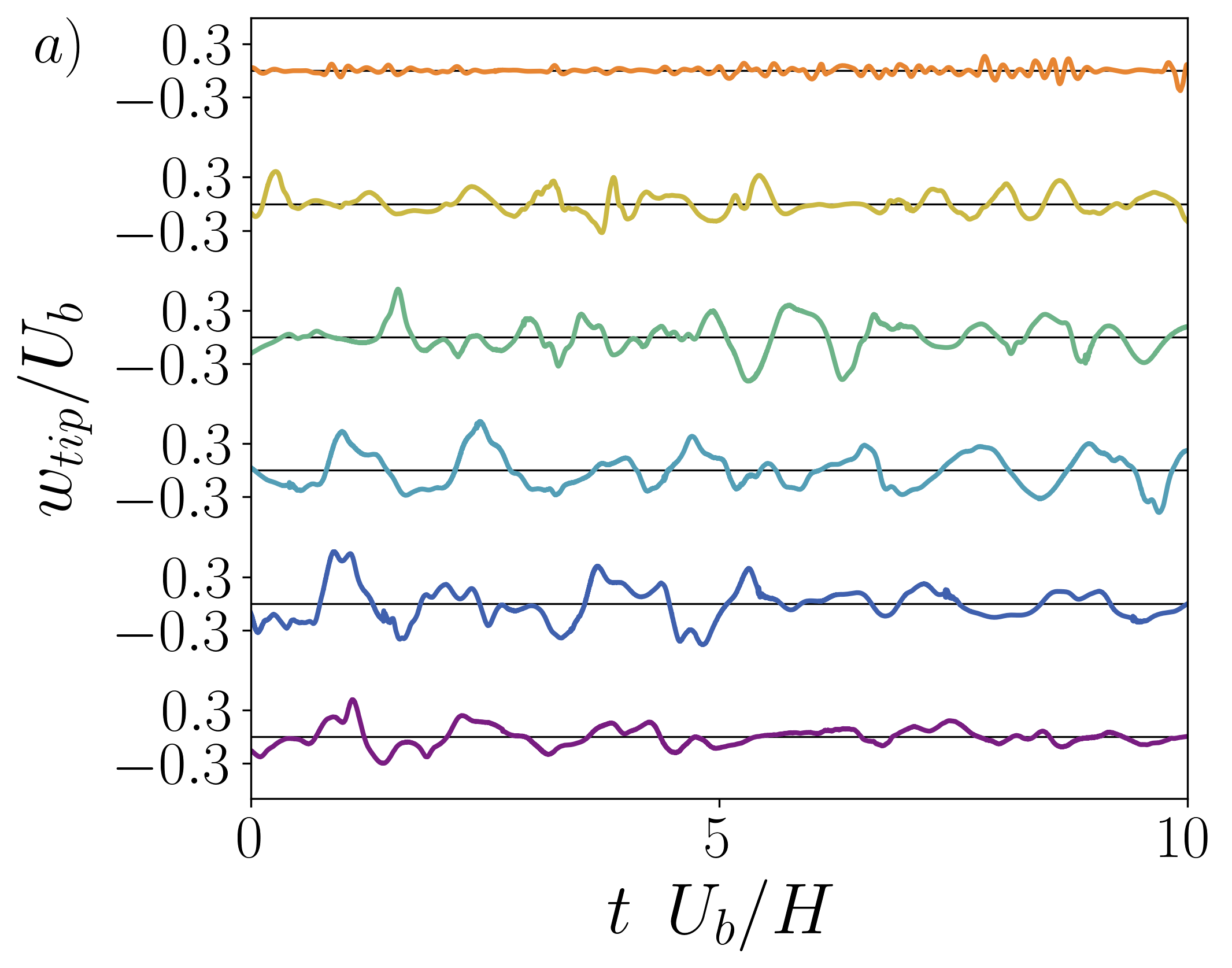}
   \hfill
   \includegraphics[width=0.48\textwidth]{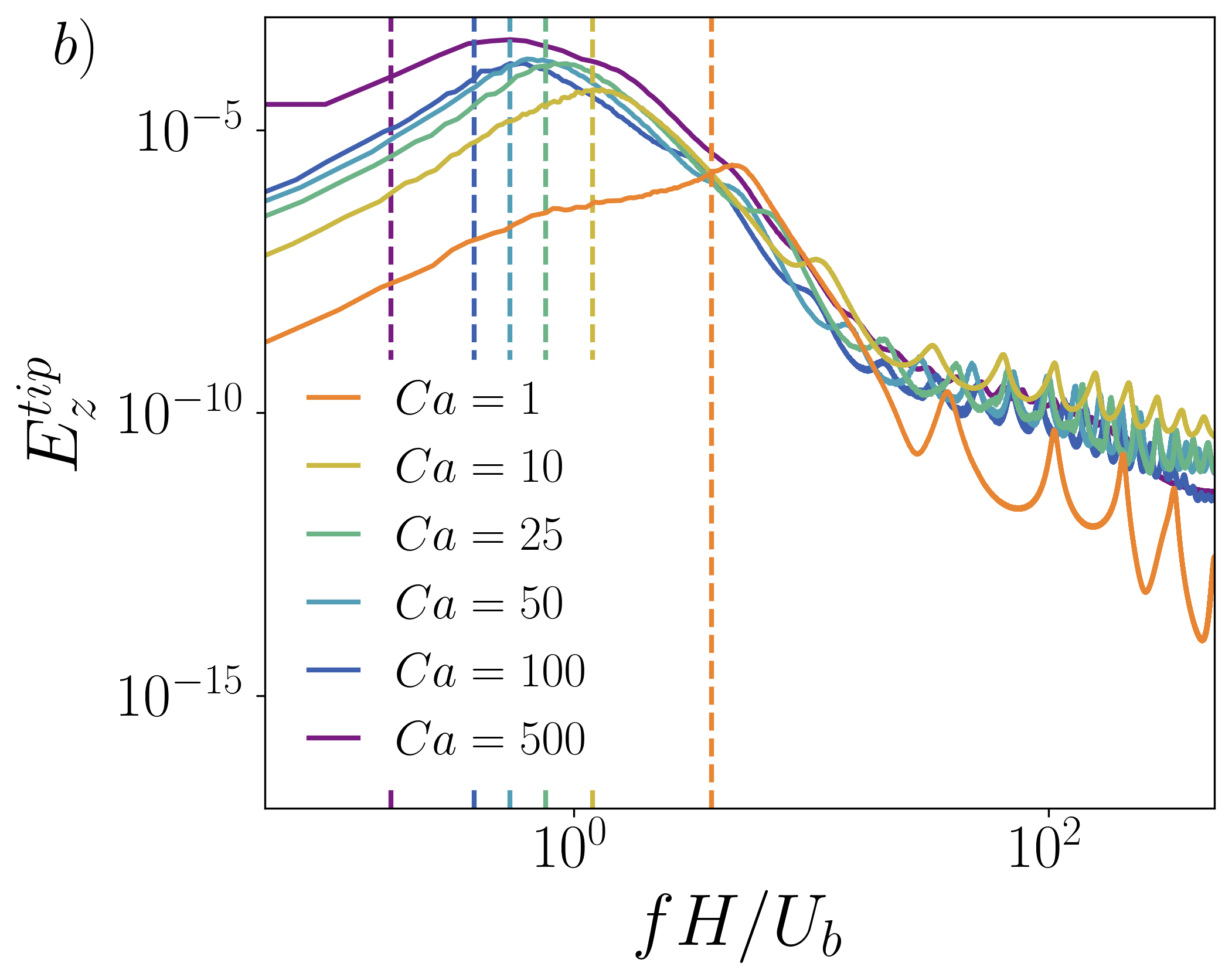}
   \caption{(panel $a$) Lagrangian velocity of the filament tips in the spanwise direction. We report the signal from one selected filament per value of $Ca$ over ten bulk time units and (panel $b$) the energy spectrum of the signal, ensemble-averaged over all the filaments at a given $Ca$. The colour-scale is the same between the two panels.}
   \label{fig:tipMotion}
\end{figure}

Multiple quantities can be observed in order to characterise the motion of the flexible filaments constituting the canopy.
Here, we focus on the spanwise velocity of the Lagrangian points at the tip of the filaments: we prefer the measurement of a velocity to that of a displacement since it is more directly related to the turbulent fluctuations of the flow, and we favour the spanwise direction as it is the one less affected by the constraints of the problem. 
The streamwise and vertical dynamics of the filaments are the most influenced by the constraints, i.e., the inextensibility and the presence of the wall. 
We therefore collect the Lagrangian velocity signal in the spanwise direction at the tip of all the filaments throughout all the values of $Ca$ considered in this study, and highlight its features both in the time and frequency domains. 
For the purpose of visualisation, we also choose one canopy filament per value of $Ca$ and show in panel $a$ of figure \ref{fig:tipMotion} the velocity signal over ten bulk time units extracted after the attainment of a fully developed flow state.
Here we also discuss the results of the simulation at $Ca=500$ to encompass the case in which the filaments are fully compliant to the flow.

We immediately observe that the amplitude of the fluctuations increases with the flexibility of the filaments up to $Ca=25\sim50$, where it appears to saturate. 
Furthermore, the small amplitude fluctuations of the most rigid case ($Ca=1$) are characterised by a nearly sinusoidal shape, modulated in amplitude by slower dynamics. 
Such behaviour contrasts with the motion of the most flexible filaments ($Ca=100\sim500$), which instead exhibit fluctuations with a less definite shape and period.
In these cases it is also possible to appreciate abrupt and fast small scale fluctuations on top of the slower dynamics: those arise from the collisions of the filament with the wall, and indeed are not observed for the lowest values of $Ca$ as the filament never gets sufficiently deflected. 
In between these two behaviours ($Ca=50$), the large amplitude fluctuations are almost sinusoidal in shape and exhibit a more definite period; they therefore appear compatible with a resonance between the natural structural response of the filaments (responsible for the nearly sinusoidal oscillations of the most rigid cases) and the turbulent fluctuations of the flow (dominating the motion of the filaments in the most flexible cases).

To confirm the picture above, we investigate the energy spectrum of the signal, $E_z^{tip}=\frac{1}{2}\overline{\overline{\mathcal{F}(w_{tip})\mathcal{F}^*(w_{tip})}}$, where $\mathcal{F}$ is the temporal Fourier's transform operator, $*$ is the conjugate of an arbitrary complex number and the double over bar denotes ensemble-averaging on all the filaments of each case.
After computing $E_z^{tip}$ over a sufficiently long time span cleaned of any initial transient, we report it in panel $b$ of figure \ref{fig:tipMotion} for the different values of $Ca$ considered.
We also mark with vertical dashed lines the natural frequencies associated to the first structural mode of the different filaments, $f_{nat}$, computed as described in \S\ref{sec:setupnmethods}.
For the lowest values of $Ca$ ($1$ to $50$), the spectrum peaks in correspondence of the natural frequency (and higher harmonics), thus confirming that the structural natural response dominates the dynamics of the most rigid filaments.
The value of $f_{nat}$ decreases with $\gamma$, hence it gets shifted towards the left for increasing values of $Ca$. 
The peak of the spectrum also obeys this shift in frequency up to $Ca=50$, while beyond that its position does not to change significantly with the $Ca$.
Such behaviour suggests that the most flexible filaments adjust to a motion no more dictated by their structural properties, but presumably compliant to the turbulent fluctuations of the flow. 
The transition between the two distinct regimes of motion occurs at $Ca\approx50$, and thus we justify the large amplitude nearly sinusoidal oscillations observed in that case with the resonance between the first structural mode of the filaments and a characteristic frequency of the turbulent forcing to which they are exposed.

The identification of two distinct regimes of motion is not unexpected: previous investigations addressing the motion of flexible fibres in homogeneous isotropic turbulence \citep{rosti-etal-2018, rosti-etal-2020c, olivieri-mazzino-rosti-2022} highlighted the existence of a structure-dominated and a turbulence-dominated regime, partially similar to what found when studying vortex induced vibrations \citep{bearman-1984}.
We recently characterised numerically the motion of a flexible fibre clamped in wall turbulence over a wide range of structural parameters \citep{foggirota-etal-2024}, also observing the emergence of the two regimes discussed above.
Recently, \citet{fu-etal-2023} experimentally investigated the individual motion of an isolated flexible plant model and how that changes when multiple plants are arranged in a canopy, while \citet{monti-olivieri-rosti-2023} studied the motion of the filaments in a flexible canopy, mostly to characterise their collective dynamics (honami/monami).
To reconcile the different insights offered by all these works, in the following section we further describe the flapping state of the individual filaments and better contextualise our observations.

\subsection{The flapping state}
\label{sec:fmap}

\begin{figure}
   \includegraphics[width=0.48\textwidth]{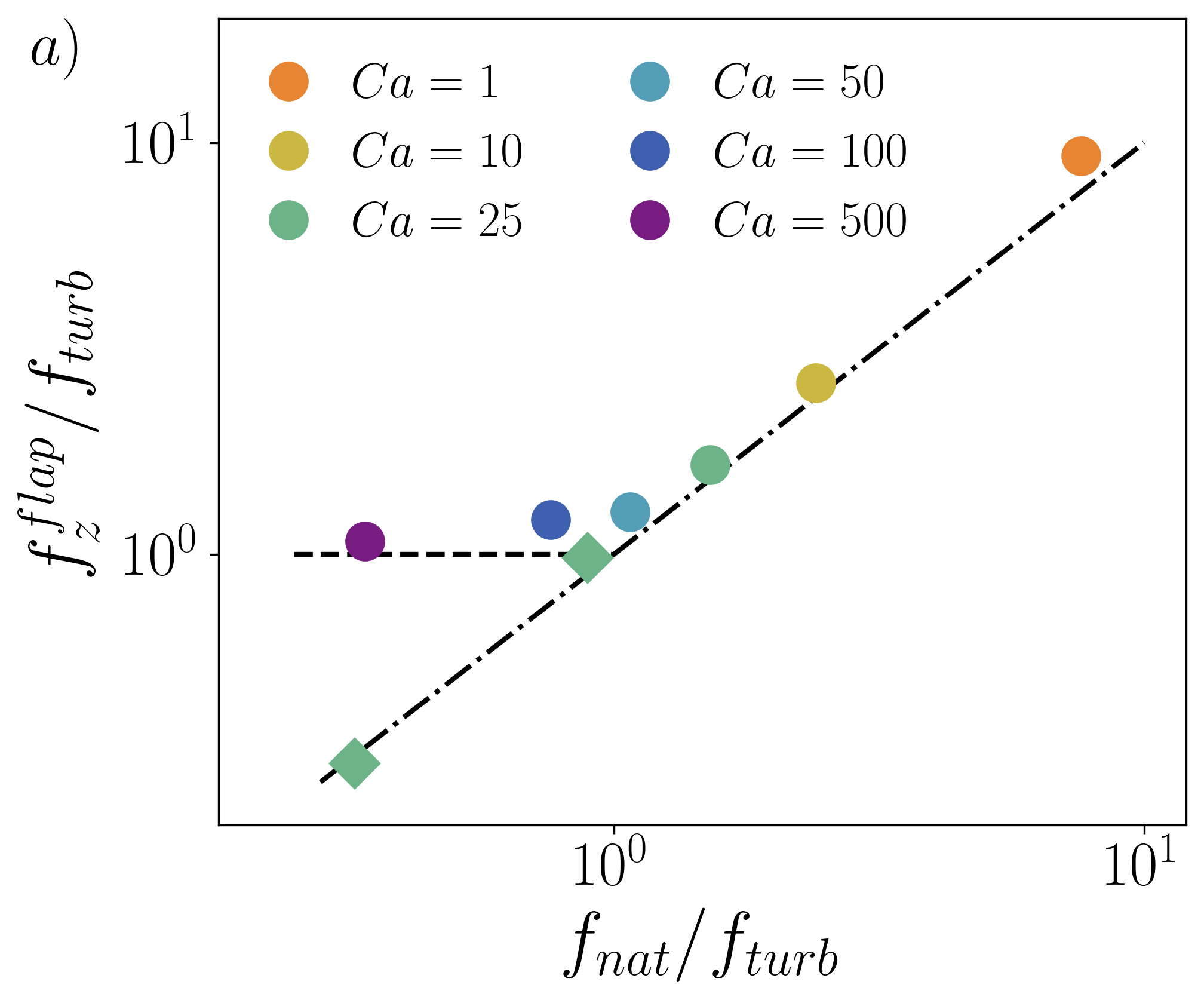}
   \hfill
   \includegraphics[width=0.48\textwidth]{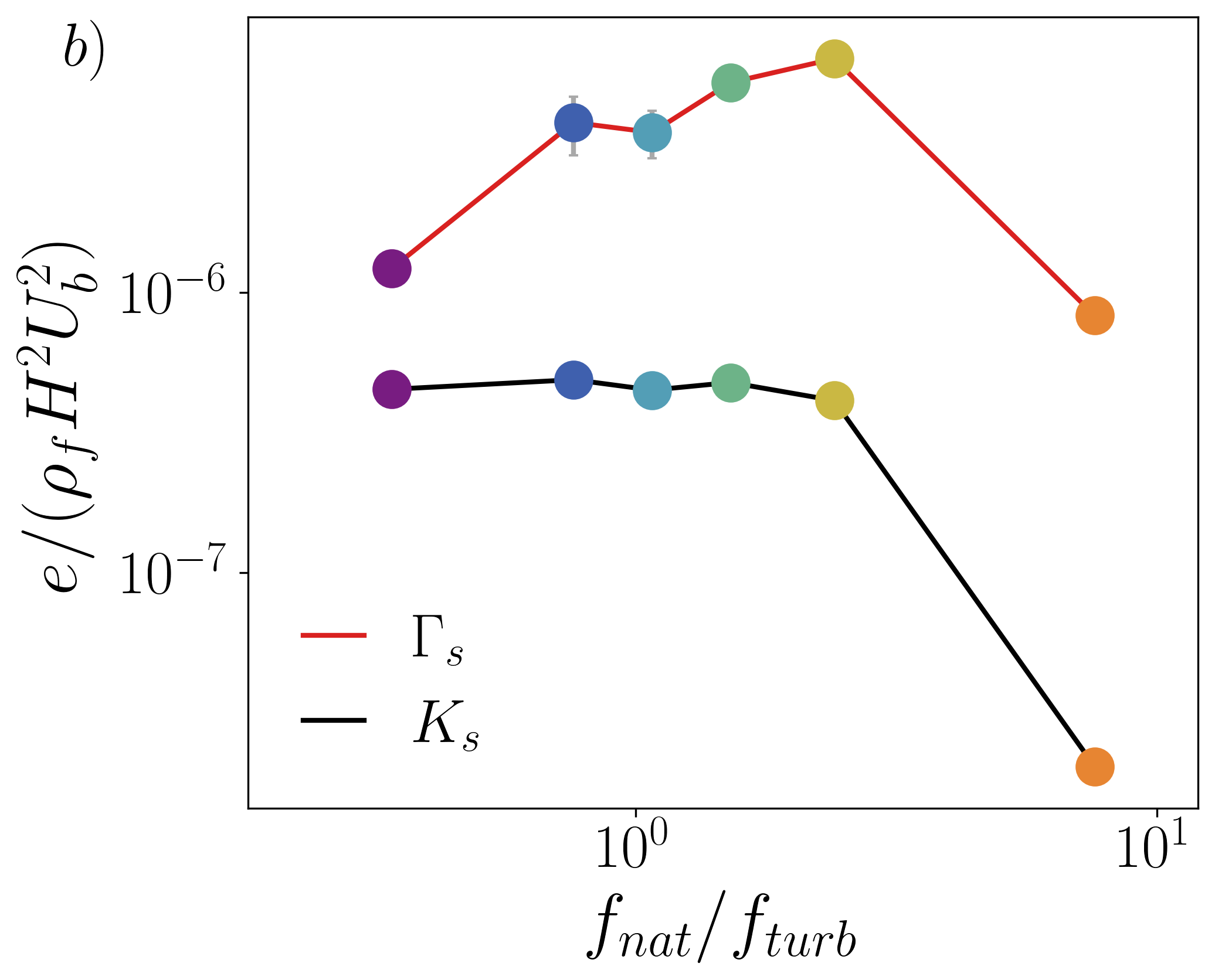}
   \caption{Flapping states of the filaments for different values of $Ca$, hence $f_{nat}/f_{turb}$. We first show (panel $a$) the dominant frequency of oscillation along the spanwise direction, $f^{flap}_z$, extracted as the peak location of the spectra in figure \ref{fig:tipMotion}. 
   Squares denote data from the cases at different density ratios.
   (panel $b$) We also report the trends of the elastic ($\Gamma_s$) and kinetic ($K_s$) energy of the filaments per unit length, averaged over time and over all the filaments. The error-bars are computed carrying out the same measurement on the time signals truncated to their first half, and  fall within the markers size for most points.}
   \label{fig:flapState}
\end{figure}

To better investigate the different dynamical regimes of the filaments, we compute their dominant flapping frequency along the spanwise direction, $f^{flap}_z$, i.e., the peak location of the spectra reported in panel $b$ of figure \ref{fig:tipMotion}, and plot it against the natural frequency associated to the corresponding value of $Ca$, i.e., with reference to section \S\ref{sec:setupnmethods}, $f_{nat}\approx\frac{3.516}{d h^2}\sqrt{ \frac{\rho_f d h^3 U_b^2}{2 Ca \rho_s \pi^3}}\approx 0.45 \frac{U_b}{h}\sqrt{\frac{\rho_f}{\rho_s}\frac{h}{d}\frac{1}{Ca}}$ .
Since for the highest values of $Ca$ the position of the spectral peak does not appear to undergo any significant change with the structural parameters of the filaments, we denote it as $f_{turb}$ (i.e., the characteristic flapping frequency of the filaments in the turbulence dominated regime) and employ it to adimensionalise both $f^{flap}_z$ and $f_{turb}$.
The outcome is reported in panel $a$ of figure \ref{fig:flapState}, where flapping frequencies close to the natural one lay on the dashed-dotted line with unitary slope, $f^{flap}_z/f_{turb}=f_{nat}/f_{turb}$, while those close to the chosen value of $f_{turb}$ approach the dashed horizontal asymptote.
We thus confirm that the most rigid filaments exhibit their natural response, while the most flexible ones deviate from it and approach $f^{flap}_z\approx f_{turb} = 0.5 U_b/H$.
Such frequency, independent from the structural characteristics of the filaments, is consistent with the outcome of our previous investigation \citep{foggirota-etal-2024} focused on the flapping states of an isolated flexible fibre clamped in wall turbulence. 
Also in that case indeed, we identified a turbulence-dominated regime for the highest values of $Ca$ where the fibre is compliant to the flow and its motion is characterised by a slow flapping with a period comparable to the bulk time-scale of the flow. 
Furthermore, despite the absence of a physical justification for  $f^{flap}_z\approx 0.5 U_b/H$, we observe that its occurrence in both the motion of an isolated flexible fibre in a full-channel flow and in the motion of the canopy filaments in the the half-channel considered here supports the reproducibility and generality of the result.

In order to assess the effect of an increase in the structural density of the filaments on their dynamics, we also report in the map of figure \ref{fig:flapState}, panel $a$, the results of the cases at ${\rho_s}/{\rho_f}=\{1+1.46\cdot10^{-2},1+1.46\cdot10^{-1}\}$, respectively associated to a higher and a lower value of the natural frequency.
Differently from what found in our previous investigation of a single filament \citep{foggirota-etal-2024}, here an increase in ${\rho_s}/{\rho_f}$ prevents the filaments from flapping at $f_{turb}$ and they thus exhibit their natural response. 
This difference originates from the fact that denser filaments yield a stronger obstruction of the mean flow in the case of a canopy (as elucidated in \S\ref{sec:drho}), and consequently get less deflected and maintain an almost upright configuration, swaying at $f_{nat}$. 
This effect is lost in the case of an isolated fibre, as the back-reaction on the mean flow is negligible; in this isolated case the fibre is less shielded and consequently gets more deflected by the flow, swaying with it. 

Interestingly, also the experimental investigation of \cite{fu-etal-2023} addresses the dynamical response of a synthetic plant submerged in a turbulent flow, comparing the isolated case to the one in which the plant is part of a vegetation patch (i.e., a canopy). There, a simplified plant model constituted by a series of five wooden buoyant pellets connected by a fine rope is exposed to a turbulent channel flow at different Reynolds numbers, thus focusing on a different scenario from that explored here. 
The synthetic plant used in the experiments, in facts, appears hinged to the bottom wall rather than clamped, and it does not have homogeneous structural properties like our filaments.
Consequently, while the motion of our filaments is dominated by a single dynamics even (and particularly) in the highest range of $f^{flap}_z$, the plant of \cite{fu-etal-2023} alternates phases of swaying at a lower frequency synchronous mode to phases at a higher frequency asynchronous one, both of them laying far above the range of $f^{flap}_z$ we considered here. 

The effect of the clamp on the dynamics of the filaments is further understood observing the variation of their elastic energy with $Ca$, while their motion is better described in terms of kinetic energy.
To those ends we therefore consider the weak form of equations \ref{eq:eulerBernoulli} and \ref{eq:inextensibility}, integrated over a time-span $\tilde{t}$ and over the length of the filaments,
\begin{equation}
	\displaystyle\int_{\tilde{t}} \displaystyle\int_h \left[ \Delta \tilde{\rho} \frac{\partial \mathbf{X}}{\partial t} \cdot \frac{\partial \mathbf{X}}{\partial t} + \gamma \frac{\partial^2 \mathbf{X}}{\partial s^2} \cdot \frac{\partial^2 \mathbf{X}}{\partial s^2} -T + \mathbf{F} \cdot \mathbf{X} \right] \mathrm{d}s \mathrm{d}t = 0,
	\label{eq:weakForm}
\end{equation}
which can be interpreted as a balance equation for the structural energy under the assumption of a beam configuration $\mathbf{X}$ compatible with the constraints.
In particular,
\begin{equation}
	\Gamma_s=\frac{1}{n_x n_z}\frac{1}{\tilde{t}}\frac{1}{h}\displaystyle\sum_{n_x n_z} \displaystyle\int_{\tilde{t}} \displaystyle\int_h \gamma \frac{\partial^2 \mathbf{X}}{\partial s^2} \cdot \frac{\partial^2 \mathbf{X}}{\partial s^2} \mathrm{d}s \mathrm{d}t,
	\label{eq:elEng}
\end{equation}
represents the structural elastic energy per unit length, averaged over time and over all the filaments, while
\begin{equation}
	K_s=\frac{1}{n_x n_z}\frac{1}{\tilde{t}}\frac{1}{h}\displaystyle\sum_{n_x n_z} \displaystyle\int_{\tilde{t}} \displaystyle\int_h \Delta \tilde{\rho} \frac{\partial \mathbf{X}}{\partial t} \cdot \frac{\partial \mathbf{X}}{\partial t} \mathrm{d}s \mathrm{d}t,
	\label{eq:kinEng}
\end{equation}
represents the structural kinetic energy per unit length, likewise averaged.
We report the trends of $\Gamma_s$ and $K_s$ with $f_{nat}/f_{turb}$, and hence with the $Ca$, in panel $b$ of figure \ref{fig:flapState}.
While the magnitude of $\Gamma_s$ is comparable to that measured by \cite{rosti-etal-2018}, the position of its peak contrasts with their findings. 
Here, in facts, $\Gamma_s$ peaks well within the regime dominated by the natural response of the filaments, confirming what can also be observed in the case of an isolated clamped fibre. 
We therefore deduce that the clamp to the wall, constraining the shape of the filaments, hinders their deflection and shifts the maximum of $\Gamma_s$ towards higher values of $\gamma$.
The kinetic energy, instead, follows a trend consistent with the amplitude of the fluctuations in figure \ref{fig:tipMotion}, panel $a$, increasing with the $Ca$ in the most rigid cases and reaching a plateau close to the resonance between $f_{nat}$ and $f_{turb}$.

Characterising the dominant dynamics of our canopy filaments, we have been able to compare our results with those attained considering a single fibre clamped in wall turbulence \citep{foggirota-etal-2024} and plants with a different geometry \citep{fu-etal-2023}. 
We have also isolated the role played by the sheltering effect of the canopy upon increasing the density of the filaments, and by the model adopted to describe them.
Nevertheless, we emphasise that the present investigation is unable to capture any sub-dominant dynamics of the filaments, such as the contribution of turbulence to their motion in the regime dominated by the natural response. 
Regardless of the value of $Ca$, in facts, the coherent motion of the filaments (honami/monami) investigated by \cite{monti-olivieri-rosti-2023} remains dictated by the turbulent structures in the flow.

\section{Conclusions}
\label{sec:conclusions}

Our numerical study systematically characterises both the turbulent flow within and above a flexible canopy and the dynamical response of its constitutive elements across different values of the filament flexibility. 
Building on top of the database generated by \cite{monti-olivieri-rosti-2023} (constituted by the first fully resolved simulations of a flexible canopy), we have considered a turbulent half-channel flow at a bulk Reynolds number of about $5000$, populated by $15552$ individually resolved flexible filaments vertically clamped to the bottom wall. 
Their dynamical response to the turbulent forcing, described by an extended version of the Euler-Bernoulli beam model \citep{yu-2005,huang-etal-2007}, is controlled by their flexibility, which we have varied over two orders of magnitude.

Filaments of increasing flexibility are more deflected by the flow, thus reducing the frontal area of the canopy, and they yield a weaker drag discontinuity at the canopy tip, associated to velocity fluctuations less intense then in the rigid case.  
Consequently, the canopy drag decreases increasing the flexibility up to saturation, attained when the fully deflected filaments start piling on top of each other parallel to the wall. 
Our data compare well to the experimental measurements of \cite{shimizu-etal-1992} and \cite{ ghisalberti-nepf-2006}.
The spectra of the turbulent kinetic energy within the canopy confirm the spectral short cut mechanism described by \cite{finnigan-2000} and \cite{olivieri-etal-2020-2}, while a more regular energy cascade is attained above the canopy tip.
Furthermore, the intensity of the velocity fluctuations within the canopy increases with its flexibility due to the increased motion of the filaments.
Outside the canopy, instead, the opposite trend is attained by virtue of the generation of a stronger shear layer in the more rigid cases. 

The characterisation of the turbulent state at different distances from the bottom wall suggests a multi-layer approach to turbulence modelling in canopy flows.
Within the canopy, turbulence is quasi 2D close to the bottom wall and it approaches an isotropic state between the inner inflection point and the virtual origin of the outer flow.
The picture becomes more blurred increasing the flexibility, as the flow in the canopy turns less isotropic. 
A strongly anisotropic state is attained approaching the canopy tip, and the flow exhibits a behaviour consistent with what observed in conventional channel flows moving above it, thus supporting outer similarity arguments. 
Turbulence inside the canopy appears to be sustained by intense events generated in the outer flow, induced by the structures populating the close proximity of the shear layer.
Those structures cause sweeps and ejections dominating the interaction between the inner and the outer flows, but only the strongest sweep events are able to penetrate the canopy due to the obstruction exerted by the deflected filaments.
The effect increases with the flexibility, yielding a more weakly driven turbulent state within the most compliant canopies even though the filaments oscillate more.
On the other hand, ejections are less obstructed by the filaments and thus dominate the inner-outer flow interactions.

For the sake of completeness we have also investigated the consequences of an increase in the filament density, which yields effects similar to those of a reduction in the flexibility. 
The blockage of the mean flow is in facts enhanced, while intense sweep events become more likely. 

In order to isolate the effects of the filament flapping on turbulence, we have ``frozen" them in one of their instantaneously deflected configurations and allowed the flow to develop.
The turbulent shear is depleted with respect to the case in which the filaments are free to flap, since less energy is ``pumped" into the turbulent fluctuations by their motion. 
Furthermore, the ``frozen" canopies also prove more effective in shielding the inner flow. 
Consequently, throughout all cases, the drag is reduced with respect to the flexible ones upon ``freezing" the canopy, and the maximum reduction is achieved for intermediate values of the flexibility, where the filaments exhibit large amplitude, regular fluctuations. 
Consistently with our previous work on the flapping states of a clamped flexible fiber in wall turbulence \citep{foggirota-etal-2024}, also here we observe two regimes of motion for the filaments: one dominated by their structural natural response and one by turbulence.
At the transition between the two, resonance occurs. 
Furthermore, in the turbulence dominated regime, all the filaments exhibit a dominant flapping frequency of abut $0.5U_b/H$ in the spanwise direction regardless of their structural parameters, once again supporting the outcome of our precursory study. 
Nevertheless, here, an increase in the density of the filaments drives them to exhibit their natural response since they better shield each other and get less deflected by the mean flow. 
Separately observing the terms of the structural energy balance we also notice that the elastic energy peaks well within the regime dominated by the natural response, differently from the case of free fibers in homogeneous isotropic turbulence \citep{rosti-etal-2018}, where the peak is found at the transition between the two regimes.
Furthermore, here, the kinetic energy saturates as soon as the filaments approach the turbulence dominated regime.  

Canopy flows are complex systems governed by a multitude of parameters responsible for coupled and non-trivial effects.
The effort to combine them together into dimensionless quantities supposedly descriptive of specific behaviours is therefore not always successful, as the individual dependencies of the results from the parameters combined in dimensionless quantities can be retained. 
This is the case, for example, of the solidity parameter \citep{monti-etal-2020,nicholas-etal-2023}, which well characterises canopies in the sparse regime, but proves unsatisfactory to describe the individual effects of the filament spacing and length in the dense regime.
To avoid such kind of issues, in our investigation we have have chosen to keep all the parameters fixed and equal to those employed in previous investigations \citep{monti-etal-2022}, but for the flexibility of the filaments, to assess its individual role. 
Proceeding in this way we have nevertheless missed the effect of a variation in the flexibility combined with any other parameter (e.g., the Reynolds number, the filament length) along with the geometry of the canopy elements adopted.
We have in facts noticed that our flexible filaments, characterised by homogeneous structural properties along their axis, exhibit a dynamical response contrasting with that of different plant models \citep{fu-etal-2023}.
Both issues can be mitigated broadening the range of parameters considered, but such analysis currently appears challenging from a computational standpoint.
Instead, it is of particular interest to investigate the effect of the motion of the filaments on turbulent mixing \citep{ghisalberti-2010, wang-etal-2023}, assessing up to what extent that is enhanced by the flexibility.
Our results provide solid grounds for such study, which we leave for a future investigation, and pave the way towards the simulation of realistic scenarios constituted by more accurate plant models. 
On the other hand, we also offer an accurate description of the flow within and above the flexible canopy, which can prove helpful in the development of accurate turbulence models for RANS and LES simulations, without the need to resolve the flow up to the filament scale.

\backsection[Acknowledgements]{
The research was supported by the Okinawa Institute of Science and Technology Graduate University (OIST) with subsidy funding from the Cabinet Office, Government of Japan. 
The authors acknowledge the computer time provided by the Scientific Computing section of the Core Facilities at OIST, and by the HPCI System Research Project with grants hp210025 and hp220402.
}

\backsection[Declaration of interests]{The authors report no conflict of interest.}

\backsection[Data availability statement]{The plotted data are available at \url{https://groups.oist.jp/cffu/foggirota2024jfluidmech}.}

\backsection[Author ORCIDs]{\\
Giulio Foggi Rota \orcidA{}: \url{https://orcid.org/0000-0002-4361-6521} \\
Alessandro Monti \orcidB{}: \url{https://orcid.org/0000-0003-2231-2796} \\
Stefano Olivieri \orcidC{}: \url{https://orcid.org/0000-0002-7795-6620} \\
Marco Edoardo Rosti \orcidD{}: \url{https://orcid.org/0000-0002-9004-2292}
}

\appendix
\section{Flow visualisations}
\label{app:flowVis}

\begin{figure}
\centering
\includegraphics[width=.98\textwidth]{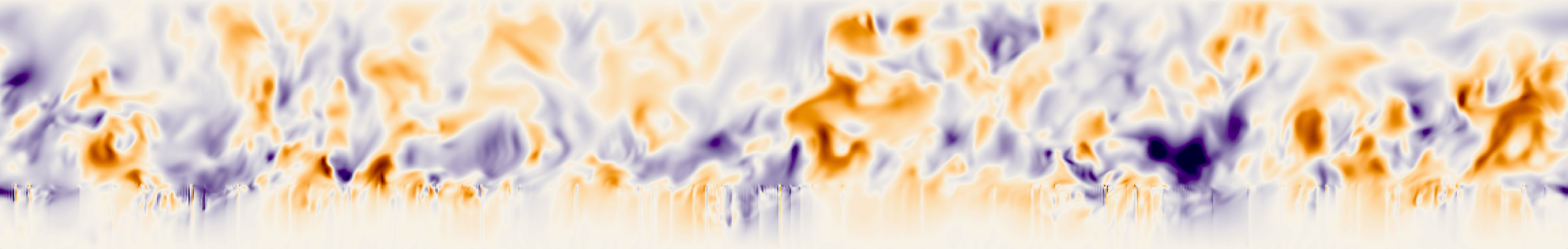}
\put(-370,50){a)}
\vspace{.03cm}
\includegraphics[width=.98\textwidth]{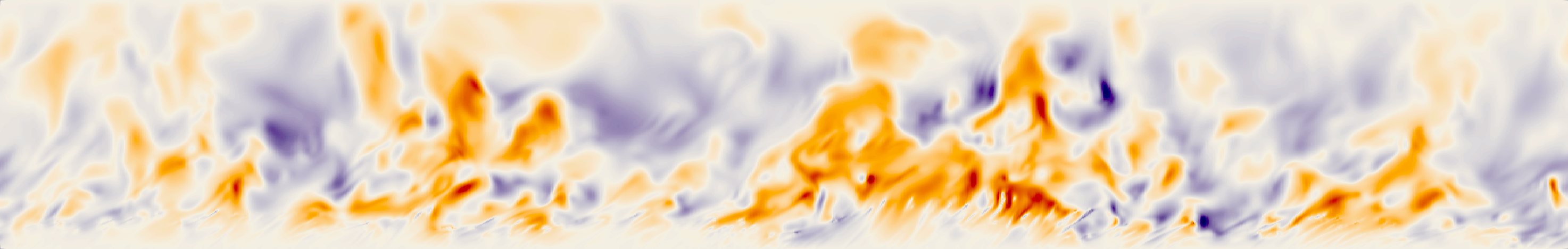}
\put(-370,50){b)}
\caption{We report vertical slices of the fluctuating wall-normal velocity in and above a rigid (panel $a$) and a flexible (panel $b$) canopy with $Ca=100$, with the mean flow directed from left to right. A colour scale going from violet to orange is adopted, ranging in $[-0.6,0.6] U_b$.}
\label{fig:vis2}
\end{figure}

\begin{figure}
\centering
\includegraphics[width=.98\textwidth]{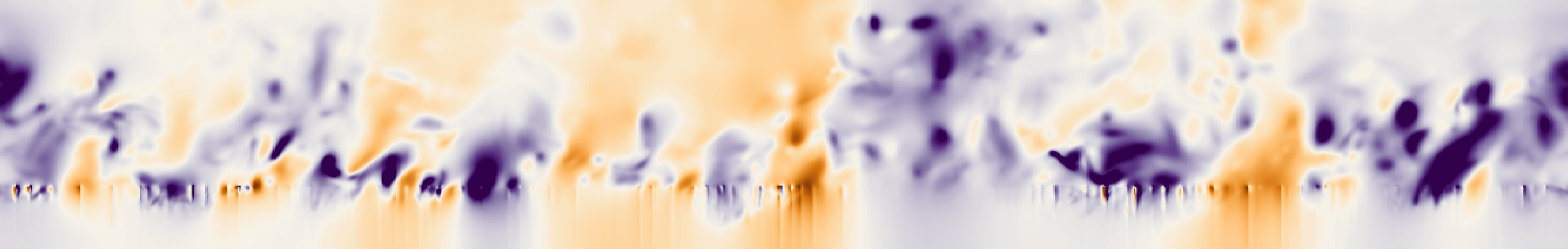}
\put(-370,50){a)}
\vspace{.03cm}
\includegraphics[width=.98\textwidth]{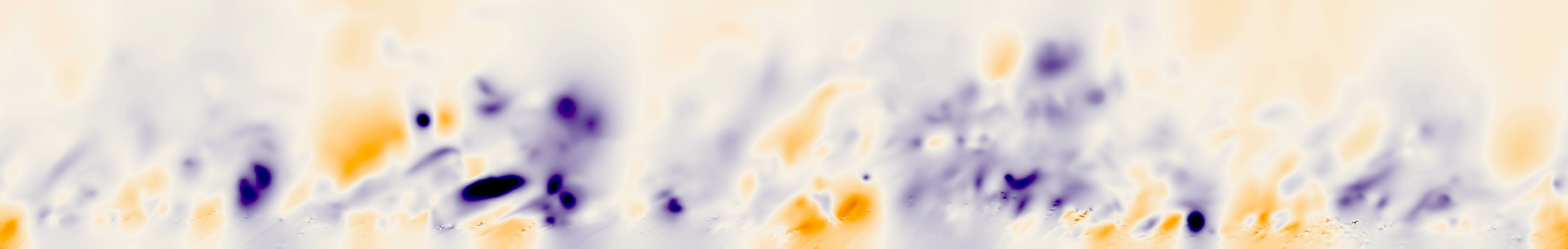}
\put(-370,50){b)}
\caption{We report vertical slices of the fluctuating pressure in and above a rigid (panel $a$) and a flexible (panel $b$) canopy with $Ca=100$, with the mean flow directed from left to right. A colour scale going from violet to orange is adopted, ranging in $[-0.2,0.2] \rho U_b^2$.}
\label{fig:vis3}
\end{figure}

\begin{figure}
\centering
\includegraphics[width=.98\textwidth]{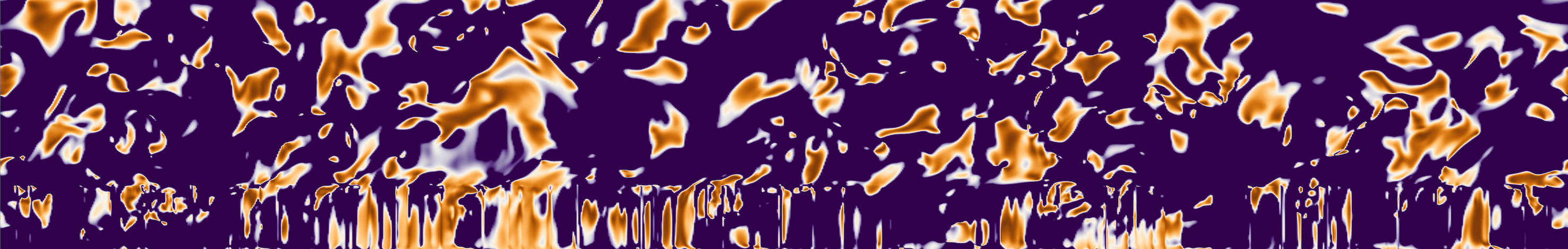}
\put(-370,50){\colorbox{white!30}{a)}}
\vspace{.03cm}
\includegraphics[width=.98\textwidth]{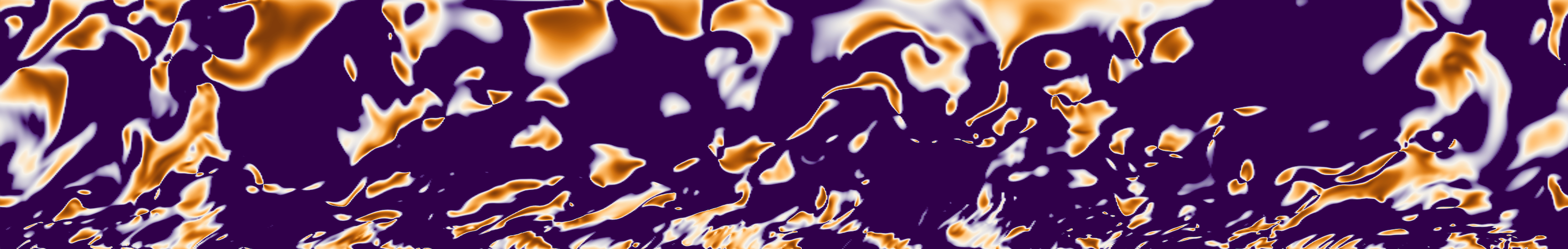}
\put(-370,50){\colorbox{white!30}{b)}}
\caption{We report vertical slices of the instantaneous flow anisotropy in and above a rigid (panel $a$) and a flexible (panel $b$) canopy with $Ca=100$, with the mean flow directed from left to right. A colour scale going from violet to orange is adopted, ranging in $[0.0,0.58]$.}
\label{fig:vis4}
\end{figure}

We devote this appendix to the visualisation of the instantaneous flows fields.
Panel $a$ of each figure reports a slice in the $x-y$ plane sampled from the rigid canopy case, while panel $b$ shows a similar slice from the flexible canopy case, at $Ca=100$.
The flow is oriented from left to right. 
We display the instantaneous values of $u'$ (figure \ref{fig:vis1}), $v'$ (figure \ref{fig:vis2}) and $p'$ (figure \ref{fig:vis3}): they indicate a less strongly driven turbulent state in the flexible canopy case, with enhanced spacial coherence compared to the rigid one. 
Furthermore, areas of markedly low pressure can be seen at the canopy tip, denoting the cores of the spanwise rollers populating that region. 
In figure \ref{fig:vis4} we also visualise the instantaneous flow anisotropy.
On average, the flow above the rigid canopy appears more anisotropic than that above the flexible one, while the opposite is true within the canopy.
Such scenario is consistent with the plots reported in figure \ref{fig:lmt}.

\begin{bottomfigure}
\centering
\includegraphics[width=.98\textwidth]{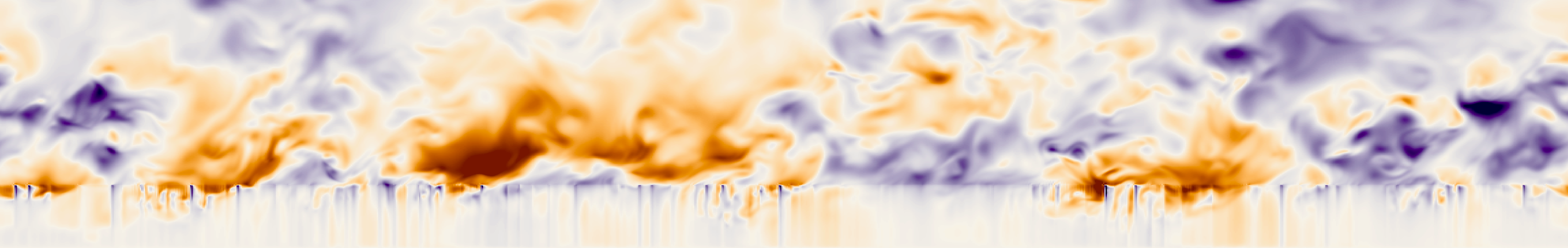}
\put(-370,50){a)}
\vspace{.03cm}
\includegraphics[width=.98\textwidth]{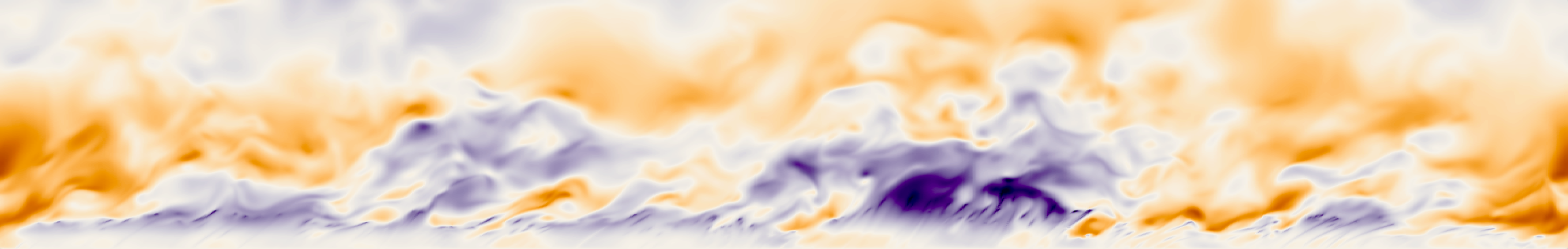}
\put(-370,50){b)}
\caption{We report vertical slices of the fluctuating streamwise velocity in and above a rigid (panel $a$) and a flexible (panel $b$) canopy with $Ca=100$, with the mean flow directed from left to right. A colour scale going from violet to orange is adopted, ranging in $[-0.8,0.8] U_b$.}
\label{fig:vis1}
\end{bottomfigure}

\clearpage

\bibliographystyle{jfm}
\bibliography{./Wallturb.bib}

\end{document}